\documentclass[twocolumn,tighten,times]{aastex631}

\usepackage{amsfonts}
\usepackage{mathrsfs}
\usepackage{bm}
\usepackage{lipsum}
\usepackage{physics}
\usepackage{multirow}
\usepackage{rotating}
\usepackage{graphicx} 
\usepackage{soul}
\usepackage{subfigure}
\usepackage{color}

\newcommand{\feii}{\ifmmode {\rm Fe\ II} \else Fe~{\sc ii}\fi}
\newcommand{\heii}{\ifmmode {\rm He\ II} \else He~{\sc ii}\fi}
\newcommand{\hei}{\ifmmode {\rm He\ I} \else He~{\sc i}\fi}
\newcommand{\oiii}{\ifmmode {\rm~[O\ III]} \else [O~{\sc iii}]\fi}
\newcommand{\nii}{\ifmmode {\rm~[N\ II]} \else [N~{\sc ii}]\fi}
\newcommand{\mgii}{\ifmmode {\rm~Mg\ II} \else Mg~{\sc ii}\fi}
\newcommand{\ciii}{\ifmmode {\rm~C\ III]} \else C~{\sc iii}]\fi}
\newcommand{\civ}{\ifmmode {\rm~C\ IV} \else C~{\sc iv}\fi}
\newcommand{\hb}{\rm{H$\beta$}}
\newcommand{\ha}{\rm{H$\alpha$}}

\newcommand{\dotm}{\ifmmode {\dot{\mathscr{M}}} \else $\dot{\mathscr{M}}$\fi}

\newcommand{\mbh}{\rm{$M_{\rm{BH}}$}}
\newcommand{\hc}{\rm{H$\gamma$}}
\newcommand{\fvar}{$F_{\rm var}$}
\newcommand{\fab}{$F_{\rm H\alpha} / F_{\rm H\beta}$}
\newcommand{\fac}{$F_{\rm H\alpha} / F_{\rm H\gamma}$}

\def\feii{Fe\,{\sc ii}}
\def\heii{He\,{\sc ii}}
\def\hei{He\,{\sc i}}
\def\oiii{O\,{\sc iii}}
\def\oi{O\,{\sc i}}
\def\sii{S\,{\sc ii}}
\def\nii{N\,{\sc ii}}
\def\nai{Na\,{\sc i}}

\shortauthors{Li et al.}

\begin{document}

\title{\bf Velocity-Resolved Ionization Mapping of Broad Line Region. I. Insights into Diverse Geometry and  Kinematics}

\author[0000-0003-3823-3419]{Sha-Sha Li}
\affiliation{Yunnan Observatories, Chinese Academy of Sciences, Kunming 650216, Yunnan, People's Republic of China}
\affiliation{Key Laboratory for the Structure and Evolution of Celestial Objects, Chinese Academy of Sciences, Kunming 650216, Yunnan, People's Republic of China}
\affiliation{Center for Astronomical Mega-Science, Chinese Academy of Sciences, 20A Datun Road, Chaoyang District, Beijing 100012, People's Republic of China}
\affiliation{Key Laboratory of Radio Astronomy and Technology, Chinese Academy of Sciences, 20A Datun Road, Chaoyang District, Beijing 100101, People's Republic of China}

\author[0000-0002-1530-2680]{Hai-Cheng Feng}
\affiliation{Yunnan Observatories, Chinese Academy of Sciences, Kunming 650216, Yunnan, People's Republic of China}
\affiliation{Key Laboratory for the Structure and Evolution of Celestial Objects, Chinese Academy of Sciences, Kunming 650216, Yunnan, People's Republic of China}
\affiliation{Center for Astronomical Mega-Science, Chinese Academy of Sciences, 20A Datun Road, Chaoyang District, Beijing 100012, People's Republic of China}
\affiliation{Key Laboratory of Radio Astronomy and Technology, Chinese Academy of Sciences, 20A Datun Road, Chaoyang District, Beijing 100101, People's Republic of China}

\author[0000-0002-2153-3688]{H. T. Liu}
\affiliation{Yunnan Observatories, Chinese Academy of Sciences, Kunming 650216, Yunnan, People's Republic of China}
\affiliation{Key Laboratory for the Structure and Evolution of Celestial Objects, Chinese Academy of Sciences, Kunming 650216, Yunnan, People's Republic of China}
\affiliation{Center for Astronomical Mega-Science, Chinese Academy of Sciences, 20A Datun Road, Chaoyang District, Beijing 100012, People's Republic of China}

\author{J. M. Bai}
\affiliation{Yunnan Observatories, Chinese Academy of Sciences, Kunming 650216, Yunnan, People's Republic of China}
\affiliation{Key Laboratory for the Structure and Evolution of Celestial Objects, Chinese Academy of Sciences, Kunming 650216, Yunnan, People's Republic of China}
\affiliation{Center for Astronomical Mega-Science, Chinese Academy of Sciences, 20A Datun Road, Chaoyang District, Beijing 100012, People's Republic of China}
\affiliation{Key Laboratory of Radio Astronomy and Technology, Chinese Academy of Sciences, 20A Datun Road, Chaoyang District, Beijing 100101, People's Republic of China}

\author[0000-0002-0539-8244]{Xiang Ji}
\affiliation{Shanghai Astronomical Observatory, Chinese Academy of Sciences, 80 Nandan Road, Shanghai 200030, People's Republic of China}

\author{Yu-Xuan Pang}
\affiliation{Department of Astronomy, School of Physic, Peking University, Beijing 100871, People's Republic of China}\affiliation{Kavli Institute for Astronomy and Astrophysics, Peking University, Beijing 100871, People's Republic of China}

\author[0000-0003-0202-0534]{Cheng Cheng}
\affiliation{Chinese Academy of Sciences South America Center for Astronomy, National Astronomical Observatories, Chinese Academy of Sciences, Beijing 100101, People's Republic of China}
\affiliation{CAS Key Laboratory of Optical Astronomy, National Astronomical Observatories, Chinese Academy of Sciences, Beijing 100101, People's Republic of China}

\author[0000-0002-2310-0982]{Kai-Xing Lu}
\affiliation{Yunnan Observatories, Chinese Academy of Sciences, Kunming 650216, Yunnan, People's Republic of China}
\affiliation{Key Laboratory for the Structure and Evolution of Celestial Objects, Chinese Academy of Sciences, Kunming 650216, Yunnan, People's Republic of China}
\affiliation{Center for Astronomical Mega-Science, Chinese Academy of Sciences, 20A Datun Road, Chaoyang District, Beijing 100012, People's Republic of China}

\author[0000-0003-4156-3793]{Jian-Guo Wang}
\affiliation{Yunnan Observatories, Chinese Academy of Sciences, Kunming 650216, Yunnan, People's Republic of China}
\affiliation{Key Laboratory for the Structure and Evolution of Celestial Objects, Chinese Academy of Sciences, Kunming 650216, Yunnan, People's Republic of China}
\affiliation{Center for Astronomical Mega-Science, Chinese Academy of Sciences, 20A Datun Road, Chaoyang District, Beijing 100012, People's Republic of China}

\author[0000-0002-3490-4089]{Rui Li}
\affiliation{School of Physics and Microelectronics, Zhengzhou University, Zhengzhou 450001, People's Republic of China}

\correspondingauthor{Hai-Cheng Feng \& H. T. Liu}
\email{hcfeng@ynao.ac.cn} \email{htliu@ynao.ac.cn}

\begin{abstract}
Broad emission lines of active galactic nuclei (AGNs) originate from the broad-line region (BLR), consisting of dense gas clouds in orbit around an accreting supermassive black hole. Understanding the geometry and kinematics of the region is crucial for gaining insights into the physics and evolution of AGNs. Conventional velocity-resolved reverberation mapping may face challenges in disentangling the degeneracy between intricate motion and geometry of this region. To address this challenge, new key constraints are required. Here, we report the discovery of an asymmetric BLR using a novel technique: velocity-resolved ionization mapping, which can map the distance of emitting gas clouds by measuring Hydrogen line ratios at different velocities. By analyzing spectroscopic monitoring data, we find that the Balmer decrement is anticorrelated with the continuum and correlated with the lags across broad emission line velocities. Some line ratio profiles deviate from the expectations for a symmetrically virialized BLR, suggesting that the red-shifted and blue-shifted gas clouds may not be equidistant from the supermassive black hole (SMBH). This asymmetric geometry might represent a formation imprint, provide new perspectives on the evolution of AGNs, and influence SMBH mass measurements.
\end{abstract}


\section{Introduction} \label{sec:intro}
Active Galactic Nuclei (AGNs) are fascinating astrophysical objects that harbor supermassive black holes (SMBHs) at their centers. Accretion of matter onto these black holes generates intense radiation across the electromagnetic spectrum \citep{Salpeter1964}. Among the multi-component structure of AGNs, the broad-line region (BLR) is a crucial component responsible for producing broad emission lines due to the high-velocity motion of gas clouds near the central black hole. Comprehending the geometry and kinematics of BLRs is essential not only for probing the physical mechanisms associated with accretion, such as inflows that fuel AGNs \citep{Grier2013, Zhou2019} or outflows related to disk winds \citep{Emmering1992, Nicastro2000, Elitzur2014}, but also for accurately measuring the masses of SMBHs, \mbh\ \citep{Pancoast2014, Mejia-Restrepo2018}. Accurate values of \mbh\ are vital for studying formation and evolution of galaxy, as well co-evolution of SMBHs and host galaxies \citep{Kormendy2013}.

In the current picture of the AGN unification scheme, the BLR is modeled as a clumpy, extended structure located between the accretion disk and dust torus \citep{Antonucci1993, Netzer2015}. Regardless of whether the structure is clumpy or smooth \citep{Laor2006}, the extended BLR geometry has been supported by a variety of observational evidence. Reverberation mapping (RM) observations, which measure the radius of the BLR based on the delayed response of emission lines to continuum variations \citep{Blandford1982}, reveal the ``breathing effect" \citep{Gilbert2003, Goad2004, Cackett2006, Wang2020}, where the luminosity-weighted radius changes in response to continuum variations. This is further supported by radial stratification of the BLR as measured by multiline RM \citep{Clavel1991, Bentz2010, Feng2021b, Lu2022}. More direct evidence comes from GRAVITY phase interferometry which shows the line-emitting gas occupying the area from the center out to the hot dust radius in three objects \citep{Gravity2018, Gravity2020, Gravity2021}. Although the overall scale of the BLR has been unambiguously confirmed, the details of geometry and kinematics remain an ongoing debate.

A number of velocity-resolved RM studies prefer symmetrical configurations of the BLR, either disk-like or spherical. Shorter delays in the line wings are typically expected from a virialized flow, while longer responses in the line wings are often attributed to the presence of additional outflows or inflows \citep{Welsh1991, Bentz2009, Grier2013, Waters2016, Lu2019, Bao2022, Villafana2022}. Nevertheless, asymmetrical geometric models are widely adopted when fitting the complex broad line profile. \citet{Chen1989a} and \citet{Chen1989b} demonstrated that the double-peaked broad emission lines could be well-fitted by a relativistic accretion disk. Subsequently, some non-axisymmetric accretion disk models, such as an elliptical disk \citep{Eracleous1995} or perturbations in the disk \citep{Newman1997, Gilbert1999, Storchi-Bergmann2003}, were proposed to explain the stronger red peak. Moreover, accretion disk winds \citep{Flohic2012}, bowl-shaped BLRs \citep{Goad2012}, and BLRs illuminated by anisotropic continuum source \citep{Wanders1995, Goad1996} can also yield double-peaked profiles, while theoretical calculations indicate that a spiral-arm geometry can generate the velocity-delay maps obtained in RM observations \citep{Du2023}. 

The broad hydrogen Balmer decrements (BDs) are widely employed to probe the internal reddening of AGNs \citep{Reynolds1997, Dong2008}, where various values can be associated with different degrees of dust extinction. This internal reddening can influence the BDs by absorbing and scattering light, thereby altering the observed BDs. On the other hand, photoionization model simulations predict that the intrinsic BD should cover a range of values within a parameter grid \citep{Korista2004, Guo2020, Wu2023}. These values are dependent on physical conditions like gas density, ionization parameter, and the spectral energy distribution of the ionizing continuum. Moreover, considering that higher-order lines are theoretically expected to originate at smaller radii on average \citep{Korista2004}, the BD may undergo changes due to reverberation effects within the spatially extended BLR.

Unfortunately, the actual dust and gas environments may vary substantially between individual AGNs, making it extremely challenging to differentiate the influences of these two physical mechanisms on the BD using statistical samples of AGNs. However, the short-term (weeks to months) variability of BD in single objects could potentially test the influence of ionizing photon intensity on intrinsic BD, considering that changes in the obscuring materials and gaseous state have much longer timescales. As the incident continuum flux declines in a power-law manner with distance for an isotropic BLR, it becomes possible to deduce a distance-dependent gas distribution through the velocity-resolved BD. This technique essentially breaks the degeneracy between geometry and kinematics according to the ionization distribution of the gas, so we named it “velocity-resolved ionization mapping (IM)”. Combining velocity-resolved IM and RM results can shed more light on the geometry and kinematics of the BLR.

Data from multiline RM campaigns, typically collected over a several-month period of spectroscopic monitoring, are ideal for the aforementioned purposes. Here, we present a joint analysis of lags and BDs as a function of line-of-sight velocity to deduce the geometry of the BLR. The paper is structured as follows. The sample is introduced in Section~\ref{sec:sample}. In Section~\ref{sec:analysis}, we present analysis processes, including spectral fitting, time series analysis, measurement of emission line flux ratios, and transfer function. The results are given in Section~\ref{sec:result}. Discussions of the results are provided in Section~\ref{sec:discu}. Section~\ref{sec:summary} is a summary. In the Appendix, we examine the impact of reverberation effects on velocity-resolved BD and the variability of dust extinction.

\begin{figure*}[!ht]
\centering 
\includegraphics[width=0.33\textwidth]{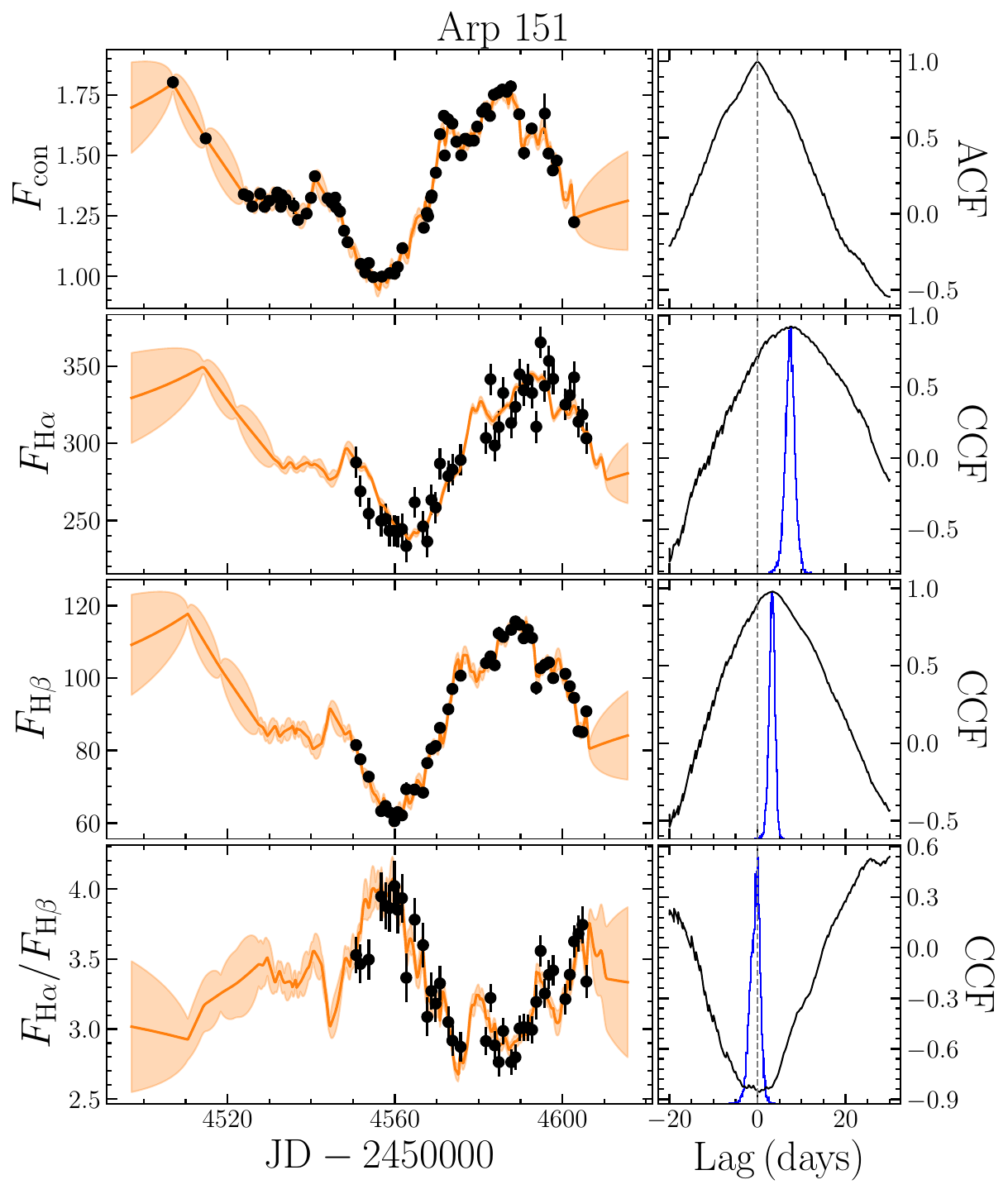}
\includegraphics[width=0.33\textwidth]{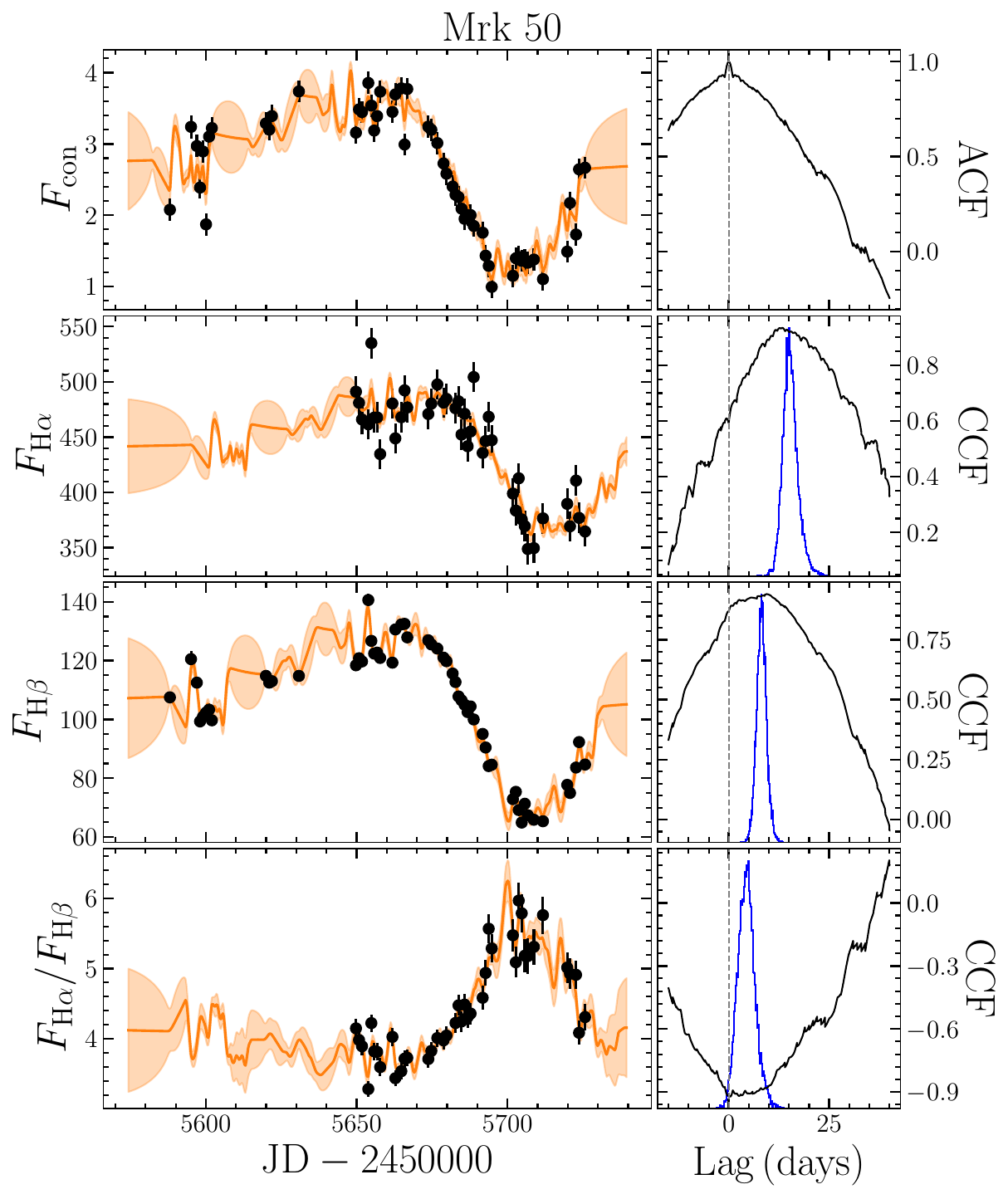}
\includegraphics[width=0.33\textwidth]{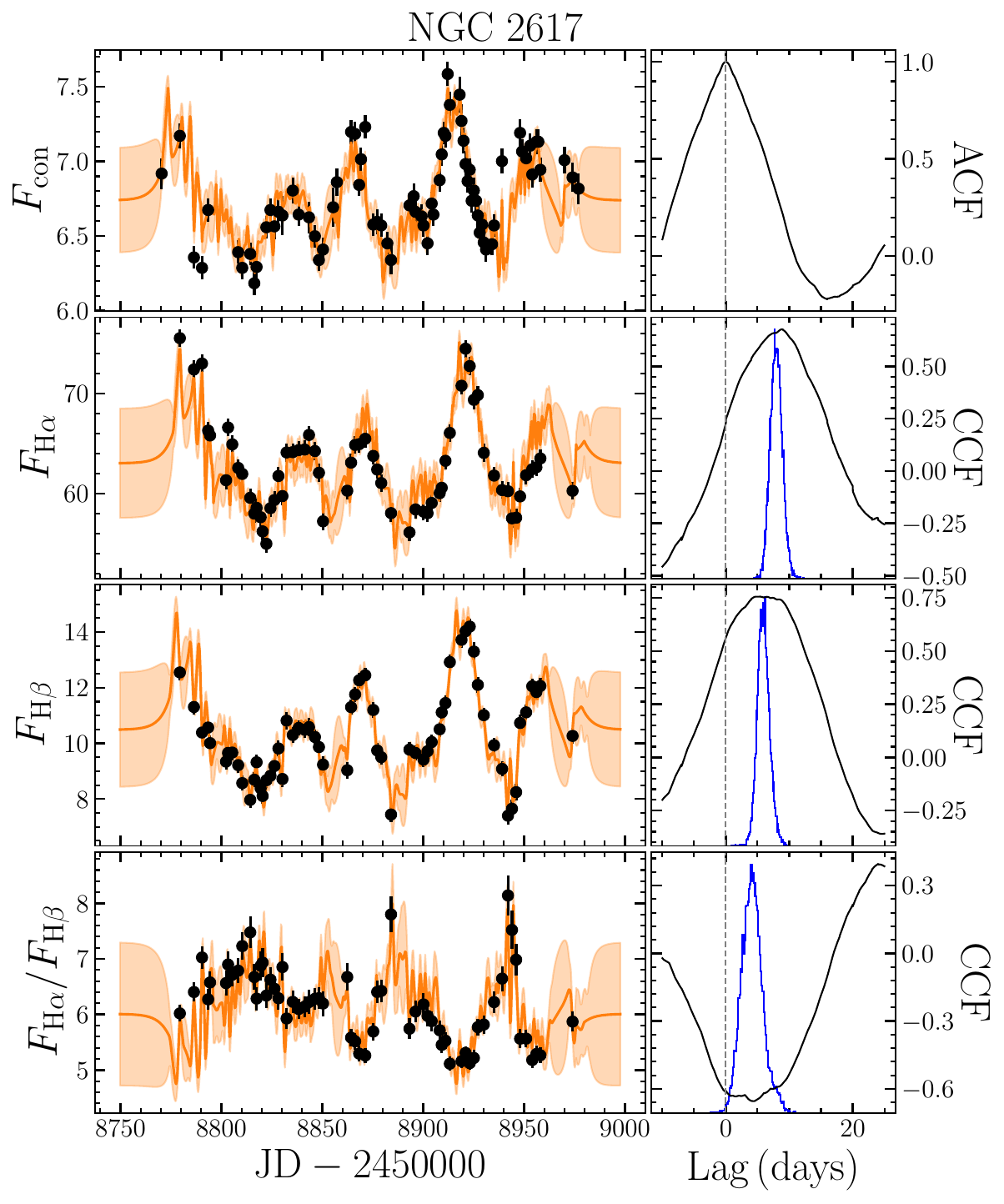}
\includegraphics[width=0.33\textwidth]{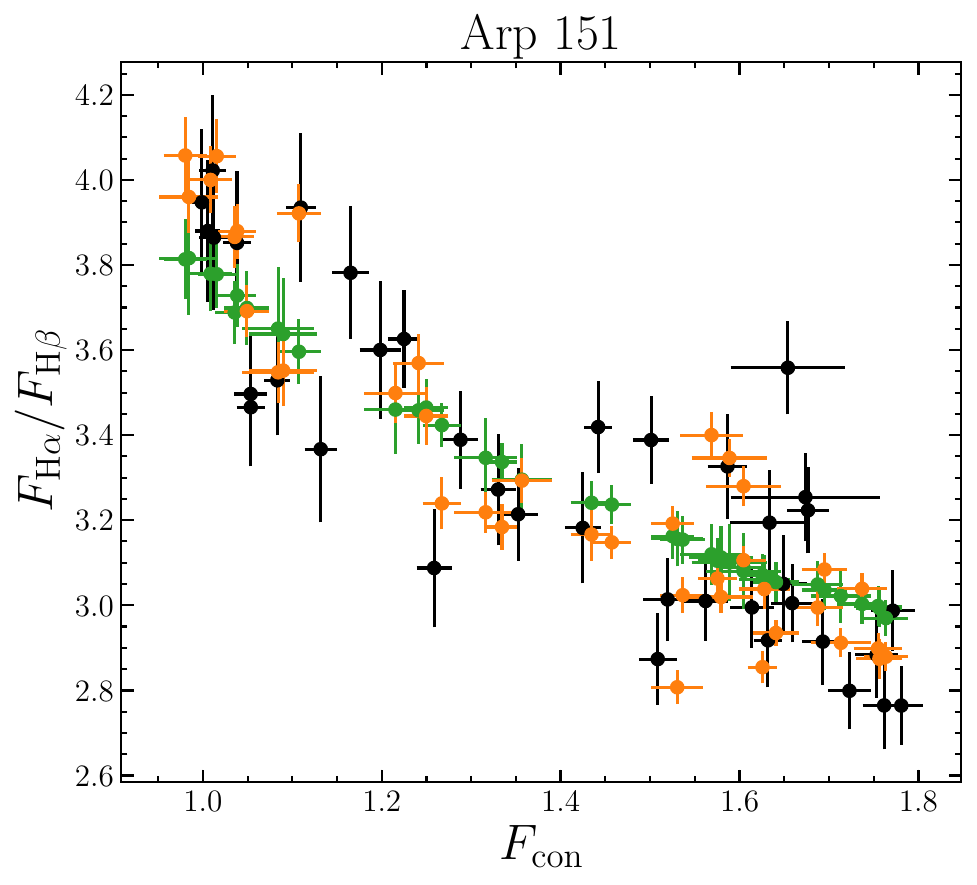}
\includegraphics[width=0.33\textwidth]{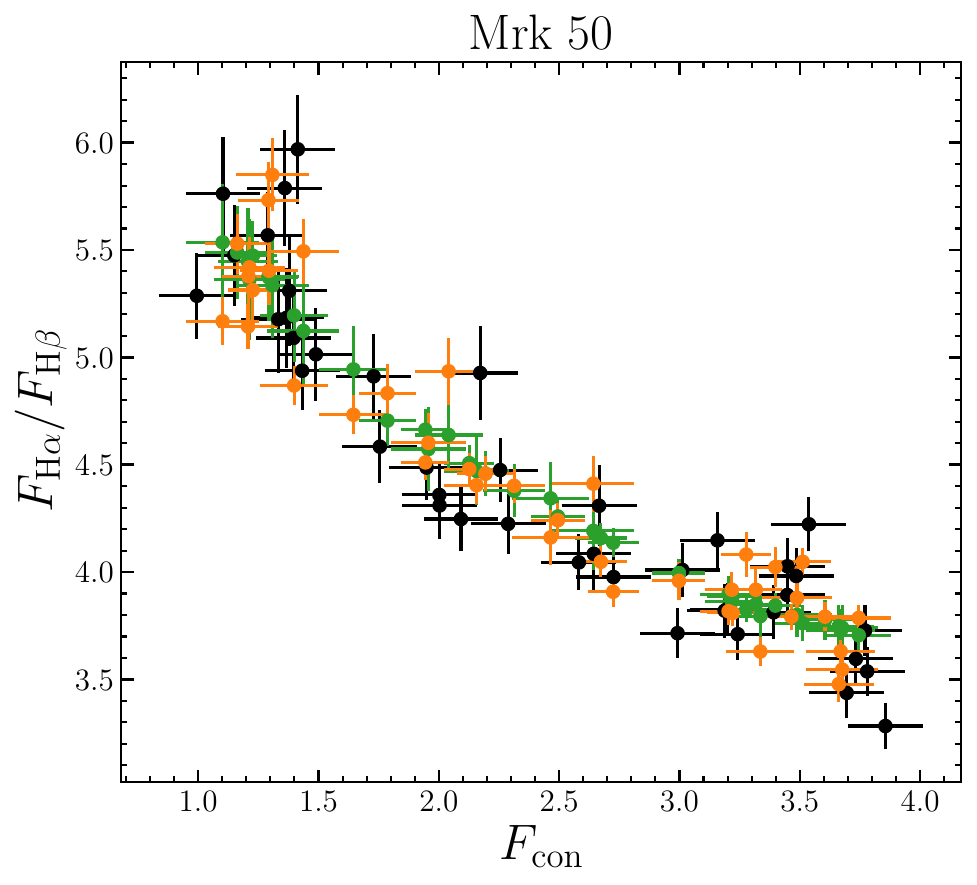}
\includegraphics[width=0.33\textwidth]{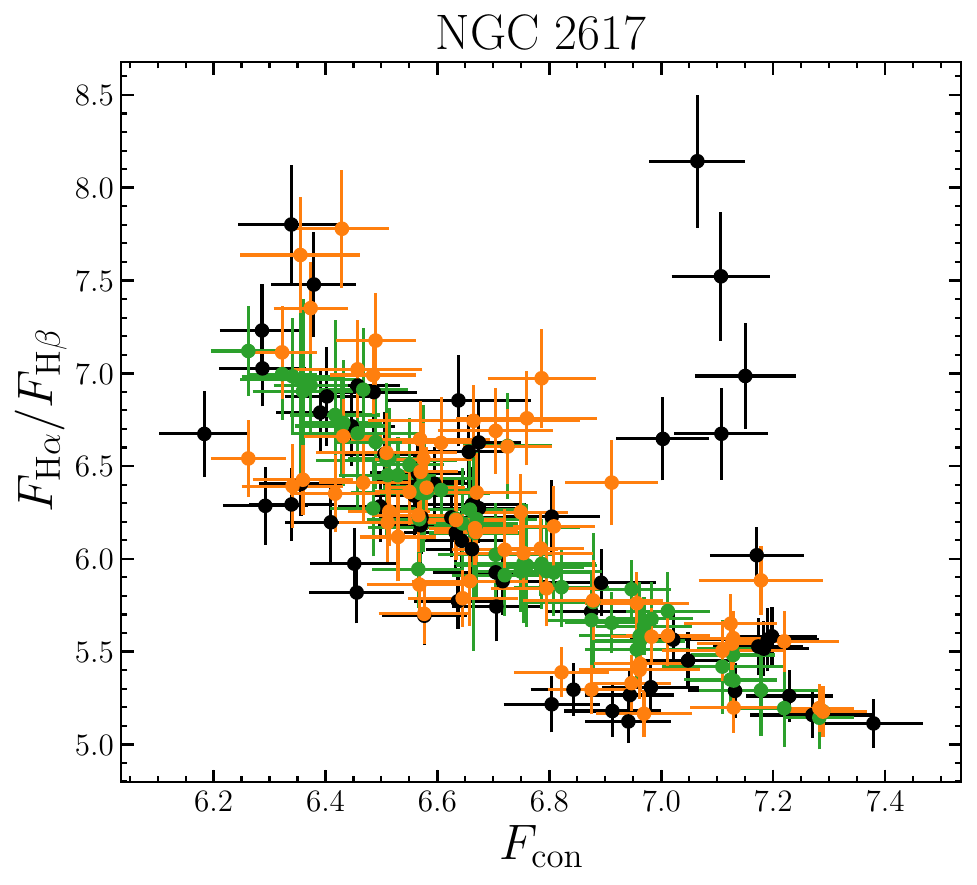}
\caption{Light curves, CCF results, and correlation between continuum and emission line flux ratios for three targets. Arp~151, Mrk~50, and NGC~2617 belong to LAMP2008, LAMP2011, and Lijiang campaigns, respectively. The top panels show light curves of the continuum, \ha, \hb, and flux ratios of \ha\ and \hb\ (\fab), and CCF results. Black points denote the observed data, while the orange-shaded regions illustrate the results reconstructed using JAVELIN. The \fab\ values, derived from both observed data and JAVELIN reconstructed results, are depicted by black points and orange bands, respectively. The ACFs of the continuum are also present. The black lines denote CCFs between the continuum and \ha, \hb, and \fab, with their corresponding CCCDs portrayed by blue lines. Vertical grey dashed lines represent instances where time lags are 0 days. The bottom panels show the relationship between the continuum and \fab. The \fab\ values, obtained from the direct observations (depicted as black points), JAVELIN reconstruction (represented as orange points), and time-lag-corrected JAVELIN reconstruction (shown as green points), exhibit an anti-correlation with the continuum. 
} 
\label{fig:lc}
\end{figure*}

\section{Sample}\label{sec:sample}
We selected AGN samples with RM observations to investigate the flux ratios of \ha\ and \hb\ emission lines, their RM results, and their joint analysis. However, simultaneous measurement of time delays for both \ha\ and \hb\ emission lines has been realised by only a handful of AGNs observations. Fewer than 10 AGN have simultaneous velocity-resolved RM measurements for both \ha\ and \hb. Here we focus on the Lick AGN Monitoring Projects in 2008 and 2011 (LAMP2008 \citep{Bentz2009} and LAMP2011 \citep{Barth2015}), along with our recently observed changing-look AGN (CL-AGN) samples from the Lijiang RM spectroscopic monitoring project campaign for our research \citep{Feng2021a, Feng2021b, Li2022}.

The LAMP2008 campaign, which was designed to measure \mbh\ for 13 nearby Seyfert galaxies, spanned over 64 nights from March to June 2008 using the 3-meter telescope at Lick Observatory \citep{Bentz2009}. As this project targeted low-redshift AGNs and the spectral data are publicly available, we are able to perform a reanalysis of the broad \ha\ and \hb\ emission lines. The campaign also included photometric observations in the $B$ and $V$ bands, which showed no significant time lags between them \citep{Walsh2009}. To improve the photometry sampling, we utilized PyCALI\footnote{\url{https://github.com/LiyrAstroph/PyCALI}}, a Python package for intercalibrating light curves in the $B$ and $V$ bands, which applies the damped random walk (DRW) process to model observed variations of AGNs \citep{Kelly2009, Li2014}. We then combined the calibrated results within 0.1-day intervals to represent the AGN continuum, as shown in Figure~\ref{fig:lc}.

The LAMP2011 campaign, a follow-up project utilizing the same telescope, was conducted from March to June 2011 at Lick Observatory. This campaign was focused on the study of 15 low-redshift Seyfert 1 galaxies \citep{Barth2015}. The sample with an apparent magnitude of $V \le 17$ were chosen to ensure high-quality spectra and to determine velocity-resolved RM results. For the analysis of LAMP2011, we directly used the light curves provided in Table~4 of \citet{Barth2015}, as the public spectra were unavailable. 

Additionally, we obtained data through our CL-AGN program from 2018 to 2021. This program was intended to investigate the BLR nature of CL-AGNs, which included NGC~3516, NGC~2617, and NGC~4151 \citep{Feng2021a, Feng2021b, Li2022}. These observations were made using the 2.4-meter telescope at Lijiang Observatory. These low-redshift samples allowed us to simultaneously detect \ha\ and \hb\ emission lines. In summary, the selected samples with low redshifts and high spectral resolution enable us to reveal the RM results associated with the broad emission lines.

\begin{figure*}[!ht]
\centering 
\includegraphics[width=0.459\textwidth]{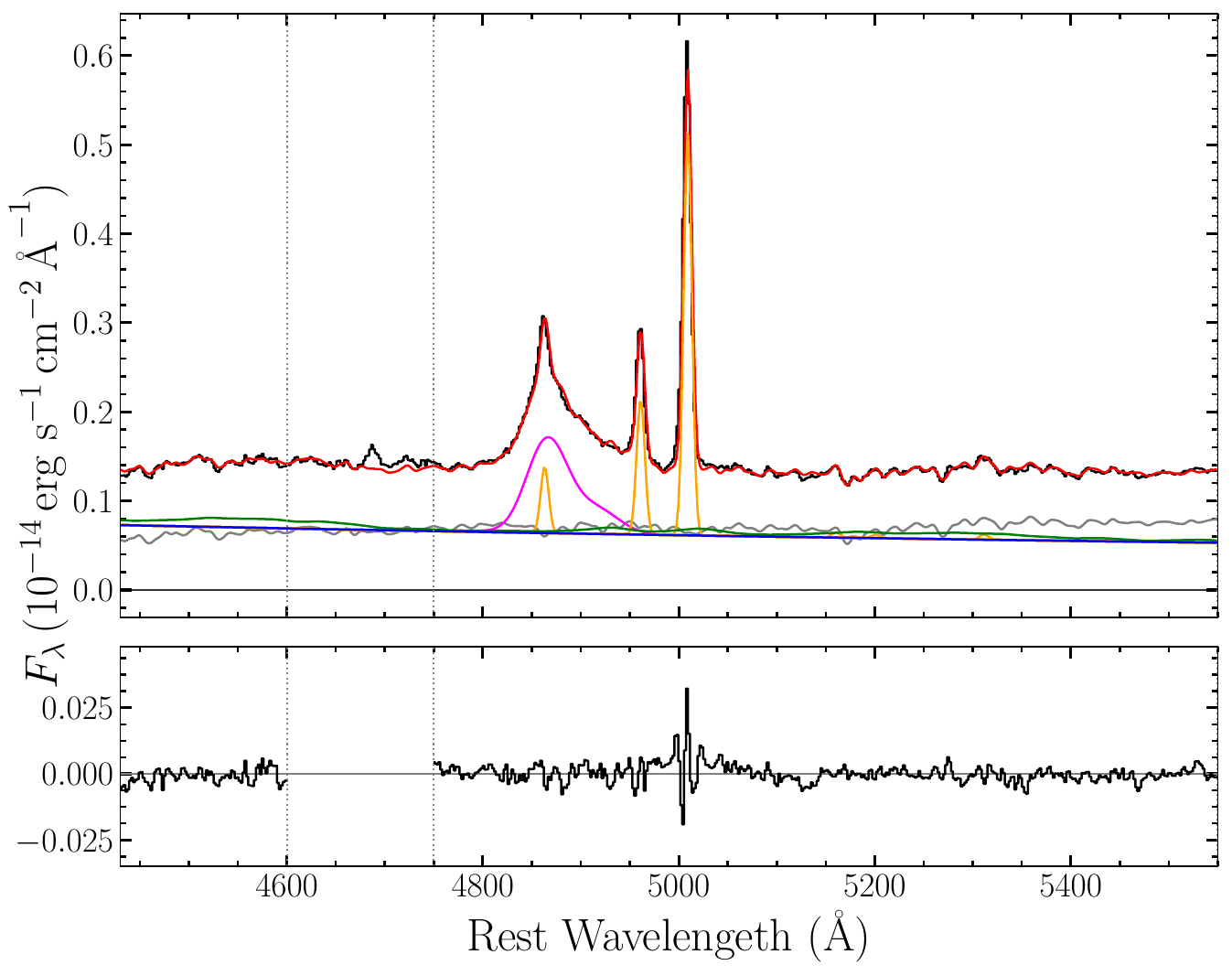}
\includegraphics[width=0.459\textwidth]{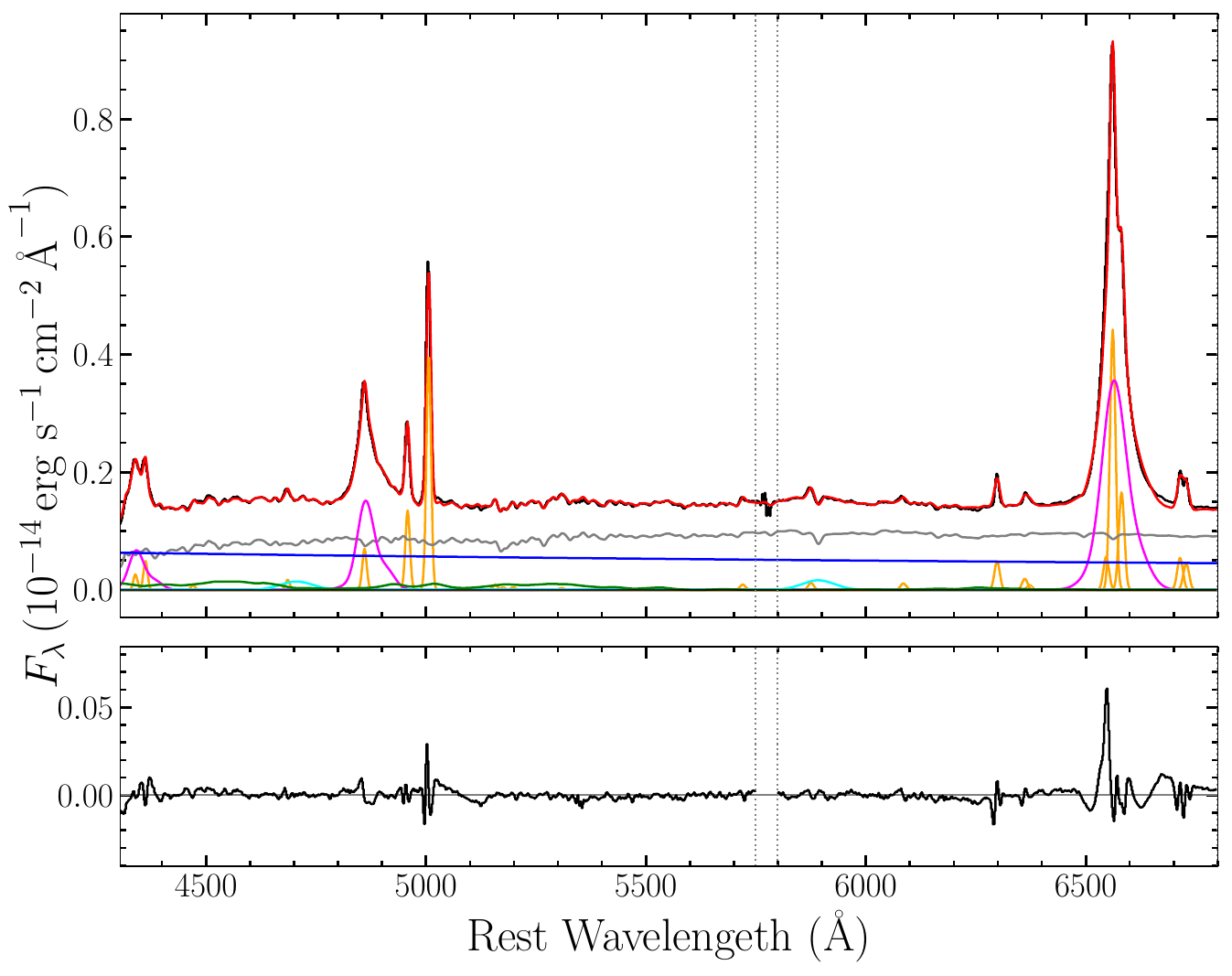}
\caption{Spectral fitting results. The upper-left panel shows the initial fitting result for an individual spectrum. The upper-right panel shows the second round fitting result for the mean spectrum of Arp~151. In both cases, the black lines are the original spectra; red lines are the best-fitting results; grey lines represent the host galaxy template; blue lines depict the AGN power law; green lines represent the \feii\ template; magenta lines represent the broad Balmer lines; orange lines represent the narrow lines; cyan lines represent broad \hei\ and \heii; the wavelength ranges appearing between two dashed vertical lines represent regions that were not considered during the fitting process. The lower panels display the fit residuals.}
\label{fig:fitspec}
\end{figure*}

\section{Analysis} \label{sec:analysis}
\subsection{Spectral Fitting} \label{sec:spectrafit}
At optical wavelengths broad emission lines in AGN spectra are blended with other components like the host galaxy, AGN power-law continuum, Paschen continuum, narrow emission lines, and optical \feii\ emission. Therefore, to extract information on these broad emission lines, we have applied multi-component spectral fitting techniques, utilizing DASpec\footnote{\url{https://github.com/PuDu-Astro/DASpec}}, a tool based on the Levenberg-Marquardt algorithm. For the CL-AGN sample, their spectra were flux calibrated using comparison stars, with details of data reductions provided in \citet{Feng2021a, Feng2021b, Li2022}. We adopted previous fitting models for these objects \citep{Feng2021a, Feng2021b, Li2022}, and the details of the models have been omitted from this text.

For the LAMP2008 sample, both reduced and scaled spectra are publicly available. The scaled spectra were calibrated according to the method described in \citet{vGW1992}, which relies on the flux of [\oiii] $\lambda$5007 measured from the reduced spectra. However, \citet{hu2016} provided a more accurate flux calibration through spectral fitting to measure the same [\oiii] $\lambda$5007 fluxes from reduced spectra. Therefore, we recalibrated the LAMP2008 spectra following their method, conducting two fitting rounds to obtain the final broad-line information. Before fitting, we corrected the spectra for Galactic extinction and redshift using parameters from the NASA/IPAC Extragalactic Database and the $R_{\rm v} = 3.1$ extinction law \citep{Cardelli1989, ODonnell1994}. 

In the first fitting round, we selected narrow wavelength windows spanning 4430-4600 \AA\ and 4750-5550 \AA, encompassing \hb\ but excluding \heii\ $\lambda$4686. Our model consisted of: (1) a power law representing the AGN continuum (with a minor contribution from the Paschen continuum), (2) an optical \feii\ template from \citet{BG1992}, (3) a simple Stellar Population model from \citet{BC2003} to fit the host galaxy, assuming solar metallicity and an age of 11 billion years, (4) a single Gaussian for narrow lines like [\oiii] $\lambda\lambda$4959,5007, and the narrow \hb\ line, (5) a fourth-order Gauss-Hermite function for broad \hb. Additionally, we fixed the widths and shifts of narrow lines to [\oiii] $\lambda$5007 and set the [\oiii] $\lambda\lambda$5007,4959 flux ratio at 3. We first used the above model to fit spectra observed on April 10, 2008 (with steady photometric night as mentioned in \citealt{Bentz2009}), deriving [\oiii] $\lambda$5007 fluxes (an example is shown in left panels of Figure~\ref{fig:fitspec}). The ratio of this measurement to the value listed in Table~3 of \citet{Bentz2009} (after correcting for extinction) was used as the calibration factor. Finally, these factors were applied to the reduced spectra for calibration. 

In the second fitting round, we began by fitting the calibrated spectra. Before fitting, we selected spectra with a signal-to-noise ratio (S/N) $\geq$ 25 pixel$^{-1}$ around 5100 \AA, which were used for analysis and to create the mean spectrum. The fitting windows covered the ranges of 4305-5750 \AA\ and 5800-6800 \AA, excluding the region around 5800~\AA\ contaminated by \nai\ D residual noise (as mentioned in \citealt{Bentz2010}). Within these windows, we could fit nearly all broad emission lines from \hc\ to \ha\ and determine their characteristics. The templates of the host galaxy and \feii\ were similar to those used in the initial fitting. We also used fourth-order Gauss-Hermite functions to model broad Balmer lines, including \ha, \hb, and \hc\ (for NGC 5548, three Gaussians were separately used to fit them), but the profile of broad \hc\ was bound with broad \hb. Broad \heii\ and \hei, as well as narrow lines, were individually modeled by a single Gaussian. The widths and shifts of narrow lines with wavelengths below 6000 \AA, e.g., narrow \hb, are tied to [\oiii] $\lambda$5007, while those of narrow emission lines with wavelengths above 6000 \AA\ are tied to [\oi] $\lambda$6300 (or [\sii] $\lambda$ 6718 if [\oi] $\lambda$6300 is weak). Also, the flux ratios of [\oiii] $\lambda\lambda$5007,4959 and [\nii] $\lambda\lambda$6583,6548 were separately fixed to 3.0 and 2.96. We first utilized these models to fit the mean spectrum (shown in right panels of Figure~\ref{fig:fitspec}) and obtained the widths and shifts of broad \heii\ and \hei\ lines, as well as the flux ratios of narrow lines relative to [\oiii] $\lambda$5007 and [\oi] $\lambda$6300 (or [\sii] $\lambda$ 6718), respectively. Then, the profiles of broad \heii\ and \hei, and these narrow-line flux ratios measured from the mean spectrum were fixed to fit individual spectra. Lastly, we measured the fluxes and widths of each emission line in all spectra. 

Based on [\oiii] $\lambda$5007 widths, we convolved each spectrum to match the maximum [\oiii] $\lambda$5007 width to alleviate the impact of different spectral spreads. We then applied the aforementioned fitting method to the convolved spectra. After the fitting process, the broad emission line profiles extracted from these convolved spectra were used in the subsequent analysis. 

When analyzing the light curves, we adopted fitting errors for the LAMP2008 and Lijiang data. For LAMP2011, we directly used the errors provided in \cite{Barth2015}. We also added a systematic error to all light curves following the method in \cite{Li2021}. Using the fluxes and errors, we computed the relative errors for each light curve, with mean values listed in Table~\ref{table1}. To assess the characteristic variability of the light curves, we computed the fractional variability amplitude (\fvar), as governed by the methodology outlined in \citet{Rodr1997}:
\begin{equation}
F_{\rm var}=\frac{(S^2-\triangle^2)^{1/2}}{\langle F \rangle},
\end{equation}
where $S^2$ is the flux variance, $\triangle^2$ is mean squared error, and $\langle F \rangle$ is the mean flux over the light curve. The \fvar\ uncertainty is given by \citet{Edelson2002}:
\begin{equation}
\sigma_{\rm var}=\frac{1}{F_{\rm var}}\bigg(\frac{1}{2N}\bigg)^{1/2}\frac{S^2}{{\langle F \rangle}^2}.
\end{equation}
where $\it N$ is the total number of flux measurements within the light curve. The computed results for each light curve are compiled in Table~\ref{table1}.

\begin{deluxetable*}{lcccccccccc}[!htbp]
\tablecolumns{8}
\tablewidth{0.49\textwidth}
\tabletypesize{\scriptsize}
\tablecaption{Measurements of variability amplitude and relative error}\label{table1}
\tablehead{& \multicolumn{3}{@{}c@{}}{\fvar\ (\%)} & \multicolumn{3}{@{}c@{}}{Relative error (\%)}  \\ 
\cline{2-4} \cline{5-7} 
\colhead{Object} & \colhead{${\rm con}$} & \colhead{${\rm H\alpha}$} & \colhead{${\rm H\beta}$} & \colhead{${\rm con}$} & \colhead{${\rm H\alpha}$} & \colhead{${\rm H\beta}$ } &\colhead{$\rm \Delta BD$}}
\startdata
\multicolumn{8}{@{}c@{}}{LAMP2008 campaign} \\
\hline
Mrk 142 & $5.01 \pm 0.84$ & $4.47 \pm 0.62 $ & $ 2.46 \pm 0.27 $ & 1.34 & 4.05 & 2.65 & 0.03 \\
SBS 1116+583A & $6.30 \pm 0.83$ & $8.78 \pm 1.31 $ & $ 9.96 \pm 0.98 $ & 2.92 & 2.87 & 5.33 & 0.27 \\
Arp 151 & $12.76 \pm 1.51$ & $19.60 \pm 2.17 $ & $ 15.98 \pm 1.38 $ & 1.44 & 3.49 & 1.29 & 0.78 \\
Mrk 1310 & $7.78 \pm 1.17$ & $15.14 \pm 1.88 $ & $ 10.97 \pm 1.04 $ & 2.85 & 5.20 & 6.78 & 0.77 \\
Mrk 202 & $12.18 \pm 1.35$ & $16.97 \pm 1.85 $ & $ 3.77 \pm 0.41 $ & 1.60 & 3.03 & 3.58 & 0.33 \\
NGC 4253$^a$ & $2.93 \pm 0.63$ & $3.58 \pm 0.72 $ & $ 2.85 \pm 0.31 $ & 1.16 & 3.10 & 3.52 & 0.26 \\
NGC 4748 & $8.82 \pm 1.04$ & $6.87 \pm 0.80 $ & $ 4.98 \pm 0.56 $ & 2.31 & 2.68 & 1.93 & 0.11 \\
IC 4218$^a$ & $7.40 \pm 1.15$ & $21.16 \pm 3.27 $ & $ 7.23 \pm 0.84 $ & 3.77 & 4.04 & 12.09 & 1.96 \\
MCG -06-30-15$^a$ & $5.18 \pm 0.76$ & $4.75 \pm 2.01 $ & $ 3.37 \pm 0.40 $ & 1.72 & 2.94 & 7.99 & 0.09 \\
NGC 5548 & $6.15 \pm 0.77$ & $8.53 \pm 1.12 $ & $ 8.95 \pm 0.80 $ & 1.12 & 2.82 & 4.50 & 0.20 \\
Mrk 290$^a$ & $3.44 \pm 0.55$ & $3.97 \pm 0.52 $ & $ 3.61 \pm 0.39 $ & 1.33 & 2.60 & 2.13 & 0.12 \\
IC 1198$^a$ & $3.17 \pm 0.74$ & $6.81 \pm 1.72 $ & $ 3.27 \pm 0.44 $ & 2.46 & 3.46 & 8.05 & 0.21 \\
NGC 6814 & $5.21 \pm 0.74$ & $9.14 \pm 1.18 $ & $ 16.97 \pm 1.81 $ & 3.18 & 2.97 & 4.07 & 0.39 \\
\hline
\multicolumn{8}{@{}c@{}}{LAMP2011 campaign} \\ 
\hline
Mrk 40 (Arp 151) & $13.99 \pm 1.61$ & $19.60 \pm 2.22 $ & $ 29.14 \pm 3.43 $ & 6.25 & 1.97 & 0.82 & 0.91 \\
Mrk 50 & $10.39 \pm 1.24$ & $19.76 \pm 1.90 $ & $ 33.91 \pm 3.34 $ & 7.12 & 3.15 & 1.88 & 1.73 \\
Mrk 279 & $3.77 \pm 0.66$ & $7.22 \pm 0.94 $ & $ 10.88 \pm 1.77 $ & 6.46 & 2.50 & 1.97 & 0.56 \\
Mrk 704 & $6.39 \pm 0.85$ & $4.58 \pm 0.58 $ & $ 11.64 \pm 1.73 $ & 6.47 & 2.51 & 1.47 & 0.30 \\
Mrk 1511 & $7.33 \pm 1.34$ & $11.67 \pm 1.33 $ & $ 18.40 \pm 2.28 $ & 5.91 & 5.82 & 1.02 & 0.41 \\
NGC 4593 & $15.01 \pm 1.80$ & $22.68 \pm 2.48 $ & $ 36.89 \pm 4.24 $ & 9.80 & 5.16 & 0.65 & 1.62 \\
Zw 229-015 & $12.23 \pm 1.94$ & $24.07 \pm 3.43 $ & $ 24.20 \pm 3.39 $ & 6.82 & 4.33 & 7.57 & 1.05 \\
\hline
\multicolumn{8}{@{}c@{}}{Lijiang CL-AGN campaign} \\
\hline
NGC 2617 & $7.29 \pm 0.67$ & $15.27 \pm 1.40 $ & $ 4.45 \pm 0.39 $ & 1.25 & 1.43 & 2.81 & 1.79 \\
NGC 3516 & $26.33 \pm 2.31$ & $40.80 \pm 3.80 $ & $ 2.59 \pm 0.27 $ & 1.00 & 4.37 & 15.20 & 3.25 \\
NGC 4151 & $3.49 \pm 0.42$ & $5.30 \pm 0.62 $ & $ 7.53 \pm 0.52 $ & 1.59 & 1.09 & 1.38 & 0.17 \\
\enddata
\tablecomments{The objects analyzed in this study were selected from the LAMP2008, LAMP2011, and Lijiang campaigns. The LAMP2011 datasets incorporate contributions from narrow lines.\\
$^a$ Those five objects, whose variability characteristics prevent lag measurements from being conducted, are also illustrated in Figure~6 of \citet{Bentz2009}.}
\end{deluxetable*}

\begin{deluxetable}{lccccc}[!htbp]
\tablecolumns{5}
\tablewidth{0.9\textwidth}
\tabletypesize{\scriptsize}
\tablecaption{Cross-correlation measurements}\label{table2}
\tablehead{\colhead{Object} & \colhead{$r_{\rm max,H\alpha}$} & \colhead{$\tau_{\rm H\alpha} \rm(days)$} & \colhead{$r_{\rm max,H\beta}$} & \colhead{$\tau_{\rm H\beta}\rm(days)$ }}
\startdata
\multicolumn{5}{@{}c@{}}{LAMP2008 campaign} \\
\hline
Mrk 142 & 0.54 & $2.87_{-1.72}^{+7.49}$ & 0.58 & $9.44_{-5.19}^{+2.35}$ \\ 
SBS 1116+583A & 0.58 & $3.33_{-0.74}^{+1.58}$ & 0.70 & $1.85_{-0.57}^{+1.50}$ \\ 
Arp 151 & 0.93 & $7.30_{-1.12}^{+0.97}$ & 0.98 & $3.10_{-0.54}^{+0.76}$ \\ 
Mrk 1310 & 0.84 & $3.76_{-0.57}^{+0.95}$ & 0.78 & $2.98_{-1.02}^{+0.71}$ \\ 
Mrk 202 & 0.75 & $16.73_{-6.74}^{+2.98}$ & 0.76 & $10.69_{-2.54}^{+3.29}$ \\ 
NGC 4253 & 0.28 & $6.19_{-4.86}^{+1.21}$ & 0.66 & $5.59_{-1.71}^{+1.53}$ \\ 
NGC 4748 & 0.78 & $10.16_{-2.59}^{+0.99}$ & 0.79 & $6.08_{-1.73}^{+1.52}$ \\ 
IC 4218 & 0.37 & $-1.10_{-4.97}^{+4.14}$ & 0.48 & $2.71_{-4.72}^{+2.58}$ \\ 
MCG -06-30-15 & 0.48 & $-17.75_{-0.81}^{+8.36}$ & 0.39 & $-0.68_{-6.32}^{+6.99}$ \\ 
NGC 5548 & 0.60 & $10.52_{-1.90}^{+1.31}$ & 0.68 & $5.63_{-1.62}^{+1.36}$ \\ 
Mrk 290 & 0.42 & $12.83_{-18.17}^{-0.06}$ & 0.72 & $3.32_{-2.19}^{+3.28}$ \\ 
IC 1198 & 0.39 & $-19.11_{+0.14}^{+20.82}$ & 0.26 & $0.88_{-9.34}^{+6.43}$ \\ 
NGC 6814 & 0.67 & $9.28_{-1.64}^{+1.08}$ & 0.77 & $3.96_{-0.77}^{+0.90}$ \\ 
\hline
\multicolumn{5}{@{}c@{}}{LAMP2011 campaign} \\ 
\hline
Mrk 40 (Arp 151) & 0.83 & $8.45_{-1.68}^{+1.91}$ & 0.94 & $5.26_{-0.92}^{+0.77}$ \\ 
Mrk 50 & 0.93 & $14.66_{-1.40}^{+2.08}$ & 0.94 & $8.09_{-1.34}^{+1.05}$ \\ 
Mrk 279 & 0.73 & $17.53_{-5.29}^{+2.41}$ & 0.83 & $16.85_{-2.66}^{+1.50}$ \\ 
Mrk 704 & 0.73 & $3.15_{-1.27}^{+3.18}$ & 0.89 & $-0.20_{-1.95}^{+1.16}$ \\ 
Mrk 1511 & 0.77 & $7.30_{-1.93}^{+2.90}$ & 0.84 & $5.99_{-0.75}^{+1.49}$ \\ 
NGC 4593 & 0.67 & $6.54_{-2.12}^{+3.72}$ & 0.88 & $3.19_{-2.10}^{+1.69}$ \\ 
Zw 229-015 & 0.86 & $16.24_{-4.60}^{+5.41}$ & 0.89 & $3.81_{-0.96}^{+3.37}$ \\
\hline
\multicolumn{5}{@{}c@{}}{Lijiang CL-AGN campaign} \\
\hline
NGC 2617 & 0.68 & $7.88_{-0.93}^{+0.89}$ & 0.76 & $5.82_{-0.88}^{+1.00}$ \\ 
NGC 3516 & 0.81 & $19.89_{-7.85}^{+2.01}$ & 0.85 & $7.36_{-1.59}^{+3.73}$ \\ 
NGC 4151 & 0.81 & $5.00_{-3.77}^{+0.89}$ & 0.88 & $4.95_{-1.20}^{+2.10}$ \\ 
\enddata
\tablecomments{The objects analyzed in this study were selected from the LAMP2008, LAMP2011, and Lijiang campaigns. Using the CCF method, we calculated the maximum cross-correlation coefficients ($r_{\rm max}$) and the corresponding time lags for \ha\ and \hb\ relative to the continuum. These values are denoted as $r_{\rm max,H\alpha}$, $r_{\rm max,H\beta}$, $\tau_{\rm H\alpha}$, and $\tau_{\rm H\beta}$, respectively.}
\end{deluxetable}

\begin{figure*}[!ht]%
\centering
\includegraphics[width=0.35\textwidth]{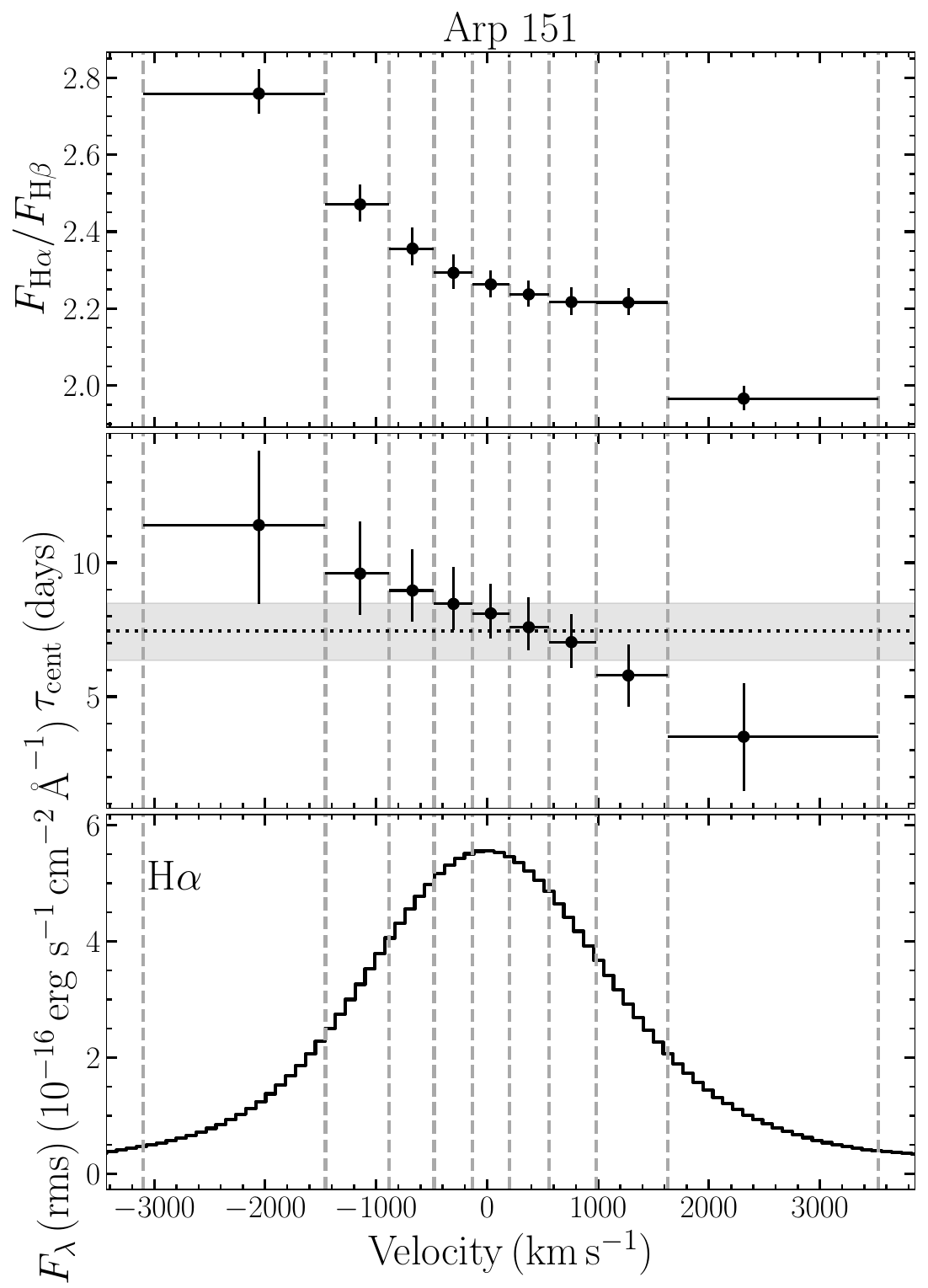}
\includegraphics[width=0.35\textwidth]{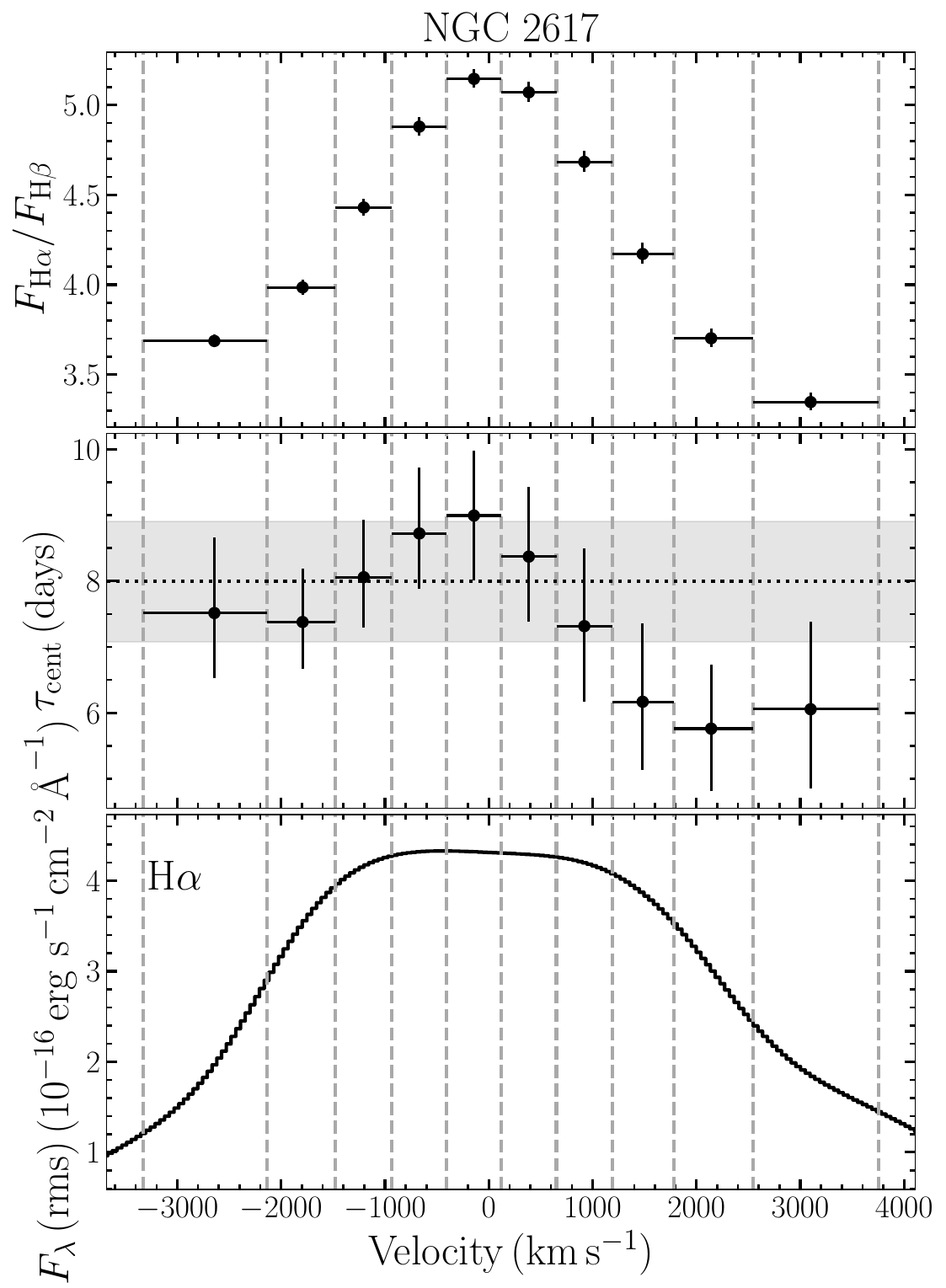}
\includegraphics[width=0.35\textwidth]{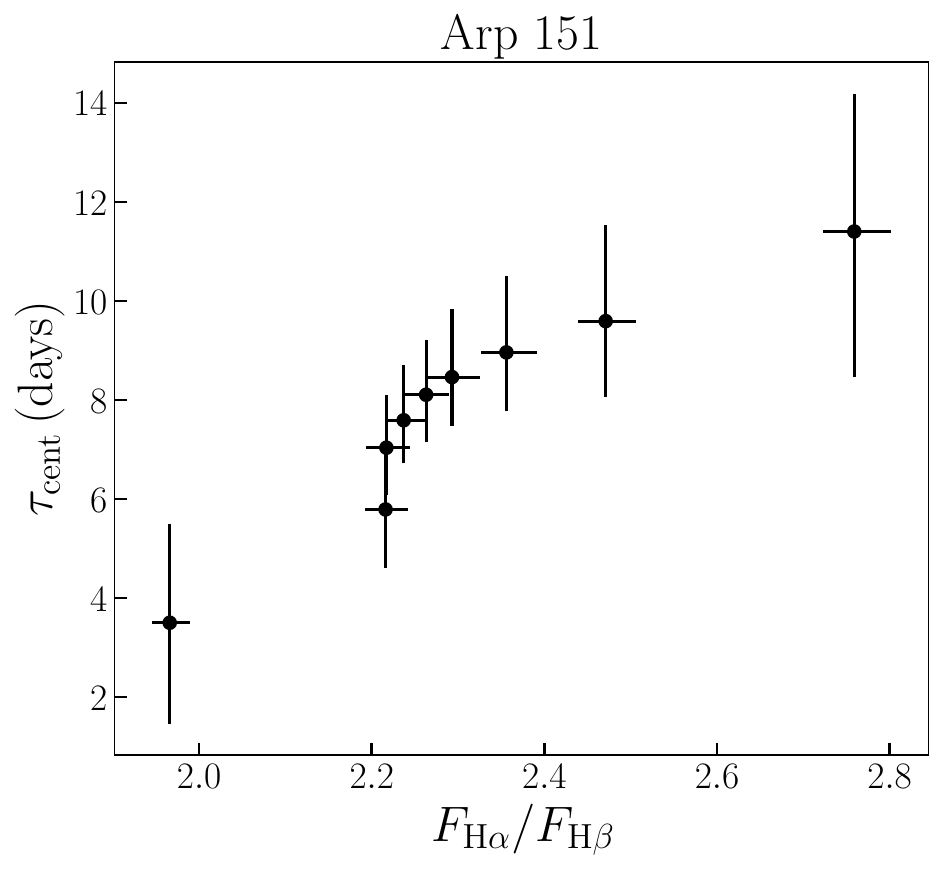}
\includegraphics[width=0.35\textwidth]{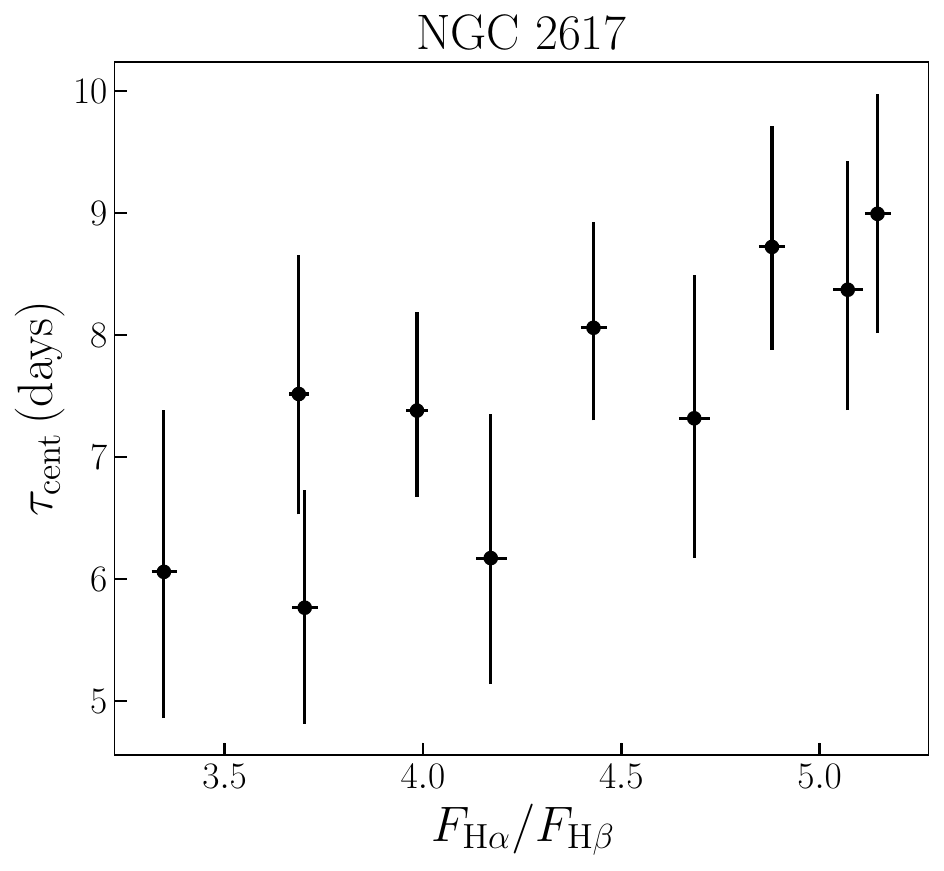}
\caption{Correlation between velocity-resolved emission line flux ratios and time lags of Arp~151 (from LAMP2008) and NGC~2617. In the top panels, the black step lines represent the rms spectra of \ha. The rms spectra are divided into multiple velocity bins, ensuring a uniform integrated rms flux across each bin. Vertical dashed lines delineate the boundaries of these velocity bins. The \fab\ values and time lags ($\tau_{\rm cent}$) of \ha\ for each velocity bin are shown as black points, with the velocity associated with each point aligning with the centroid of the rms spectra in its respective bin. Each velocity bin was not corrected for the delay of emission line relative to the continuum before choosing the flux ratio. Horizontal dotted lines and grey bands depict the time lags of \ha\ and their associated uncertainties, respectively. In the bottom panels, the relationship between \fab\ values and time lags of \ha, as obtained from various velocity bins. 
}
\label{fig:vr}
\end{figure*}

\begin{figure*}[!ht]%
\centering
\includegraphics[width=0.33\textwidth]{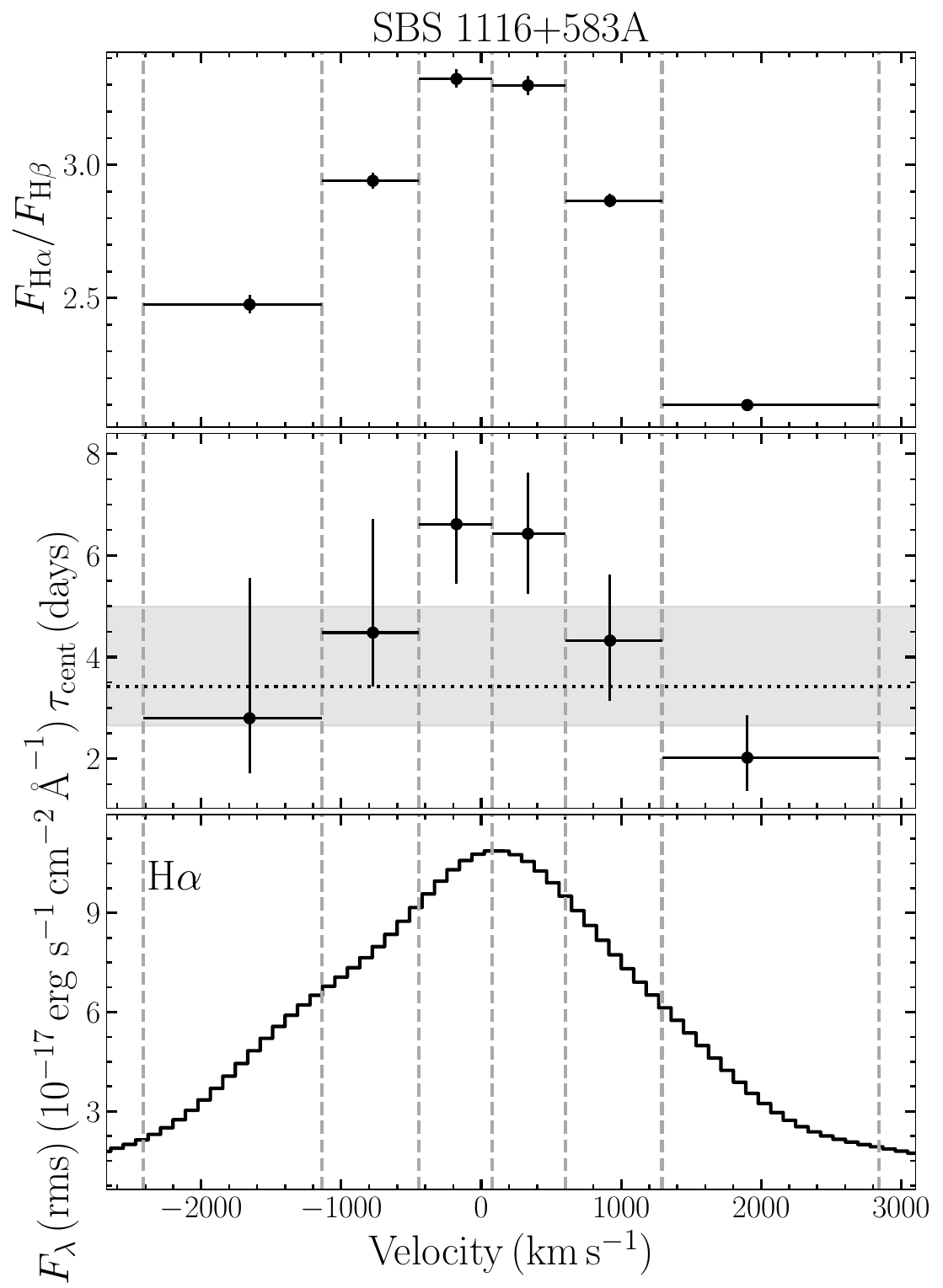}
\includegraphics[width=0.33\textwidth]{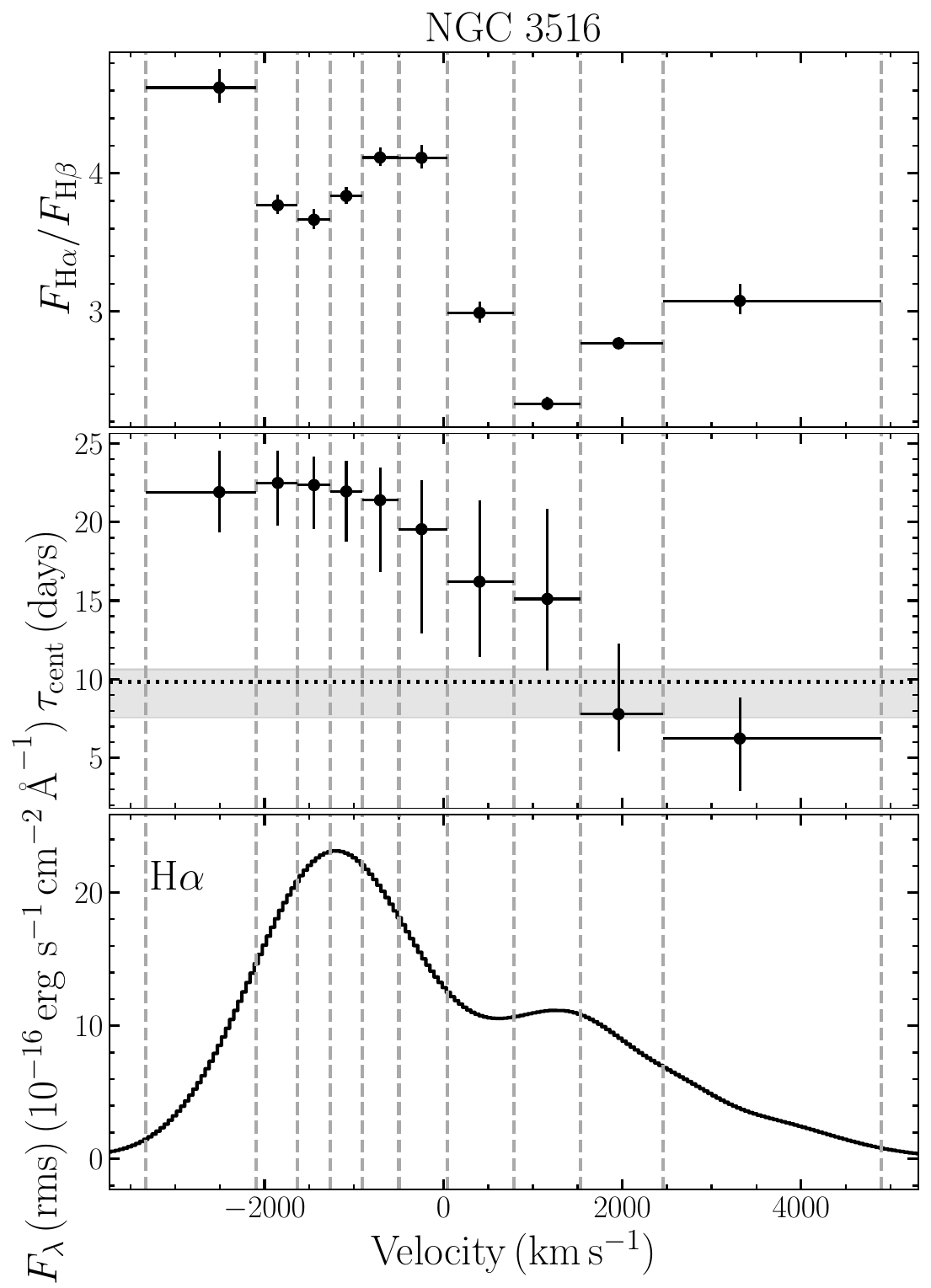}
\includegraphics[width=0.33\textwidth]{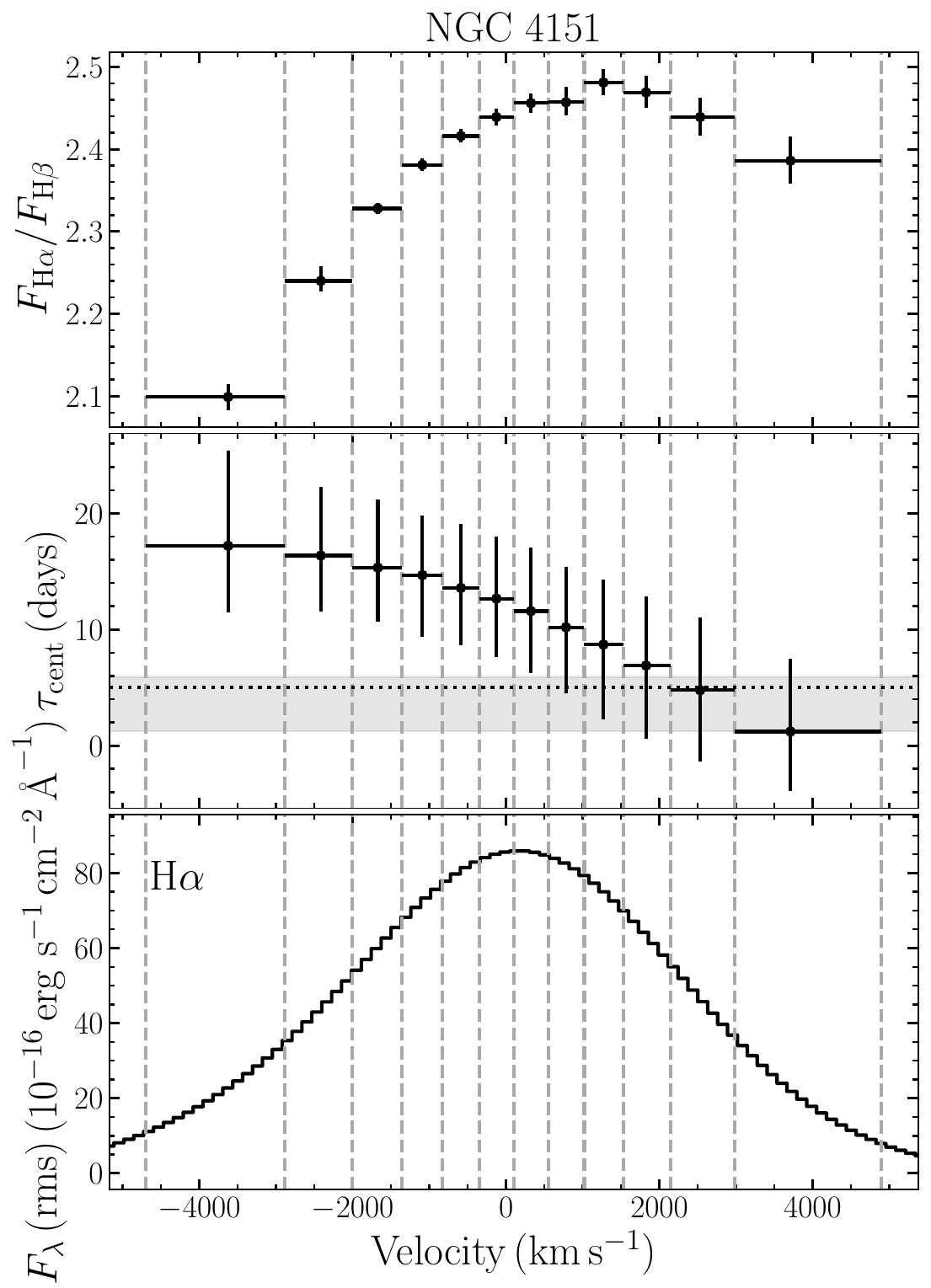}
\includegraphics[width=0.33\textwidth]{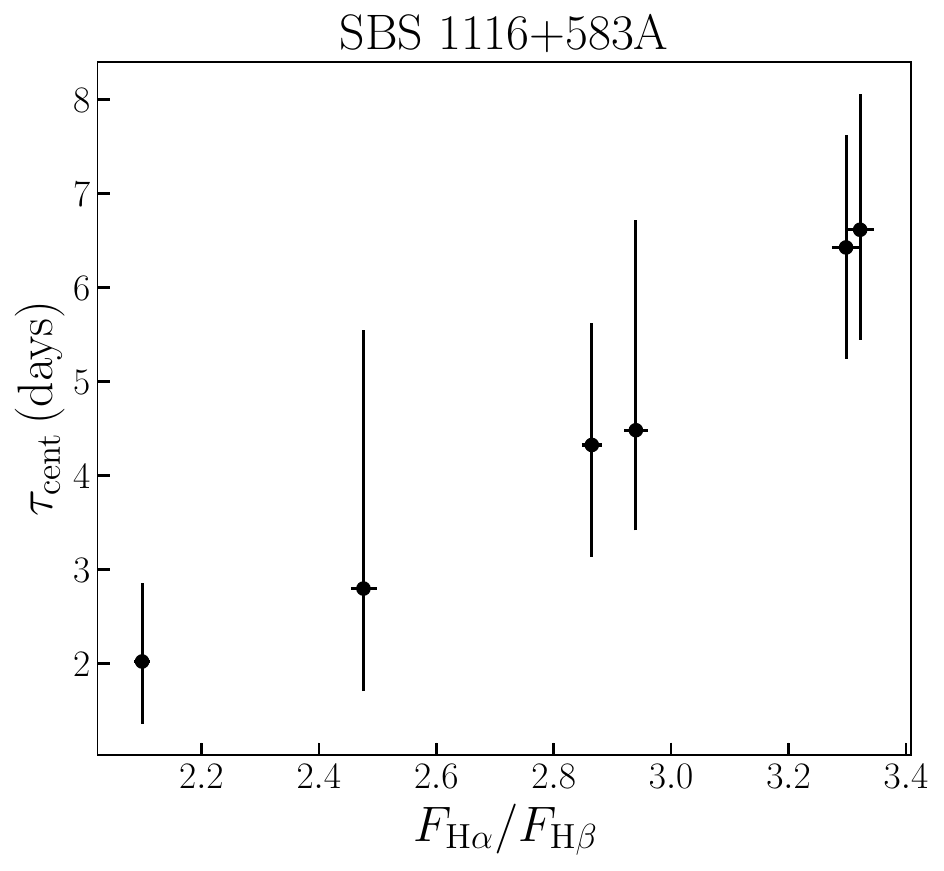}
\includegraphics[width=0.33\textwidth]{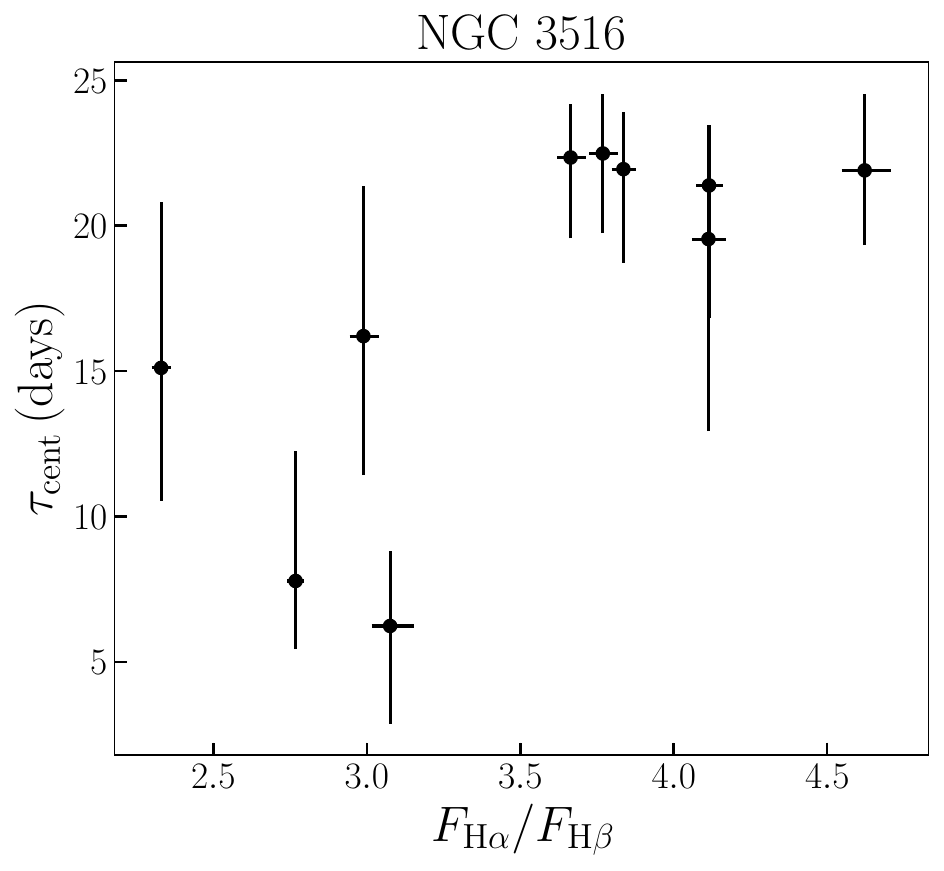}
\includegraphics[width=0.33\textwidth]{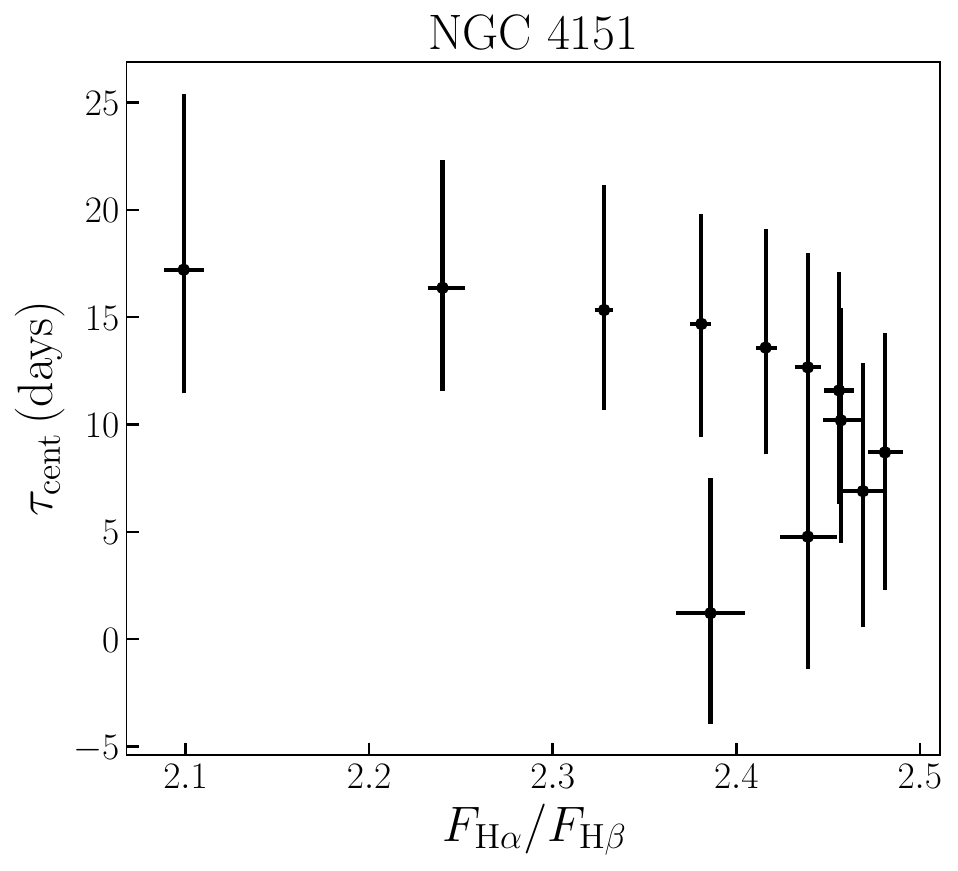}
\caption{Same as Figure~\ref{fig:vr}, but for SBS~1116+583A, NGC~3516, and NGC~4151.}
\label{fig:vr2}
\end{figure*}

\subsection{Time Series Analysis}
The light curves of the sample allow us to estimate the time delays between broad emission lines and the AGN continuum using the interpolated CCF method \citep{Gaskell1986, Gaskell1987}. These time delays (denoted as $\tau_{\rm cent}$) are determined by measuring the centroid of the segment above 80\% of the peak value in the CCF, with the peak representing the maximum cross-correlation coefficients ($r_{\rm max}$). We used Monte Carlo simulations incorporating flux randomization (FR) and random subset selection (RSS) to calculate uncertainties in the time lags, thereby accounting for flux errors and sampling uncertainties \citep{Peterson1998}. This resulted in a cross-correlation centroid distribution (CCCD), where the 15.87\% and 84.13\% quantiles represent the lower and upper bounds of time lags, respectively. These time lags are tabulated in Table~\ref{table2}. Figure~\ref{fig:lc} presents the results of auto-correlation functions (ACFs), CCFs, and CCCDs for a subset of targets within the sample. It should be noted that for 5 objects, the emission line lags cannot be reliably determined. Examination of the light curves reveals that their emission line variability amplitudes are comparable to the errors (Table~\ref{table1}), suggesting that it may be due to insufficient S/N in the spectra, or potentially attributed to the rare anomalous response of the BLR, known as an ``AGN holiday" \citep{Goad2016, Pei2017}.

Furthermore, we conducted measurements of time delays for emission lines across different velocity bins relative to continuum variations, referred to as velocity-resolved delays. The process involved the following steps. Initially, we generated the rms spectrum using the broad emission line profiles obtained from the previous section. The emission line in the rms spectrum was subsequently divided into several segments, each ensuring reliable lag measurement by selecting smaller bins with sufficient variation and containing identical integrated flux. Next, we applied these velocity bins to each spectrum, yielding light curves for different velocity intervals. We then separately calculated the time delays for \ha\ and \hb\ within each of these velocity bins, enabling an exploration of the relationship between the BD and time lag across velocity space. The results are displayed in Figure~\ref{fig:vr} and \ref{fig:vr2}.

\begin{figure*}[!ht]%
\centering
\includegraphics[width=0.45\textwidth]{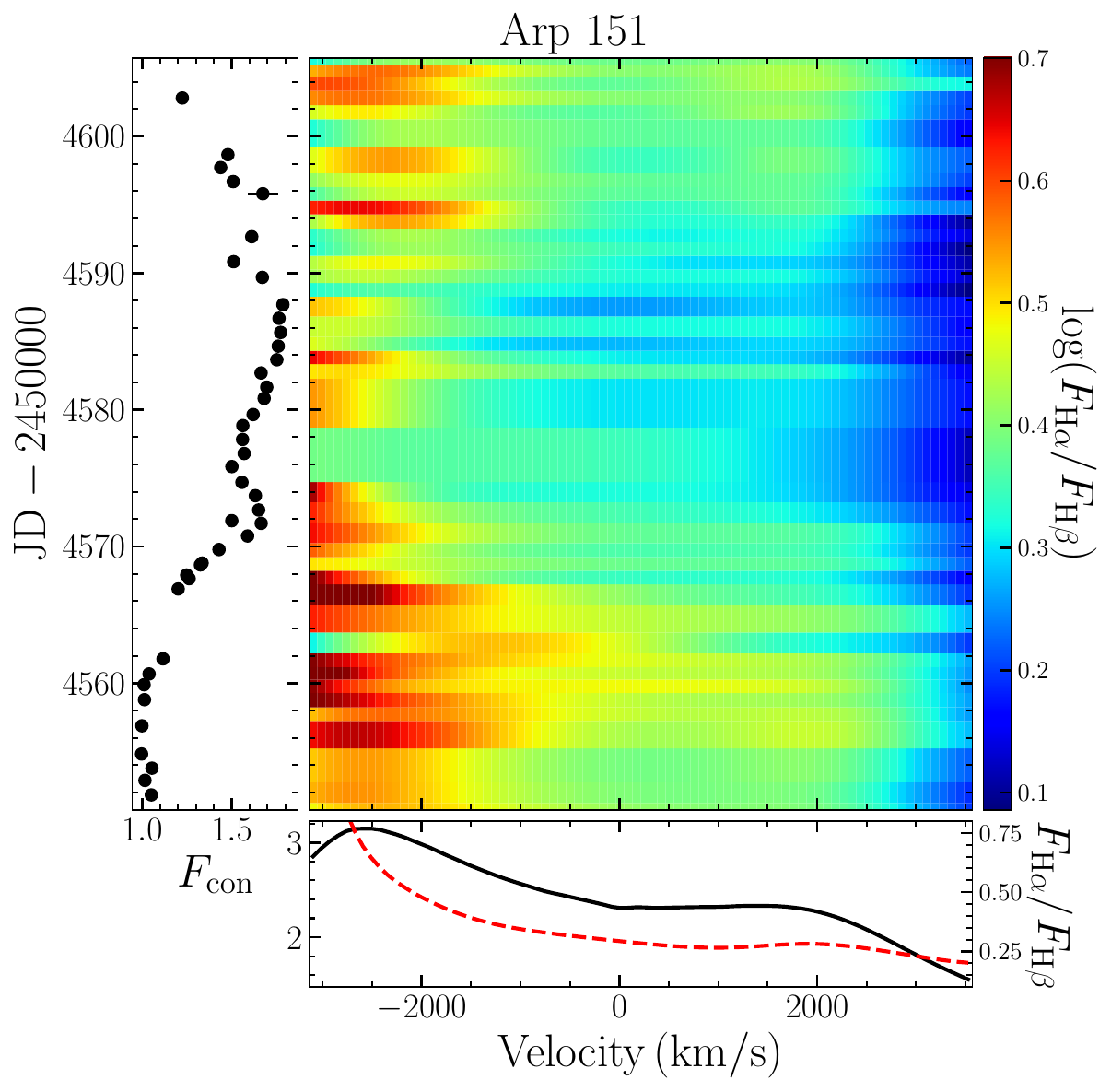}
\includegraphics[width=0.45\textwidth]{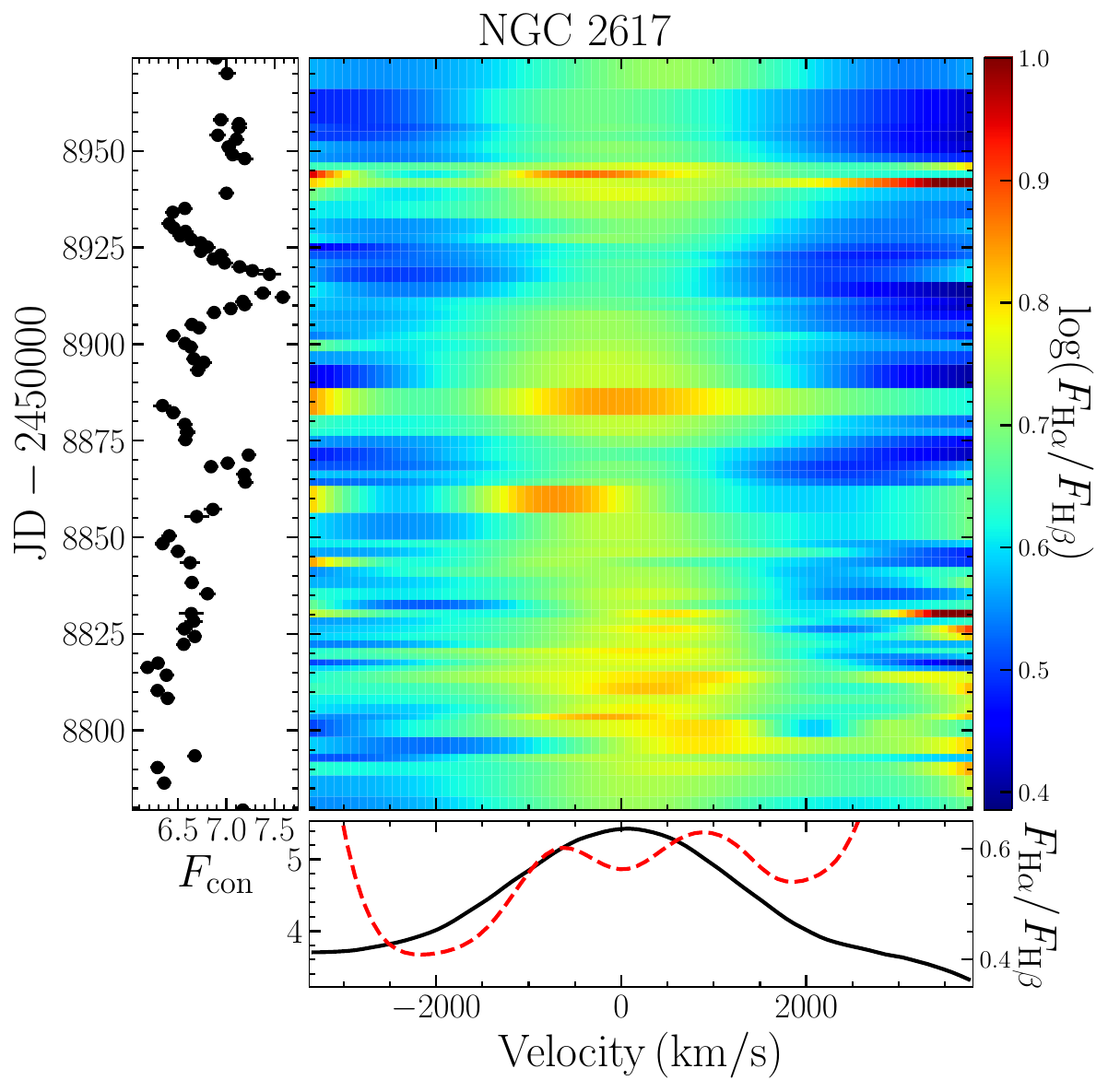}
\caption{Velocity-resolved emission line flux ratios of Arp~151 and NGC~2617. The black points represent the light curves of the continuum. The color maps show the distributions of \fab\ values at different velocities and times, where \fab\ values vary with velocity and time. The black lines indicate the velocity-resolved time-averaged values of \fab, and the red dashed lines represent the corresponding rms values. 
}
\label{fig:2D-bd}
\end{figure*}

\subsection{Measurement of Emission Line Flux Ratios} \label{sec:hahb}
Apart from determining time delays of emission lines, to understand the relationship between the continuum and emission line ratios, we also computed flux ratios between emission lines, such as \fab\ and \fac. To demonstrate the robustness of the relationship, we calculated the flux ratios of the emission lines using three distinct methods. The first results were derived from direct observations carried out on the same night, with ratio errors estimated using the uncertainties of spectral fitting. We used linear interpolation to align the dates of these ratio measurements with the continuum light curve, resulting in a newly adjusted continuum light curve. Considering the time delay between \ha\ and \hb, this could weaken the correlation between the directly measured BD and the continuum, as illustrated in Figure~\ref{fig:lc}, where the minimum value of the CCF correlation is delayed. To achieve a more accurate correlation, correcting the lag of each emission line is necessary. To mitigate the effects of both interpolation and time delay, we further examined the correlation between flux ratios of emission lines and \hb. 

Direct observation results neither consider the influence of outliers in the original light curves nor the difference in time delay. To test such factors in our analysis, we utilized the code JAVELIN\footnote{\url{https://github.com/nye17/javelin}}, which models the continuum and emission line variations using the DRW process \citep{Zu2011}, to reconstruct the light curve. We derived emission line flux ratios and uncertainties from these modeled curves, both at the same observation time and after time delay correction. We then conducted a comparison of the line ratios as determined by the direct observational method and JAVELIN-based methods, which are depicted in Figure~\ref{fig:lc} (showing three AGNs examples). The relationship remains robust across all three methods. Nevertheless, the presence of time delays and outliers will contribute to a dispersion in the results, reducing the apparent strength of the correlation. We also measured $\rm \Delta BD$, which was calculated by taking the median of the five highest values and subtracting the median of the five lowest values using the JAVELIN modeled curves after time lag correction. These values are detailed in Table~\ref{table1}.

In addition, we measured the flux ratios of emission lines as a function of velocity to investigate variation trend with line-of-sight velocity. We first resampled broad \ha\ and \hb\ profiles onto a shared velocity grid. We then computed flux ratios from the velocity-binned line profiles at each epoch. These velocity-dependent ratios are displayed against velocity and time in Figure~\ref{fig:2D-bd} to illustrate the more detailed variations. We also calculated flux ratios of emission lines in velocity bins that correspond to the velocity-resolved delays (as shown in the top panels of Figures~\ref{fig:vr} and \ref{fig:vr2}). By integrating the resampled \ha\ and \hb\ profiles at each velocity edge, we obtained light curves of \fab\ in these bins. Finally, we derived the weighted average velocity-resolved flux ratios by utilizing the \fab\ light curves at each velocity edge. It is important to note that these ratios were not corrected for the time delays of \ha\ and \hb\ relative to the continuum. The reason for this is that in our analysis, we used the average BD value for each velocity bin, which can eliminate the influence of time delay (see tests in Appendix~\ref{sec:appendix2}). To estimate the uncertainties of these ratios, we employed a bootstrap approach, similar to the approach for CCF error calculations, with 10,000 iterations. This process accounted for errors in the emission line flux ratios at each data point and the influence of the sampling. These results are illustrated in Figures~\ref{fig:vr} and \ref{fig:vr2}. 

\begin{figure*}[ht]%
\centering
\includegraphics[width=0.4\textwidth]{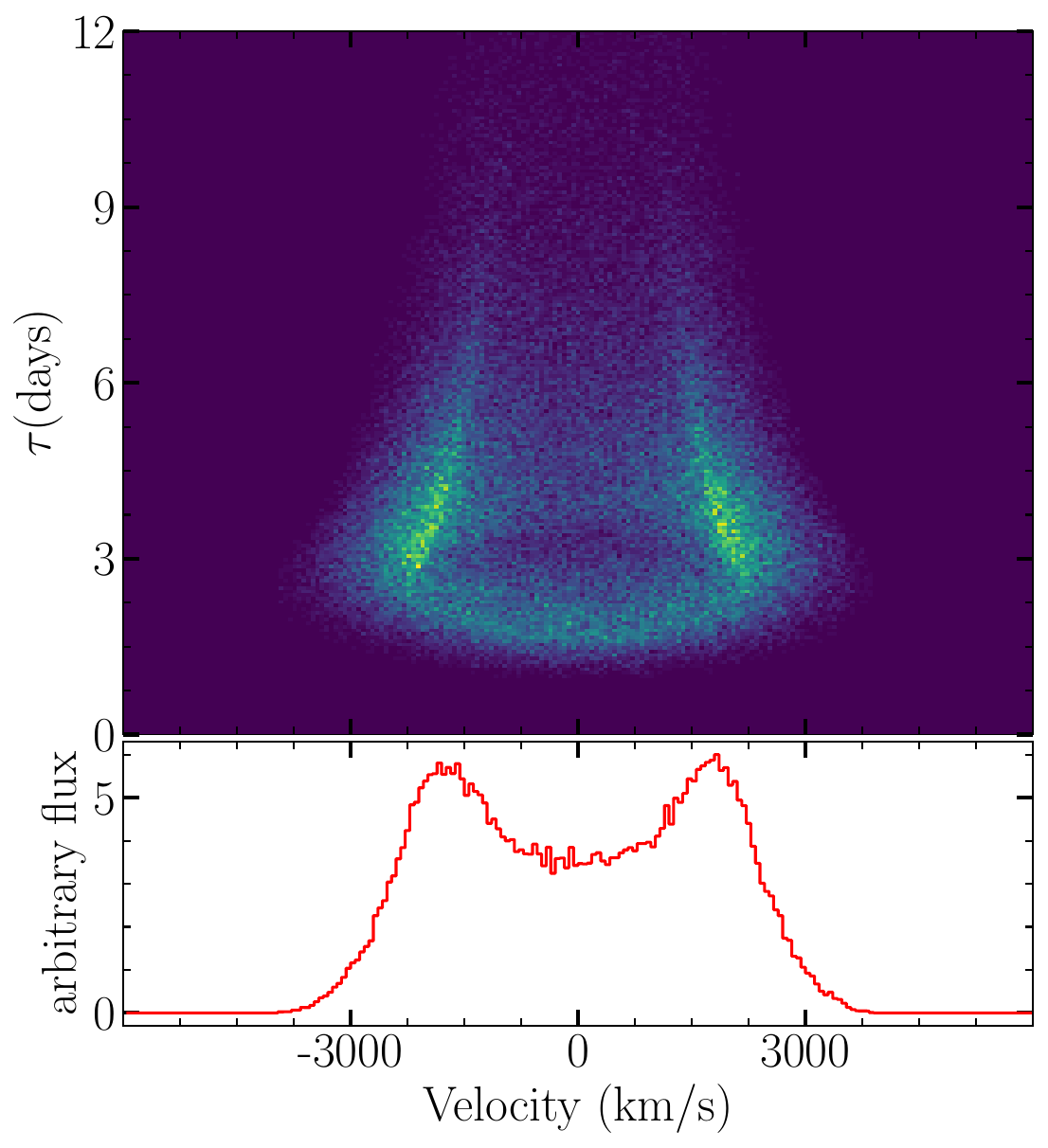}
\includegraphics[width=0.4\textwidth]{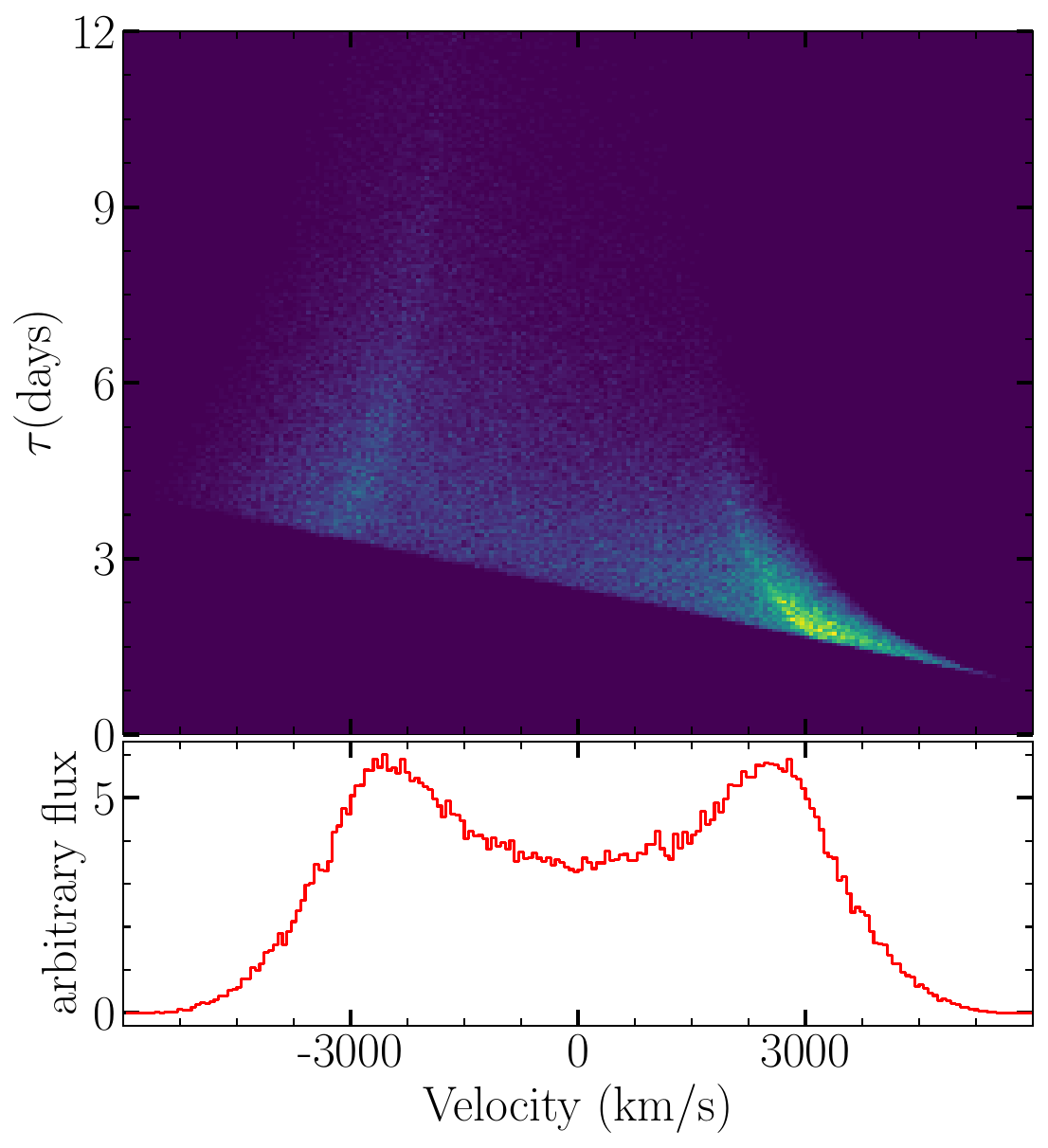}
\includegraphics[width=0.4\textwidth]{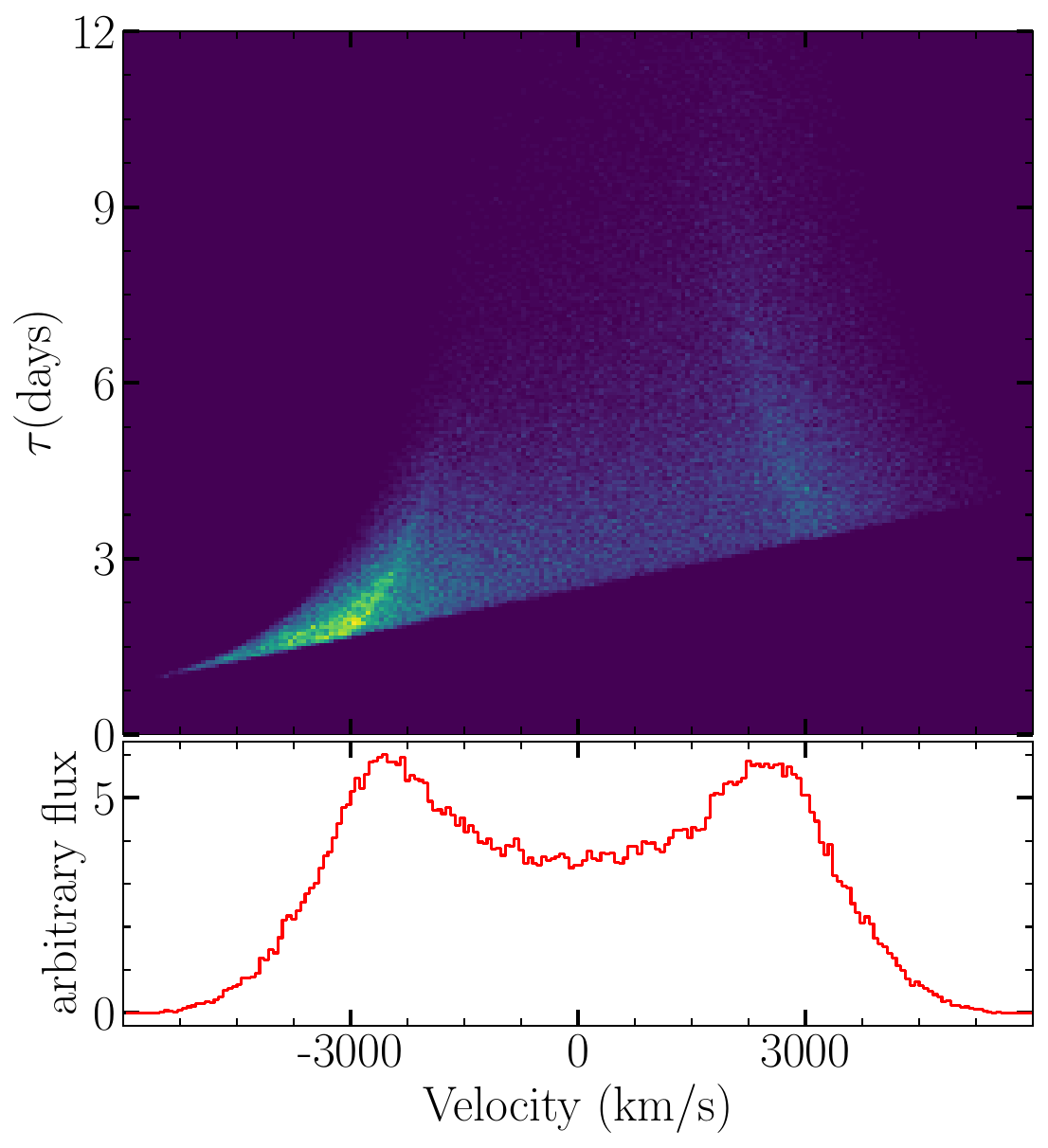}
\includegraphics[width=0.4\textwidth]{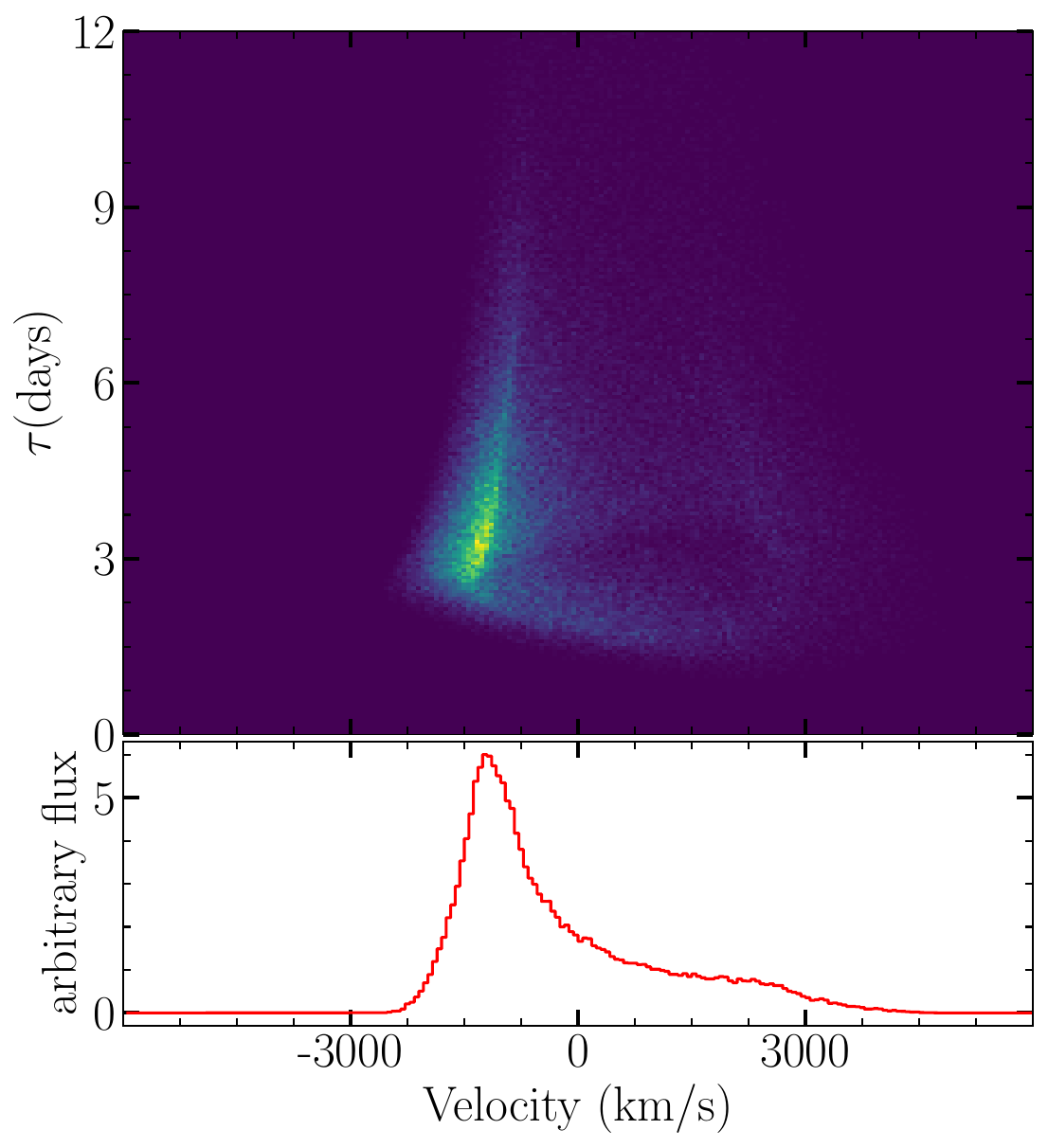}
\caption{Simulated TFs for modeled BLR configurations. The upper-left panel shows circular BLR with Keplerian motion. The upper-right panel shows circular BLR with Keplerian and inflow motion. The lower-left panel shows circular BLR with Keplerian and outflow motion.  The lower-right panel shows elliptical BLR with Keplerian motion. In each case, the red step line in the bottom panel represents the corresponding steady-state emission line profile.
}
\label{fig:tf}
\end{figure*}

\subsection{Transfer Function}
According to the fundamental principles of the RM technique \citep[e.g.,][]{Blandford1982, Welsh1991, Bentz2010}, the relationship between the light curves of the continuum and an emission line can be expressed by the following equation:
\begin{equation}
L(\nu,t) = \int_{- \infty}^{\infty}  \Psi(\nu, \tau) C(t- \tau) d\tau,   \label{eq1}
\end{equation}
where $\Psi(\nu,t)$ is the transfer function (TF) encoding the geometry and kinematics of the BLR. The terms $L(\nu,t)$ and $C(t- \tau)$ stand for the emission line and the continuum, respectively. Given the forms of the continuum and the emission lines, we can determine the corresponding TF for a specific BLR.

Studies on the BLRs of AGNs suggest that the BLR can be considered as a collection of discrete clouds surrounding a single SMBH. The radial emissivity distribution of these clouds can be well represented by a shifted $\Gamma$-distribution \citep{Pancoast2014}:
\begin{equation}
R_{\rm ga} = R_{\rm S} + FR_{\rm BLR} + g(1-F) \beta^{2} R_{\rm BLR},
\end{equation}
where $R_{\rm S}=2 G M_{\rm{BH}}/c^2$ is the Schwarzschild radius of a black hole with a mass \mbh, $R_{\rm BLR}$ is the mean BLR radius, $F$ is the ratio of the minimum radius to the mean BLR radius, $\beta$ is the shape parameter, and g is randomly drawn from a $\Gamma$-distribution. For a positive variable $x$, the $\Gamma$-distribution is described as:
\begin{equation}
P(x|\alpha,\theta) \propto x^{\alpha -1}\rm exp(-x/\theta),     
\end{equation}
where $\theta$ and $\alpha$ represent the scale parameter and shape parameter, respectively. In this paper, we set $\beta$ =0.9, $F$ =0.6, $\theta$ = 1, and $\alpha = 1/\beta^{2}$ for simplicity. Additionally, we take into account the angular distribution of the BLR clouds, which can be described by: 
\begin{equation}
\theta_{\rm mom} = {\rm arccos}[{\rm cos} \, \theta_{\rm o} + (1-{\rm cos} \, \theta_{\rm o}) \times U^{\gamma}],
\end{equation}
where $\theta_{\rm mom}$ represents the angle between the angular momentum of a cloud in the BLR relative to the BLR mid-plane. $U$ is a random number, and $U^{\gamma}$ is considered to describe the concentration of BLR clouds \citep{Pancoast2014}. In this work, we set $U^{\gamma}$ as a random number between 0 and 1, indicating that the BLR clouds are uniformly distributed but confined within $\pm \theta_{\rm o}$ range.

Here, we primarily model four disk-like BLRs with an opening angle of $\theta_{\rm o}=$ 20 degrees. This configuration suggests that the clouds are distributed within 20 degrees above and below the disk plane, representing a flattened disk geometry that more accurately reflects the realistic BLR than infinitely thin disk or spherical BLR geometry. Initially, we simulate the Keplerian elliptical and circular BLR to test the effect of geometry, followed by incorporating inflow and outflow motion into the Keplerian circular BLR. For each case, simulations are conducted with 100,000 particles (clouds) under a black hole mass of $2 \times 10^7 M_{\odot}$, similar to estimates for NGC~3516 \citep{Feng2021a}. The resulting TFs and steady-state emission line profiles are presented in the following panels of Figure~\ref{fig:tf}.

\begin{deluxetable}{lcccccc}[!htbp]
\tablecolumns{5}
\tablewidth{\textwidth}
\tabletypesize{\scriptsize}
\tablecaption{Correlation measurements}\label{table3}
\tablehead{ & \multicolumn{2}{@{}c@{}}{${\rm con}~^a$} & \multicolumn{2}{@{}c@{}}{${\rm H\beta~^b}$} \\ 
\cline{2-3} \cline{4-5}
\colhead{Object} & \colhead{$r_{\rm s}$} & \colhead{$p_{\rm null}$} & \colhead{$r_{\rm s}$} & \colhead{$p_{\rm null}$}}
\startdata
\multicolumn{5}{@{}c@{}}{LAMP2008 campaign} \\
\hline
Mrk 142 & 0.16 & 0.28 & -0.27 & 0.07 \\
SBS 1116+583A & -0.34 & 0.03 & -0.74 & $2 \times 10^{-08}$ \\
Arp 151 & -0.84 & $5\times 10^{-11} $& -0.80 & $5 \times 10^{-10}$ \\
Mrk 1310 & -0.54 & $2\times 10^{-04} $& -0.83 & $6 \times 10^{-13}$ \\
Mrk 202 & -0.56 & $5\times 10^{-05} $& -0.68 & $2 \times 10^{-07}$ \\
NGC 4253 & -0.21 & 0.19 & -0.77 & $2 \times 10^{-10}$ \\
NGC 4748 & -0.20 & 0.19 & -0.09 & 0.56 \\
IC 4218 & -0.36 & 0.04 & -0.95 & $5 \times 10^{-18}$ \\
MCG -06-30-15 & -0.06 & 0.71 & -0.80 & $6 \times 10^{-10}$ \\
NGC 5548 & -0.49 & $6\times 10^{-04} $& -0.68 & $1 \times 10^{-07}$ \\
Mrk 290 & -0.35 & 0.02 & -0.64 & $1 \times 10^{-06}$ \\
IC 1198 & -0.11 & 0.48 & -0.85 & $2 \times 10^{-13}$ \\
NGC 6814 & -0.56 & $1\times 10^{-04} $& -0.63 & $5 \times 10^{-06}$ \\
\hline
\multicolumn{5}{@{}c@{}}{LAMP2011 campaign} \\ 
\hline
Mrk 40 (Arp 151) & -0.82 & $4\times 10^{-10} $& -0.75 & $6 \times 10^{-08}$ \\
Mrk 50 & -0.93 & $4\times 10^{-19} $& -0.94 & $8 \times 10^{-20}$ \\
Mrk 279 & -0.18 & 0.33 & -0.76 & $2 \times 10^{-07}$ \\
Mrk 704 & -0.13 & 0.45 & -0.09 & 0.59 \\
Mrk 1511 & -0.63 & $2\times 10^{-05} $& -0.62 & $4 \times 10^{-05}$ \\
NGC 4593 & -0.88 & $1\times 10^{-14} $& -0.73 & $5 \times 10^{-08}$ \\
Zw 229-015 & -0.63 & $9\times 10^{-04} $& -0.70 & $2 \times 10^{-04}$ \\
\hline
\multicolumn{5}{@{}c@{}}{Lijiang CL-AGN campaign} \\
\hline
NGC 2617 & -0.58 & $7\times 10^{-07} $& -0.88 & $3 \times 10^{-21}$ \\
NGC 3516 & -0.67 & $7\times 10^{-10} $& -0.92 & $9 \times 10^{-28}$ \\
NGC 4151 & -0.49 & $1\times 10^{-03} $& -0.84 & $1 \times 10^{-11}$ \\
\enddata
\tablecomments{The objects analyzed in this study were selected from the LAMP2008, LAMP2011, and Lijiang campaigns. We determined the Spearman rank correlation between BD and continuum, as well as BD and \hb. The results include the Spearman rank correlation coefficient ($r_{\rm s}$) and probability of the null hypothesis ($p_{\rm null}$).\\
$^a$ The Spearman rank correlation analysis between continuum and BD.\\
$^b$ The Spearman rank correlation analysis between \hb\ and BD.}
\end{deluxetable}

\section{Results}\label{sec:result}
\subsection{Anti-correlation between BD and Continuum}
The BD is defined as the relative intensities of the hydrogen Balmer series emission lines. \ha\ and \hb\ are among the strongest optical emission lines, and their flux ratio serves as the most frequently utilized BD. We compiled a sample of 22 AGNs with simultaneous RM observations of both \ha\ and \hb\ from the LAMP2008 \citep{Bentz2009} and LAMP2011 \citep{Barth2015} and Lijiang CL-AGN RM campaign\citep{Feng2021a, Feng2021b, Li2022}, and some of these sources also included \hc\ observations. We conducted spectral fitting for 15 objects (from LAMP2008 and Lijiang campaigns) with available spectral data to obtain pure broad emission line profiles. For the remaining sources, we adopted the spectral decomposition results from \citet{Barth2015}. By utilizing the fluxes extracted from the fitted broad emission lines, we can initiate an investigation into the variability of the BD.

Figure~\ref{fig:lc} illustrates examples of BD light curves obtained from the LAMP2008, LAMP2011, and Lijiang projects. It is evident that the BD can change by more than 1 even within a timescale of $\sim$2 months, and similar variability amplitudes are observed in most of the other sources. This suggests that the BD in AGNs can exhibit considerable variability. Comparing the BD and continuum light curves reveals almost completely opposite variability trends. To quantify their correlation, we employed the Spearman rank correlation and found a significant anti-correlation in 15 sources with 16 observations (see Table~\ref{table3}). Therefore, decreasing BD with increasing continuum intensity appears to be a common phenomenon. 
Upon further examination of the data, we noticed that the remaining 7 sources, which did not show an inverse correlation, were influenced by small variability amplitudes, non-synchronous observation times between the continuum and emission line, and outliers. Furthermore, 5 of these sources did not have a reliable time delay measurement from the CCF, because their CCFs displayed anomalous profiles with $r_{\rm max}<$ 0.5. Notably, our sample (including previous RM observation data) showed a significant time delay between \ha\ and \hb\ variations in many AGNs. Variability introduces additional errors when attempting to measure the BD values at the same incident photon level in the BLR using a single spectrum.

To assess the impact of these factors on our conclusions, we employed a method based on the DRW model to fit multiple light curves simultaneously (see details in Section~\ref{sec:hahb}). This approach can mitigate the influence of outliers on individual light curves and enable us to test the effects of time delay on correlation. We found a generally enhanced correlation between the continuum and the BD measured with the DRW-based light curves, while almost all theoretical DRW results corrected for lag display an inverse correlation. Furthermore, by substituting \hb\ for the continuum, we can somewhat reduce the effects of interpolation and lag, and found that 19 sources show a significant inverse correlation. We also conducted tests using the BD values measured from \ha\ and \hc, and the results exhibit an identical relationship. This relationship further supports the notion that the BD values in the BLR are anti-correlated with the continuum fluxes.

The inverse relation between BD and the continuum can be explained by the intrinsic Baldwin effect, which follows a power law relationship \citep[e.g.,][]{Gilbert2003}, where $F_{\rm line} \propto F_{\rm con}^{m}$. Specifically, for \ha\ and \hb, they can be expressed as powers of $m1$ and $m2$, respectively. Then the BD is determined by $F_{\rm H\alpha}/F_{\rm H\beta} \propto F_{\rm con}^{(m1 - m2)}$, and will be anti-correlated with the continuum when $m1 < m2$, where $m$ represents the effective line responsivity. This is consistent with the predictions from photoionization models \citep{Korista2004}.

\subsection{Results of Velocity-resolved IM and RM}
The inherent ionization property leads to the natural inference that gas clouds in the BLR located farther from the central ionizing source should exhibit larger BD values. Therefore, we can study the gas distribution in the radiation region by measuring the BD values at different velocities of broad emission lines. In other words, comparing BD values between different velocity bins allows us to obtain the relative distances of the emitting gas clouds from the center. This technique is similar to velocity-resolved RM but is independent of line-of-sight and kinematics. When combining these two methods for analysis, we can effectively break the degeneracy of BLR geometry and kinematics.

This kind of study requires high data quality, and we only analyzed 5 objects that can simultaneously measure both \ha\ and \hb\ velocity-resolved lags. Initially, we examined the variation in the distribution of BD along the velocity direction -- the velocity-resolved BDs -- using the fitted broad emission line profiles. Figure~\ref{fig:2D-bd} presents an example of BD as a function of velocity and time for the LAMP2008 and Lijiang observations. As expected, we do observe substantial variations in BD values at different velocities. Notably, the BDs at any given velocity consistently exhibit an inverse correlation with the continuum, and the overall shapes of the velocity-resolved BDs are almost identical across each epoch. This inverse correlation indicates that the BLR geometry should be constant throughout the monitoring period. To mitigate the influence of variability, we constructed an average of the BD values at each velocity across all observations. Such a process allowed us to examine the ultimate form of the BD profile across different velocities. Our findings revealed that the BD is symmetric around zero velocity in NGC~2617 and SBS~1116+583A. In contrast, Arp~151 and NGC~3516 display a decrease in BD with increasing velocity, while NGC 4151 exhibits the opposite trend.

To compare the relationship between the BDs and lags at different line-of-sight velocities of the broad emission line, we divided the rms spectrum of broad \ha\ into several bins of equal flux, and then measured the BD and lag values within each bin. We found that, except for NGC~4151, the remaining four sources exhibit a robust positive correlation between the BD and lag values (see Figure~\ref{fig:vr} and \ref{fig:vr2}). This correlation suggests that gas emitting with longer time delays is indeed located at a considerable distance from the ionizing source for these objects. The validity of this inference can be further confirmed by analyzing the variation in the \ha-to-\hb\ ratio at different velocities. In most cases, where gas density exceeds $10^{10}$ cm$^{-3}$ and the \fab\ ratio falls within the range of 2 to 10 \citep[as shown in Figure 5 of][]{Korista2004}, a higher hydrogen-ionizing photon flux (i.e., a smaller distance from the center) corresponds to a less pronounced response in the BD. Specially, at sufficiently high levels of ionizing photon flux, the BD becomes nearly insensitive to continuum variations. Our approach primarily involved assessing the rms values at different velocities to capture the BD variability, despite the inherent uncertainties associated with rms due to its sensitivity to S/N ratios. The distribution of the BD values is broadly consistent with the distribution of their rms values, as shown in Figure \ref{fig:2D-bd}. Intriguingly, Arp~151 and NGC~3516 exhibit distinct distances from the blue and red sides of the emission line to the center, suggesting the existence of an asymmetric geometry in the BLR. Moreover, the anti-correlation of BD and lag in NGC~4151 could serve as evidence of the complex kinematics within the BLR \citep{Chen2023}.

\section{Discussion}\label{sec:discu}
\subsection{Mechanisms of BD Variation}
It has been reported in some previous work that there is a negative correlation between BD and continuum over years to decades in several individual AGNs \citep{Shapovalova2004, Popovic2011, Ma2023, Oknyansky2023}. This phenomenon, especially when accounting for the CL mechanisms, has been partly ascribed to dust extinction variations \citep{Kollatschny2000}, whilst others propose ionization effects within the BLR as the principal cause \citep{Shapovalova2010}. It is generally believed that dust grains are located beyond the sublimation radius (around 3-5 times the BLR radius), otherwise they would be vaporized within hours \citep{Baskin2018}. Changes in dust clouds on these size scales typically require decades \citep{LaMassa2015}, which could explain some long-term BD variations. However, the objects in our selected sample were monitored for only 2-6 months, and the variability timescales of their BDs are even shorter at $\sim$ 2 months. This discrepancy suggests that it is implausible to expect significant changes in the obscuring material. To validate our hypothesis, we conducted two tests (see Appendix~\ref{sec:appendix1} for details): First, we used Zwicky Transient Facility (ZTF) data to examine variability of spectral index for these AGNs (often used as an extinction proxy), and found that their spectral indices remained nearly constant on a timescale of $\sim$5 years. Second, we simulated the BD-continuum relationship under varying extinction conditions, which did not reproduce the observed patterns, further excluding the impact of extinction.

The detection of delayed emission line responses relative to continuum changes for all substantially variable sources contradicts expectations from extinction models (see Table~\ref{table2}). Unified AGN models posit a dusty torus surrounding the outer BLR \citep{Antonucci1993}. If extinction changes drove variations, the BLR flux would respond first before the accretion disk continuum. This would result in emission line variations preceding continuum changes, which are not observed. Therefore, the anti-correlation between BD and the continuum shown in our study reflects an ionization effect rather than extinction. For a finite BLR size, the smaller mean formation radius and larger effective responsivity of \hb\ will result in a lower BD at increased flux levels. In reality, an increase in continuum emission can illuminate more distant BLR regions, which enhances the broad Balmer emission lines with larger local BDs. Despite this, the overall BD measured across the BLR is observed to decline, primarily due to the overwhelming contribution from closely orbiting, more heavily-ionized inner clouds, which possess lower intrinsic BDs. This effect becomes more pronounced with increasing central ionizing flux. The observed BD-continuum anti-correlation reinforces the fundamental relationship between local BD and radius, which is necessary to the velocity-resolved IM technique. 

\begin{figure*}[!ht]%
\centering
\includegraphics[width=0.4\textwidth]{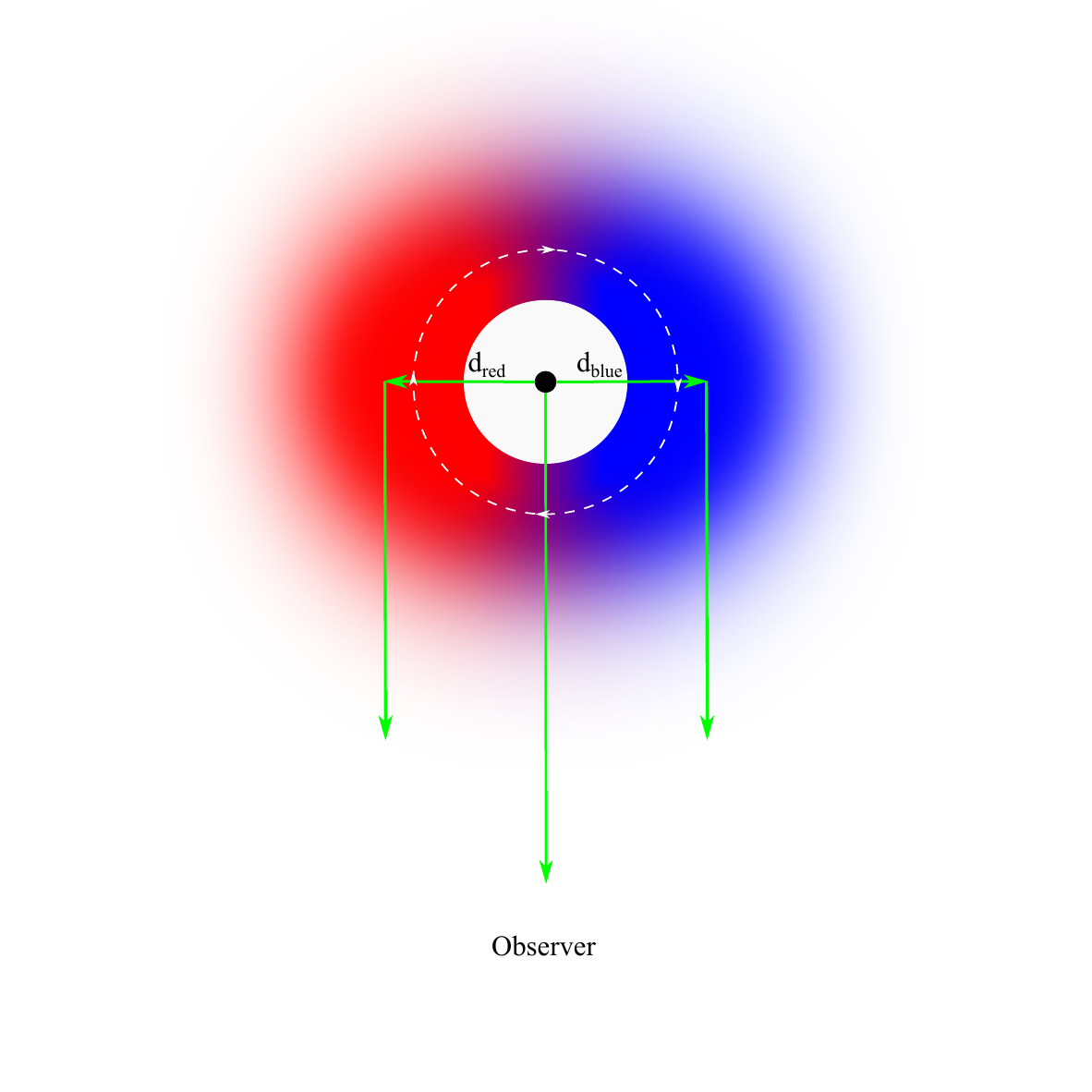}
\includegraphics[width=0.4\textwidth]{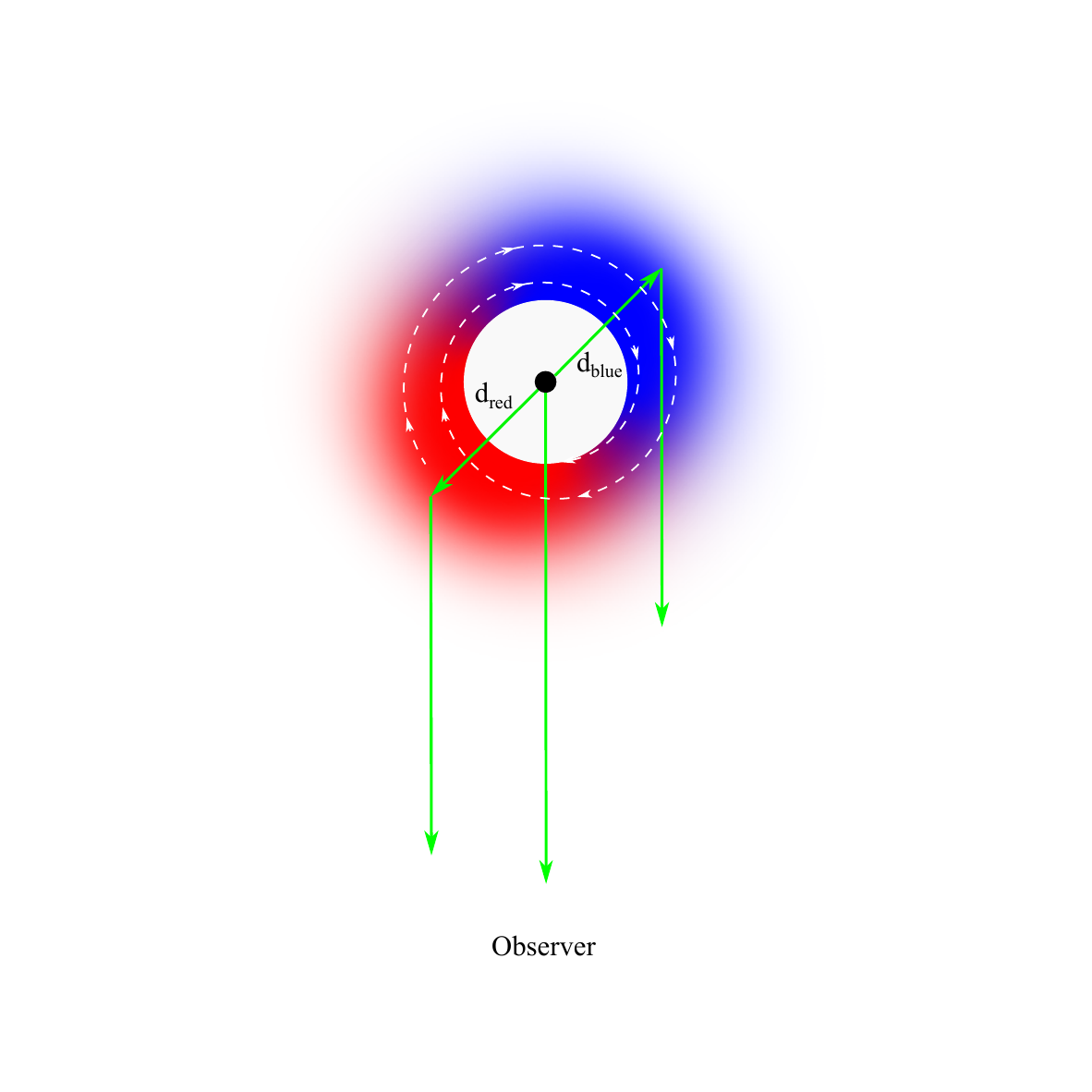}
\includegraphics[width=0.4\textwidth]{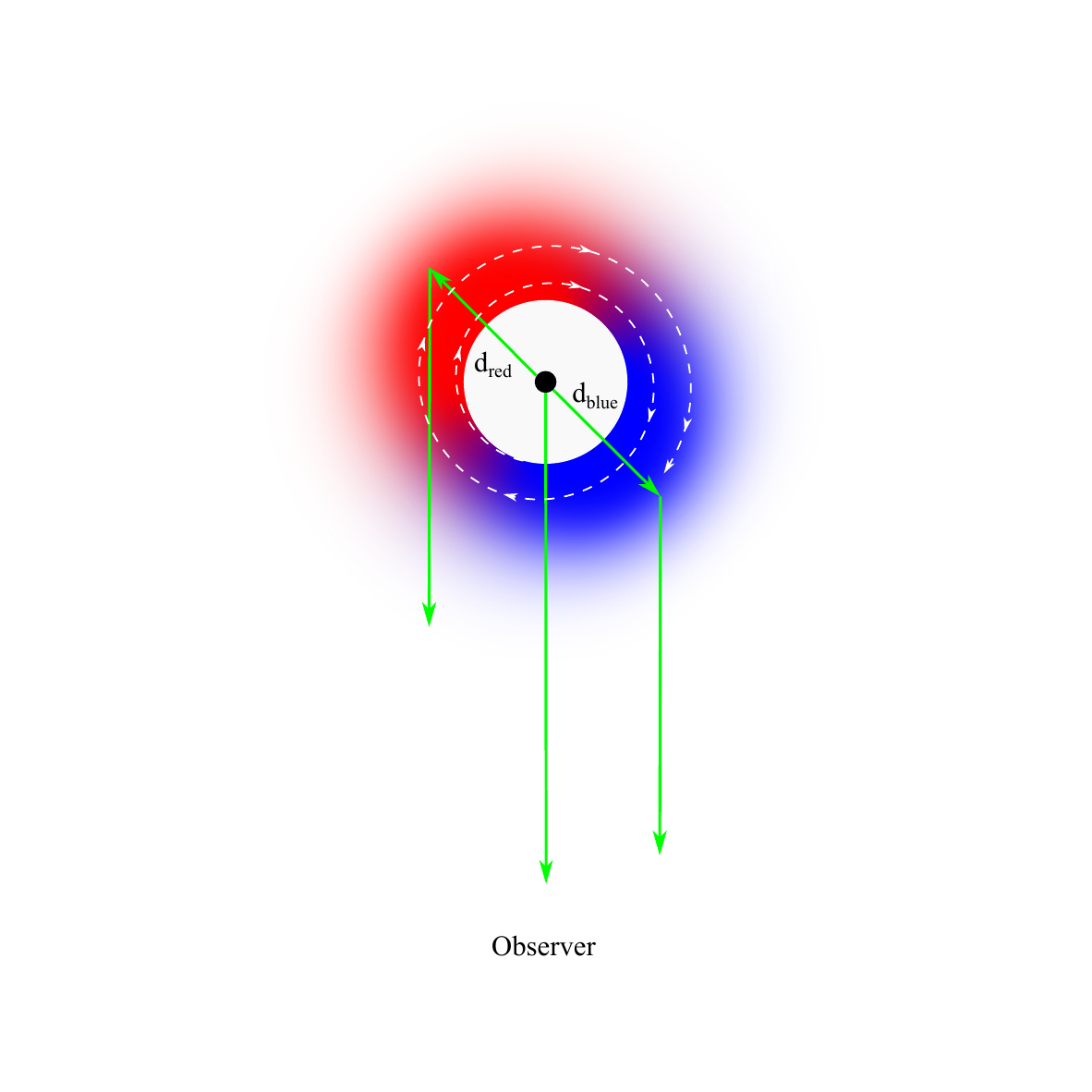}
\includegraphics[width=0.4\textwidth]{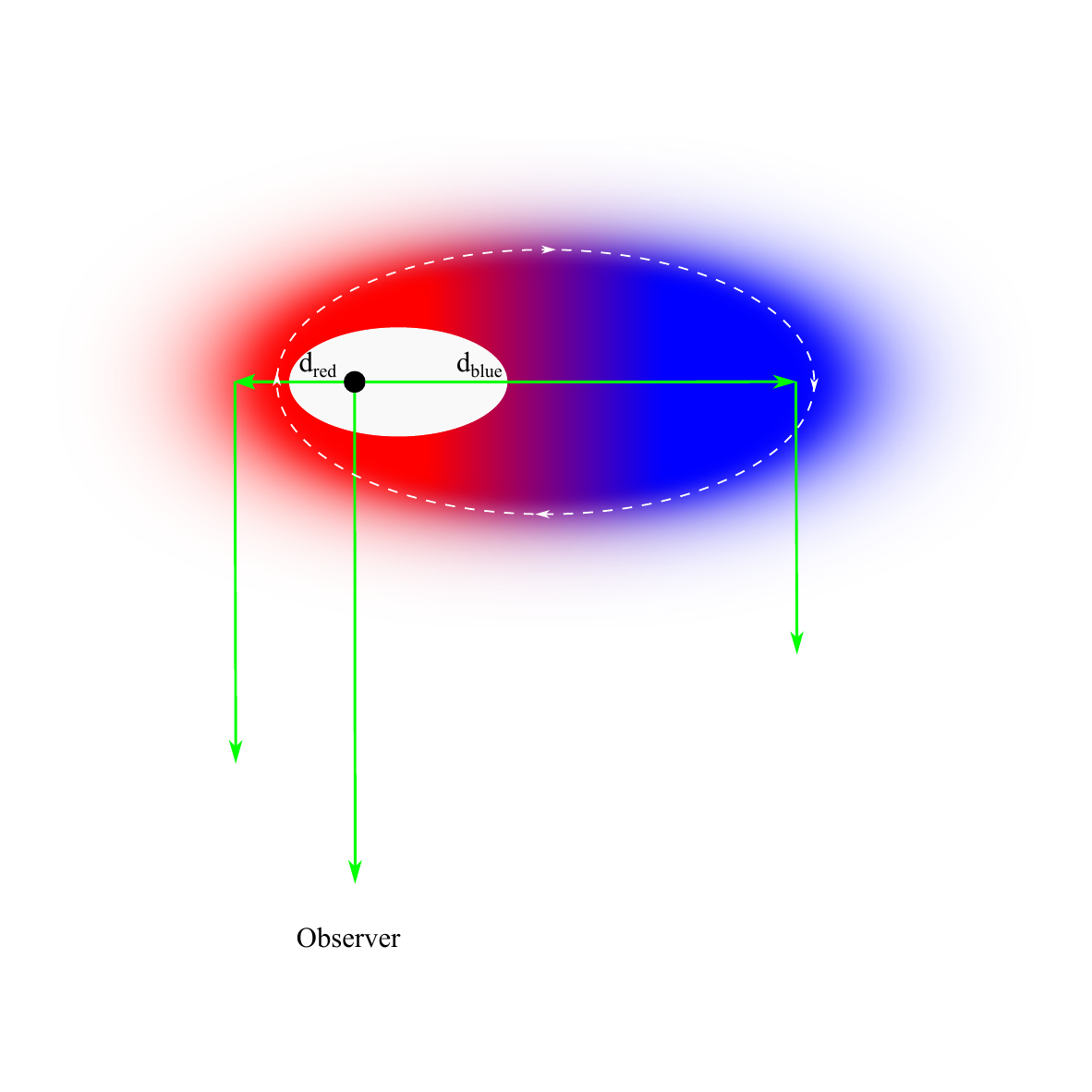}
\caption{A cartoon depiction of kinematics and geometry effects on velocity-resolved BDs and lags. The upper-left panel shows circular BLR with Keplerian motion. The upper-right shows circular BLR with Keplerian and inflow motion. The lower-left shows circular BLR with Keplerian and outflow motion. The lower-right shows elliptical BLR with Keplerian motion. In each case, the dashed line is the trajectory of a cloud, and the arrows are the motion direction of the cloud. The green line is the direction of the photon propagation. The red color represents the part of BLR where emission lines are redward shifted with respect to the black hole, and the blue color represents the part of BLR where emission lines are blueshifted.}
\label{fig:cartoon}
\end{figure*}

\subsection{Velocity-Resolved IM Technique}
Velocity-resolved IM technique assumes that these local BDs of BLR gas clouds increase monotonically with distance from the ionization source, a reasonable premise for idealized dust-free environments with homogeneous matter distributions. Real AGNs deviate from these idealized conditions - the ubiquitous presence of dust can impact observed BDs through internal extinction \citep{Czerny2011,Baskin2018}. Radial dust density gradients necessitate re-evaluating the local BD-distance correlation even if the overall extinction level holds steady. Near the outer boundary of BLR where it transitions into the dusty torus, mixing of gas and dust becomes increasingly likely as temperatures approach dust sublimation thresholds. Here, outwardly rising dust densities could amplify BDs beyond these increments predicted by dust-free models.

The ionization state of the BLR is not solely determined by the incident continuum, but is also intricately linked to the physical condition of the gas \citep{Ilic2012}. When significant changes occur in the physical parameters within the BLR, the BD value will also vary accordingly. One key parameter influencing the positive BD-distance relationship is hydrogen density. In the vicinity of the central region, where gas density is high, this leads to an increased tendency for collisional excitation, which enhances the emission of the \ha\ line and thereby increases the BD value. Meanwhile, as gas density reaches a certain threshold, the BD value tends to stabilize or even inversely correlate with density increases \citep{Korista2004}. This is attributed to the reduced efficiency of converting ionizing photons into escaping line photons at high gas densities, particularly affecting \ha\ with its higher optical depth compared to \hb. These properties may lead to a more complex distribution of BD values within the BLR, where the former effect seems to contradict the foundational premises of velocity-resolved IM and introduces some uncertainty in its application, while the latter effect still supports the assumption of BD increasing with distance. 

Current studies indicate a weak correlation between BD and hydrogen density. Even significant changes in density can induce only modest changes in BD, as illustrated by the fact that a density contrast of several orders of magnitude between the BLR and the narrow line region typically translates into a predicted BD variation of about 0.25 under Case B recombination conditions. Our results demonstrate that a 20\% fluctuation in the continuum can result in a significant change in the BD, with absolute values greater than 1 in some objects as shown in Figure~\ref{fig:lc} and Table~\ref{table1}. This indicates that within the BLR, where hydrogen density is expected to decrease with distance following a power-law of $n(R) \propto R^{-s}$ ($s$ often between 0 and 2)—the effect of the incident continuum on BD is more pronounced than the effect of varying hydrogen densities. Given that the incident continuum flux diminishes approximately as $F(R) \propto R^{-2}$, outpacing the decrease in hydrogen density, we conclude that the ionizing effect of the continuum is the dominant factor in determining the BD in AGNs. In reality, the presence of dust or higher gas density in certain regions of the BLR \citep{Netzer2015} can lead to a more rapid decline in the incident continuum flux $F(R)$ due to absorption and scattering \citep{Baskin2018}, further reinforcing our conclusion.

Photoionization simulations show that the emissivity distributions of \ha\ and \hb\ in BLR do exhibit differences (or different transfer functions), displaying a trend that $F_{\rm H\beta}$ decreases faster than $F_{\rm H\alpha}$ with the increasing distance from the center \citep{Guo2020, Wu2023}. The implications of this property are significant for velocity-resolved IM, as it suggests the BD value is a monotonic function of radius. For a symmetric Keplerian BLR \citep{Horne2004}, the detected BD and lag values should be consistent for the emitting gas clouds with the same line-of-sight velocities on the blue and red sides of broad emission line because these clouds are equidistant to the center of the BLR. Considering that the gas cloud closer to the central black hole moves at a higher velocity and receives a stronger flux of the incident continuum, the detected lag is expected to be shorter, and the BD value should be smaller \citep[see discussion in][for more details]{Korista2004}. A decreasing/increasing trend of lag from the blue side to the red side of broad emission line is a key characteristic previously attributed to inflow/outflow within the BLR. In this scenario, the BLR is still considered to be symmetric, but the viewing angle causes the blue/red-shifted gas to appear more distant, resulting in a longer lag. 

However, for a symmetric BLR, regardless of its dynamical state, the BD profile should maintain symmetry in the direction of line-of-sight velocity. This is due to the fact that the BD value is solely dependent on its distance from the central ionizing source and is independent of the propagation path of the light. Figure~\ref{fig:cartoon} gives a schematic cartoon showing the effects of kinematics on the velocity-resolved lags and BDs, and the corresponding simulated two-dimensional velocity-resolved time lag maps are shown in Figure~\ref{fig:tf}. These simulations illustrate that both inflow and outflow can result in asymmetric velocity-resolved lag profiles across the broad emission line, assuming that the ionization properties are not influenced by kinematics.

\subsection{Diverse geometry and kinematics of BLR}
Our results reveal remarkable diversity in the geometry and kinematics of the BLR among different AGNs. For NGC 2617 and SBS 1116+583A, the symmetric nature of the velocity-resolved BD profiles, together with corresponding time lags, are indicative of a BLR that predominantly exhibits Keplerian motion in a symmetric geometry (Figure~\ref{fig:cartoon}). Conversely, the scenarios presented by Arp~151, NGC~3516, and NGC~4151 imply a starkly different picture. Their longer lags in the blue wing suggest the possibility of inflowing characteristics or the blue-shifted gas being located farther from the center. In the context of Arp~151 and NGC~3516, distinct BD-lag correlations on the blue and red wings point to non-axisymmetric geometry. 

Intriguingly, NGC~4151 shows an anti-correlation between velocity-resolved BD and lag values. This peculiar signature may be attributed to special motions within the BLR. Indeed, we noticed that the shape of the velocity-resolved \hb\ lags exhibited temporal changes during the 2020-2021 period \citep{Chen2023}, but the results from additional epochs between 2018 and 2022 align with our velocity-resolved BDs conducted in 2020-2021. The transient kinematic shifts could be related to the dramatic fluctuations in radiation pressure owing to changes in AGN luminosity, or some localized perturbations within the BLR, neither of which should substantially impact on the global geometry of the BLR. It must be emphasized that the aforementioned conclusions are predicated on the assumption of symmetrical kinematics within the BLR. The presence of an asymmetric inflow or outflow could also replicate similar observational signatures under finely-tuned viewing angles. For instance, in Arp~151 and NGC~3516, we might coincidentally detect only the blueshifted inflowing gas, whereas, in NGC~4151, the situation could be the inverse. These kinematic properties would likely be inconsistent with the virialized motions and can be tested through multiple RM observations.

There are several potential configurations of geometry for such an asymmetric BLR. One possible scenario is that the BLR contains an uneven spiral arm or hot spot, creating a stronger response at some particular velocity \citep{Newman1997, Storchi-Bergmann2003}. These substructures can change as the BLR undergoes rapid motion, causing variations in the velocity-resolved signal as shown in NGC~4151. The velocity-resolved RM measurements of Arp~151 and NGC~3516 have remained virtually unchanged for several years \citep{Denney2010, Pancoast2018, Feng2021a}, requiring a BLR with stable geometry and kinematics. A Keplerian elliptical disk model may satisfy this condition \citep{Eracleous1995}. This configuration can persist for long durations and will result in unequal distances between the red and blue sides of the BLR to the central ionizing source (Figure~\ref{fig:cartoon}). Also, gas clouds in elliptical orbits would have orbital velocities varying with their positions. This would lead to an observational asymmetric profile of broad emission line due to the longer orbiting time at the apocenter. Figure~\ref{fig:tf} shows the transfer functions and emission line profiles given by our simulations of elliptical disk BLR. It is evident that the emission line on the blue side is noticeably stronger than on the red side, and the lag on the blue side is generally larger than on the red side. These findings are consistent with observational results in NGC~3516. Binary SMBH can also lead to asymmetric BLR \citep{Eracleous1997}, but this situation is generally believed to produce periodic variability, which are rarely detected reliably in AGNs.

\subsection{Implications for AGN physics}
The architecture of an asymmetric BLR might be related to the formation and evolution of AGNs, such as the merging of SMBHs, the formation of BLR, the supply of matter to the accretion disk, and the feedback mechanisms. \citet{Eracleous1995} proposed that an elliptical disk in the BLR could be associated with binary black holes or the tidal disruption of stars by an SMBH. \citet{Wang2017} suggested that tidally disrupted clumps by the central SMBH from the dusty torus can form the BLR, including spiral-in gas as inﬂow, circularized gas, and ejecta as outﬂow. In this framework, the evolutionary timescale of the BLR typically requires decades or longer, consistent with the observed stability of the BLR in Arp~151, NGC~3516. The detection of a tidal disruption event in 1ES 1927+654 also supports this theory, which suggests that the BLR originates from the torus \citep{Trakhtenbrot2019}. If the inward spiraling process is disturbed, for instance, by a smaller black hole or a star passing through, it could form a spiral arm structure or an elliptical disk. Alternatively, the BLR could simply be a result of failed radiation-driven dusty outflows from the accretion disk \citep{Czerny2017}. In this model, if the powerful winds are anisotropic, it can also lead to an asymmetric BLR. Furthermore, when there are enough clouds in the BLR, their self-gravity can also produce some non-uniform structures \citep{Du2023}.

While previous observations have suggested the possibility of asymmetry within the BLR, and various theoretical models have been proposed to explain this asymmetric geometry, there is still a tendency to utilize simpler symmetrical models when interpreting observational results. The joint analysis of velocity-resolved BDs and time lags can effectively reveal non-axisymmetric geometries of the BLR. In the future, as we confirm more instances of such asymmetry, it will enable us to investigate the correlation between the geometry of the BLR and the physical parameters of AGN, such as the accretion rate and the black hole mass. These relationships are vital to our comprehension of the formation of the BLR and the internal physical processes in AGNs. Furthermore, the geometry of BLR is essential to accurately measure \mbh, as emission line profiles and time delays resulting from asymmetric BLRs significantly differ from expectations of circular disk-shaped BLR models (Figure~\ref{fig:tf}). Therefore, further research into asymmetric BLRs is warranted to improve our understanding of AGN dynamics and measurements of \mbh.

\section{Summary} \label{sec:summary}
In this work, we utilized high-quality spectroscopic monitoring data from the LAMP2008, LAMP2011, and Lijiang projects to conduct a detailed analysis of the geometric and kinematic characteristics of the BLR in AGNs. Specifically, we examined the flux ratios and time delays of the \ha\ and \hb\ emission lines in this sample. By integrating the new technique of velocity-resolved IM with traditional velocity-resolved RM, we unveiled the complex kinematics and geometry within the BLR. Our key results are as follows.

\begin{enumerate}
\item We detected a significant anti-correlation between the BD and continuum on timescales of weeks to months in 15 out of 22 AGNs. Dust extinction variations were excluded as the primary driver, indicating that this phenomenon arises from ionization effects within the BLR. The decreasing BD with increasing ionizing flux is consistent with photoionization model predictions.

\item Tight correlations between velocity-resolved BDs and RM lags were found in 4 out of 5 AGNs with high-quality data. The BD profiles of NGC 2617 and SBS 1116+583A are symmetric around zero velocity, suggesting a Keplerian BLR with symmetric geometry. In contrast, the asymmetric BD profiles and distinct lags on the blue and red sides in Arp 151 and NGC 3516 point to non-axisymmetric BLR geometry, possibly an elliptical disk or spiral arm structure. Furthermore, NGC 4151 exhibits an anti-correlation between velocity-resolved BDs and lags, implying complex kinematics within the BLR.

\item We discussed the assumptions and limitations of the IM technique. While factors like radial dust distribution and gas density may affect local BDs, the incident continuum flux is likely the dominant driver of the BD-distance relation within the BLR. Simulations of various BLR configurations demonstrated the power of combining IM and RM to distinguish between geometric and kinematic effects.

\item The detection of asymmetric BLR geometry has important implications for AGN physics. It may represent a formation imprint related to SMBH mergers, tidally disrupted clumps from the torus, or radiation-driven disk winds. Asymmetry can also impact SMBH mass measurements. Further investigations into the correlations between BLR geometry and AGN properties are warranted.
\end{enumerate}

This study showcases the potential of the velocity-resolved IM technique in probing the BLR geometry and its promise in advancing our understanding of AGN physics when combined with RM. Future applications to larger samples will provide valuable insights into the formation and evolution of AGNs and refine SMBH mass estimates. As spectroscopic monitoring data continues to accumulate, the IM method can be extended to more objects, opening up new avenues for exploring the diversity of BLR geometry and kinematics across the AGN population.

\vspace{5mm}
We appreciate the referee for constructive comments and valuable suggestions, that greatly improved this paper. We are grateful to J.-R. Mao, J.-C. Wu, and H.-X. Guo for the comments that improved the manuscript. This work is supported by National Key R\&D Program of China (No. 2021YFA1600404), the National Natural Science Foundation of China (grants No. 12303022, 12203096, 12373018, and 11991051), Yunnan Fundamental Research Projects (grants NO. 202301AT070358 and 202301AT070339), Yunnan Postdoctoral Research Foundation Funding Project, Special Research Assistant Funding Project of Chinese Academy of Sciences, and the science research grants from the China Manned Space Project with No. CMS-CSST-2021-A06.

We acknowledge the support of the staff of the Lijiang 2.4 m telescope. Funding for the telescope has been provided by Chinese Academy of Sciences and the People’s Government of Yunnan Province. This work has made use of data from the Lick AGN Monitoring Project public data release. The Zwicky Transient Facility Collaboration is supported by U.S. National Science Foundation through the Mid-Scale Innovations Program (MSIP). This research has made use of the NASA/IPAC Extragalactic Database (NED), which is operated by the Jet Propulsion Laboratory, California Institute of Technology, under contract with NASA.

\vspace{5mm}

\facility{YAO:2.4m}
\software{DASpec (\url{https://github.com/PuDu-Astro/DASpec}), PyCALI \citep{Li2014}, JAVELIN \citep{Zu2011}.}

\appendix \label{sec:appecdix}
\section{Examining the Impact of Reverberation Effect on Velocity-resolved BD} \label{sec:appendix2}

\begin{figure*}[!ht]%
\centering
\includegraphics[width=0.33\textwidth]{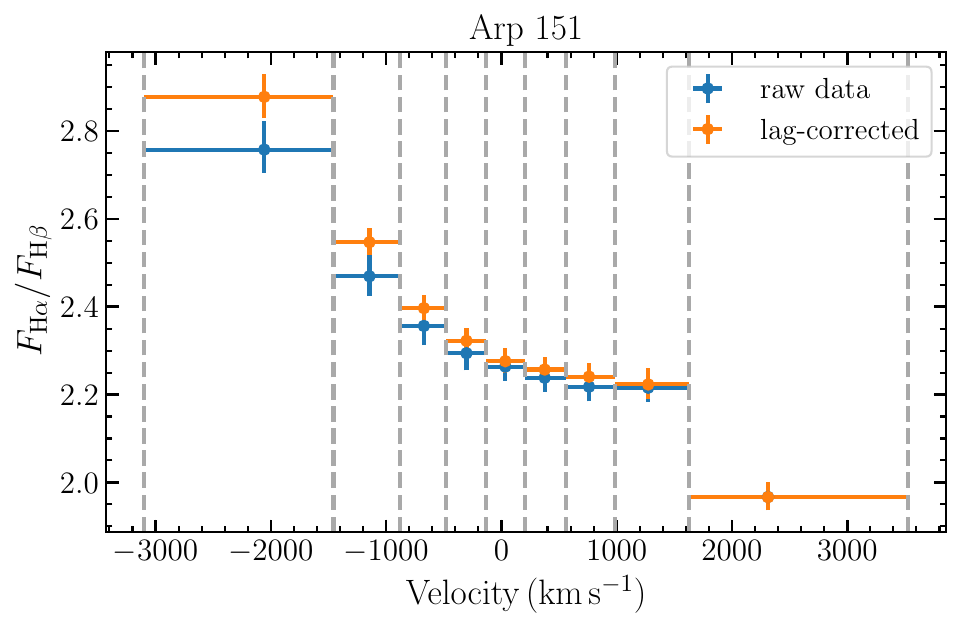}
\includegraphics[width=0.33\textwidth]{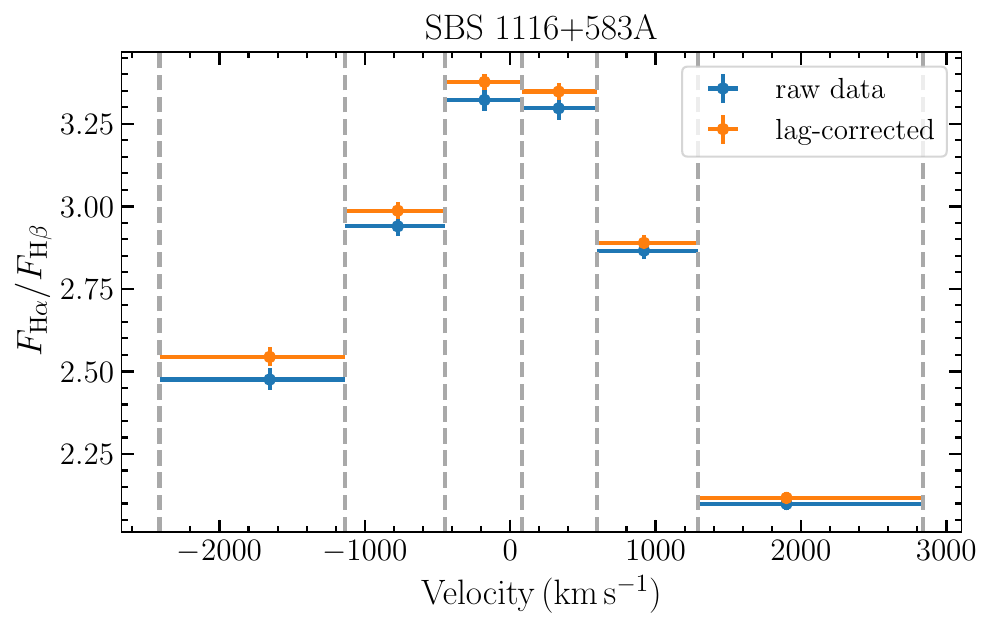}
\includegraphics[width=0.33\textwidth]{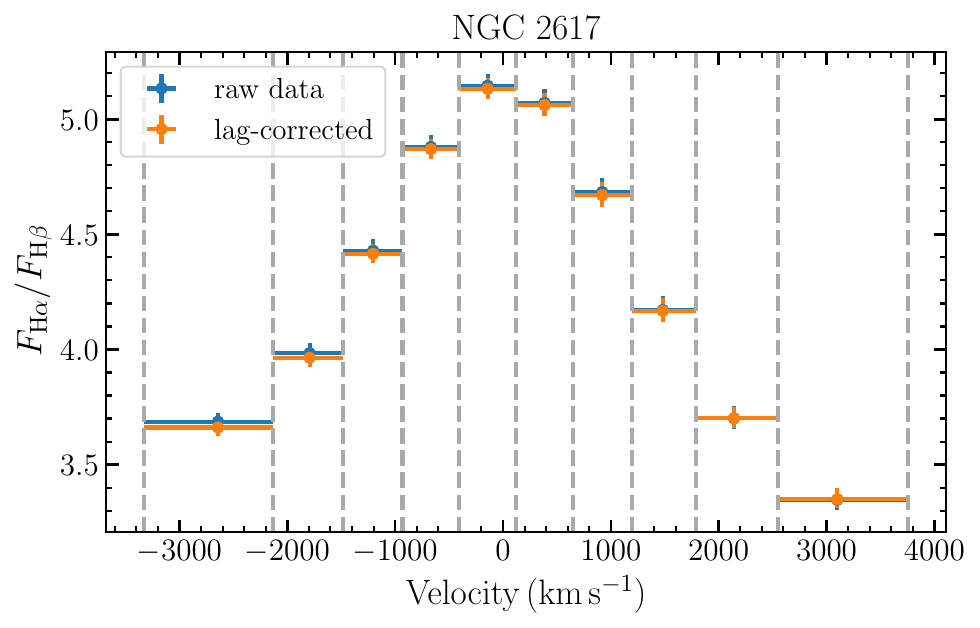}
\includegraphics[width=0.33\textwidth]{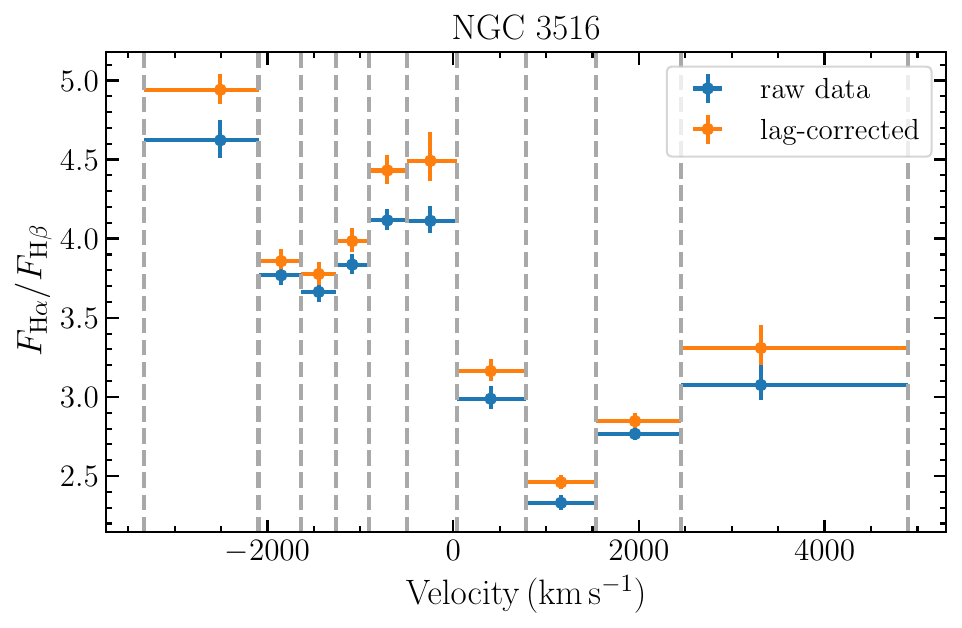}
\includegraphics[width=0.33\textwidth]{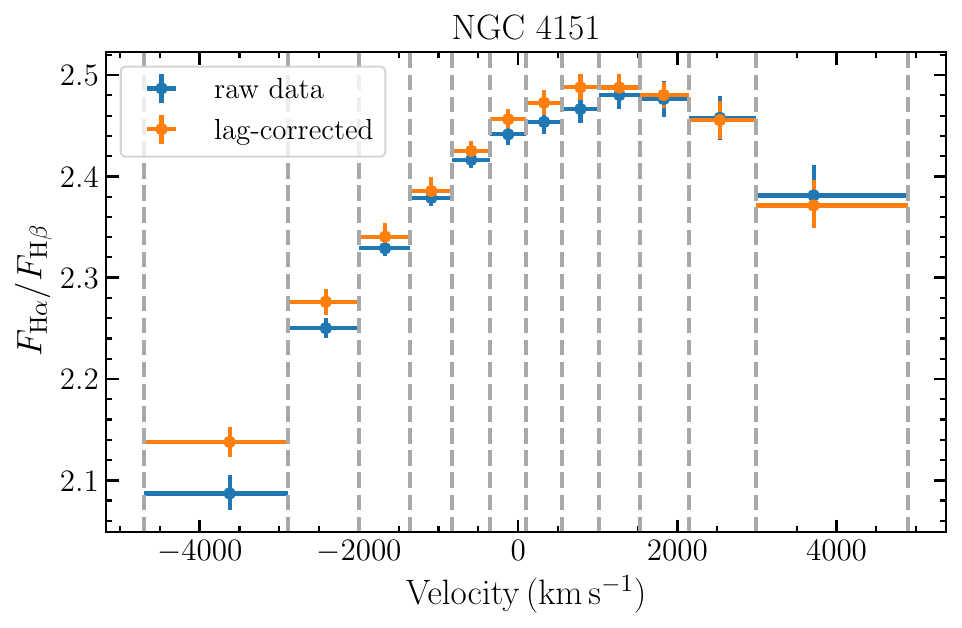}
\caption{Velocity-resolved emission line flux ratios for Arp 151, SBS 1116+583A, NGC 2617, NGC 3516 and NGC 4151. In each panel, the blue points represent the flux ratios directly measured from \ha\ and \hb\ light curves within each velocity bin. The orange points show the flux ratios of light curves that have been adjusted for the time lags of \ha\ and \hb\ relative to the continuum within each velocity bin. The vertical dashed lines indicate the boundaries of the velocity bins, same as shown in Figures~\ref{fig:vr} and~\ref{fig:vr2}.}
\label{fig:11}
\end{figure*}

To examine the reverberation effect in velocity-resolved flux ratios (BDs), we conducted the relevant tests. Initially, we measured the light curves of \ha\ and \hb\ within individual velocity bins. The direct flux ratios were determined from these light curves, and the average BDs for each bin are depicted as blue points in Figure~\ref{fig:11}. Subsequently, we analyzed the time lags of \ha\ and \hb\ in each velocity bin relative to the continuum using the CCF method. This allowed us to correct the time lags of each emission line relative to the continuum in each velocity bin, resulting in lag-corrected emission line light curves. The corresponding flux ratios were then calculated in the same manner as before, depicted as orange points in Figure~\ref{fig:11}. Our findings revealed that the average BD results and trends remained largely consistent between the two methods.

\section{Examination of Dust Extinction Variability} \label{sec:appendix1}
\begin{figure*}[!ht]
\centering
\includegraphics[width=0.245\textwidth]{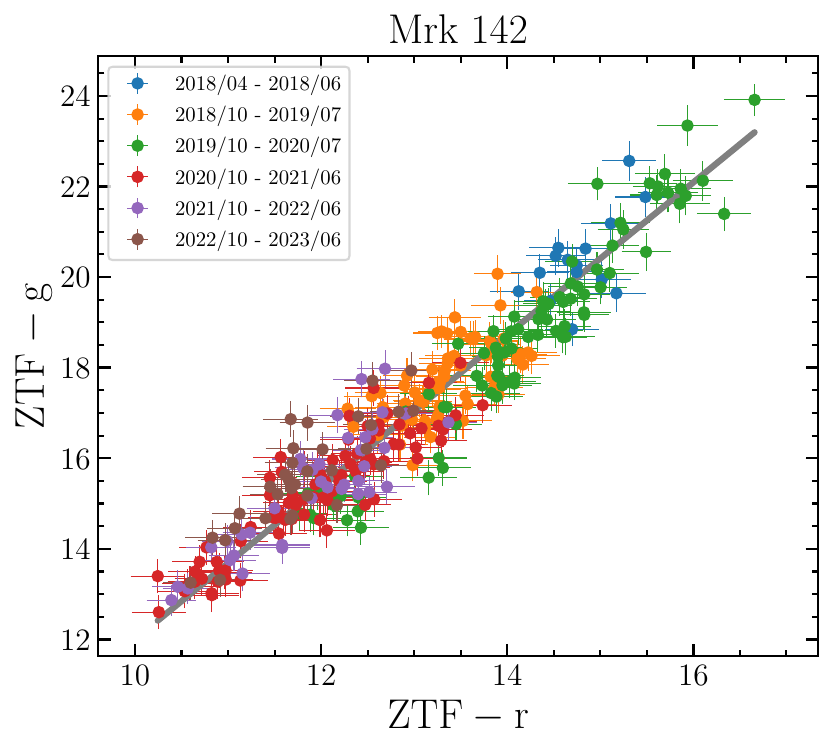}
\includegraphics[width=0.245\textwidth]{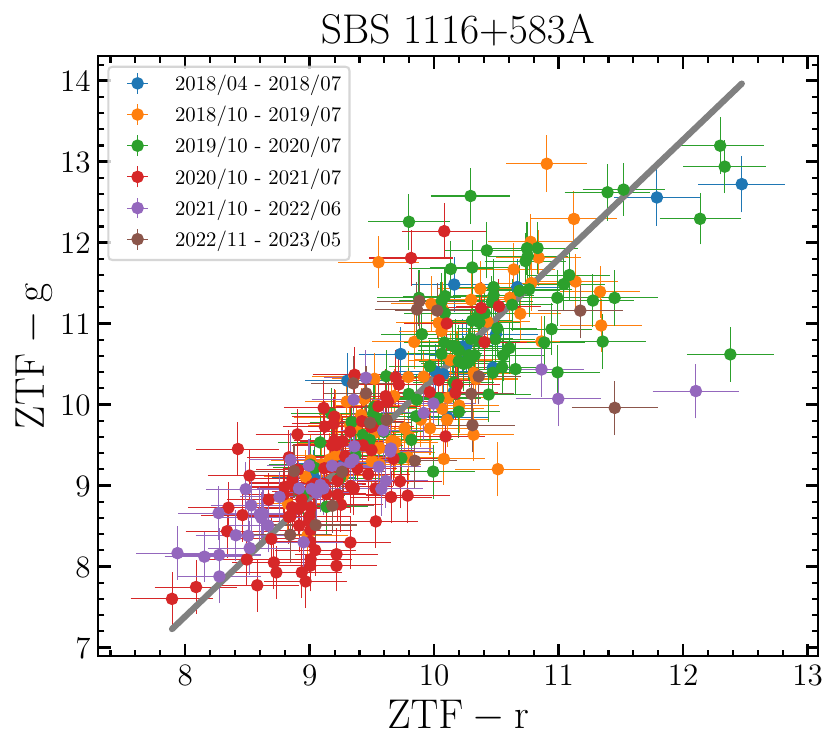}
\includegraphics[width=0.245\textwidth]{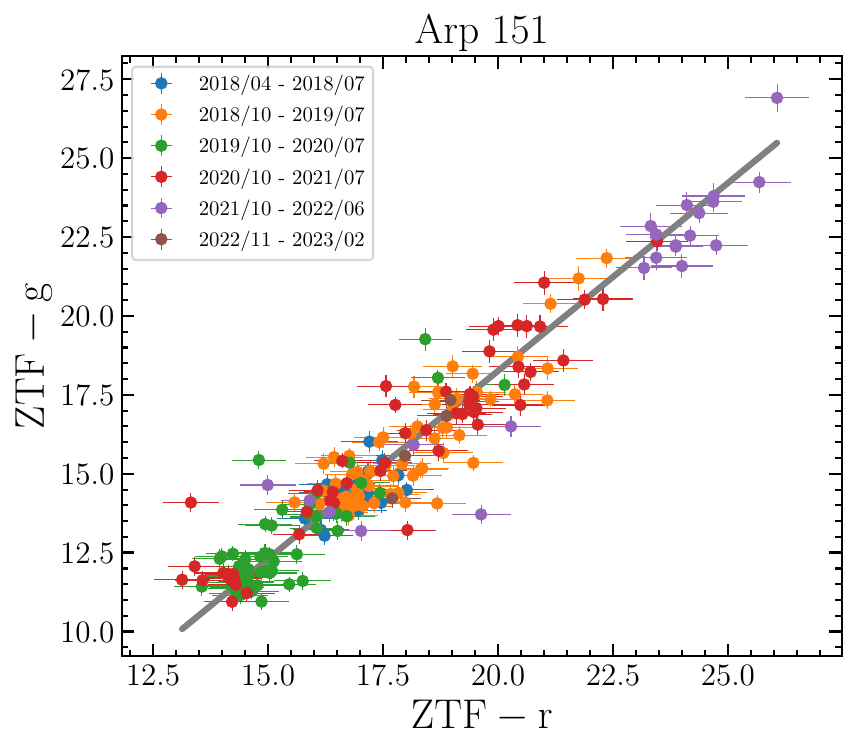}
\includegraphics[width=0.245\textwidth]{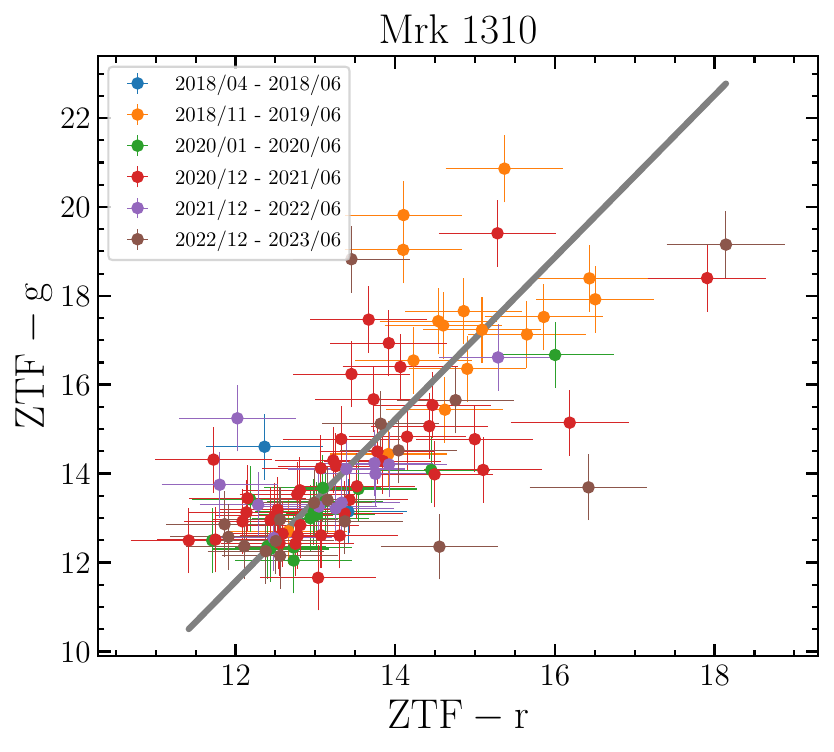}
\includegraphics[width=0.245\textwidth]{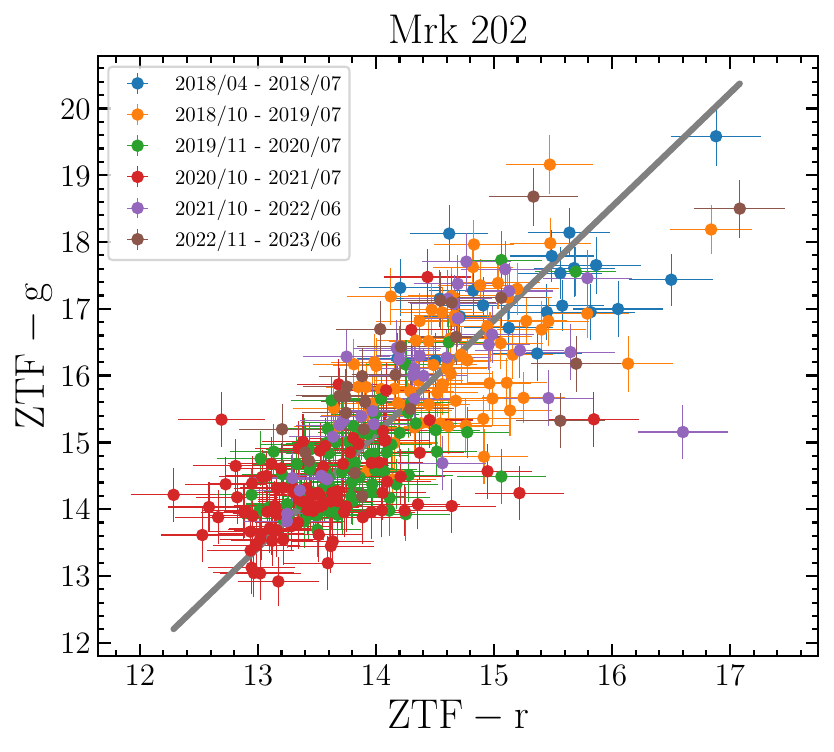}
\includegraphics[width=0.245\textwidth]{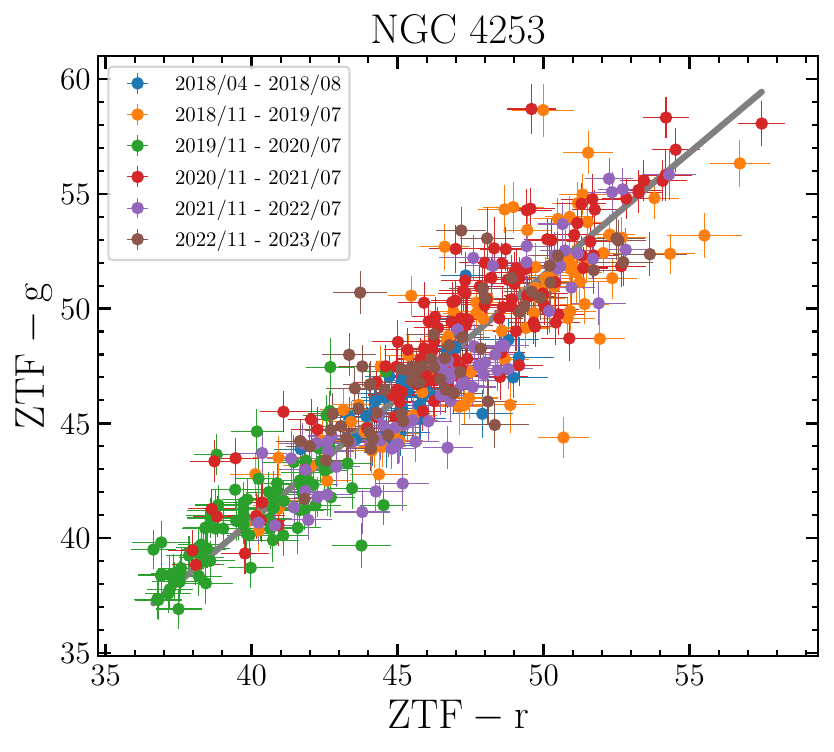}
\includegraphics[width=0.245\textwidth]{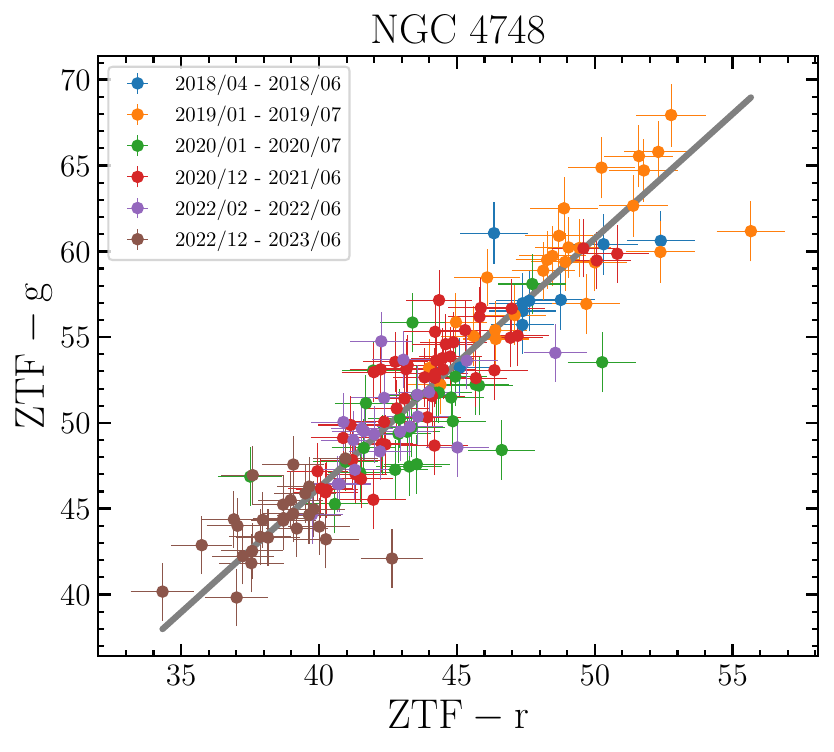}
\includegraphics[width=0.245\textwidth]{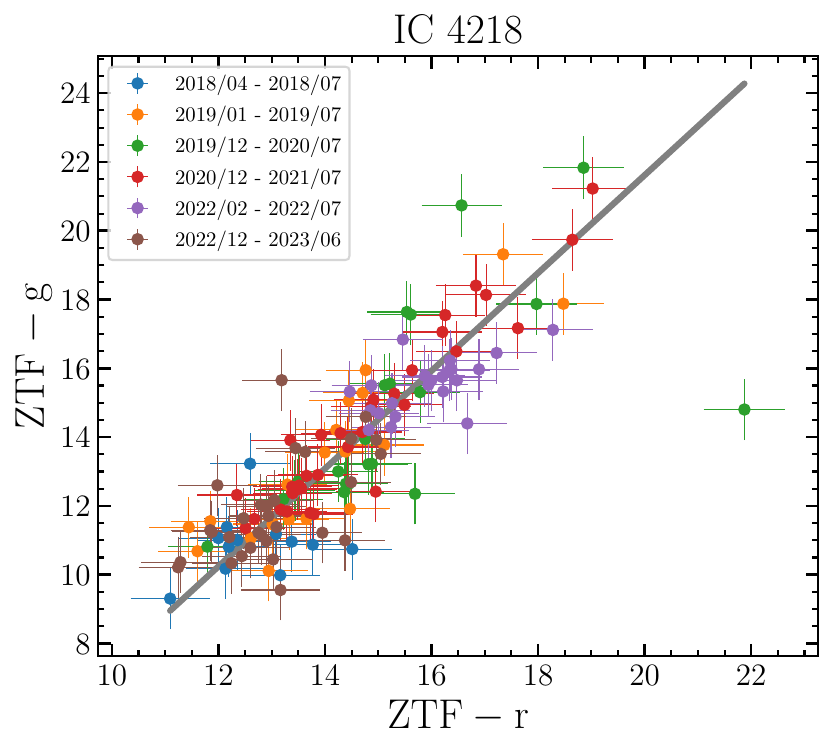}
\includegraphics[width=0.245\textwidth]{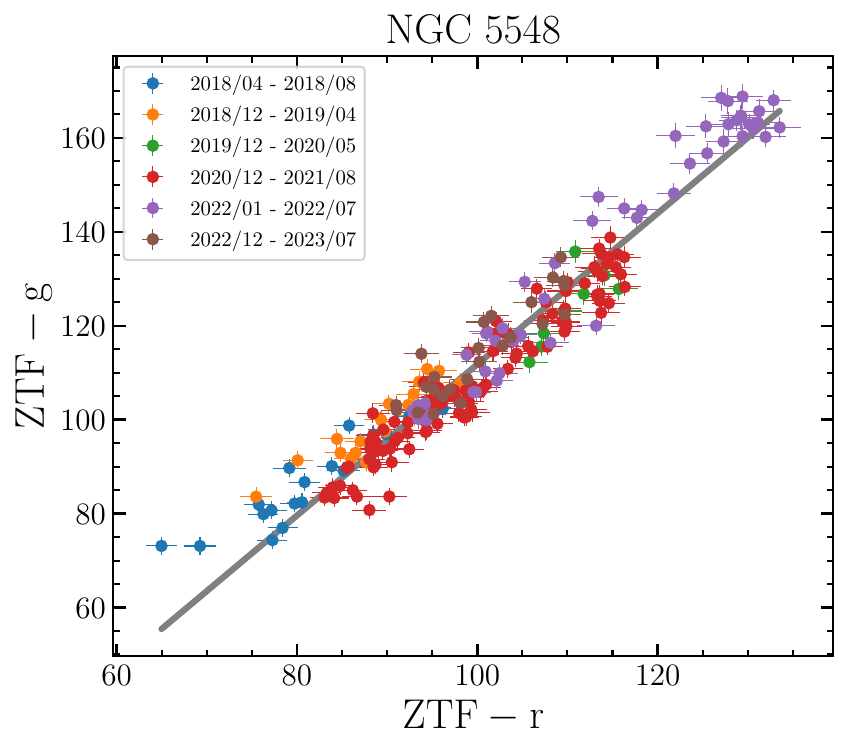}
\includegraphics[width=0.245\textwidth]{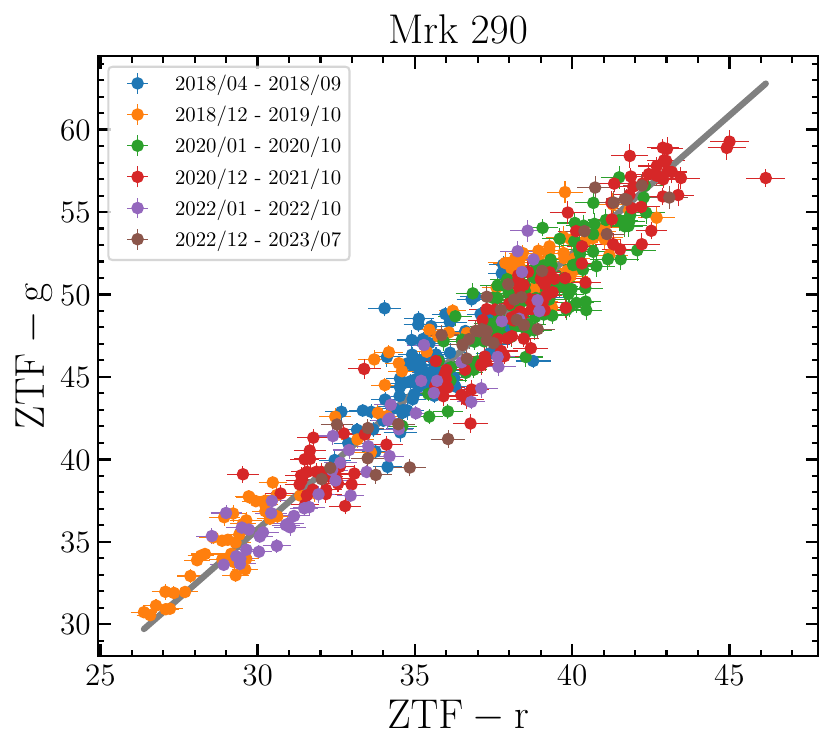}
\includegraphics[width=0.245\textwidth]{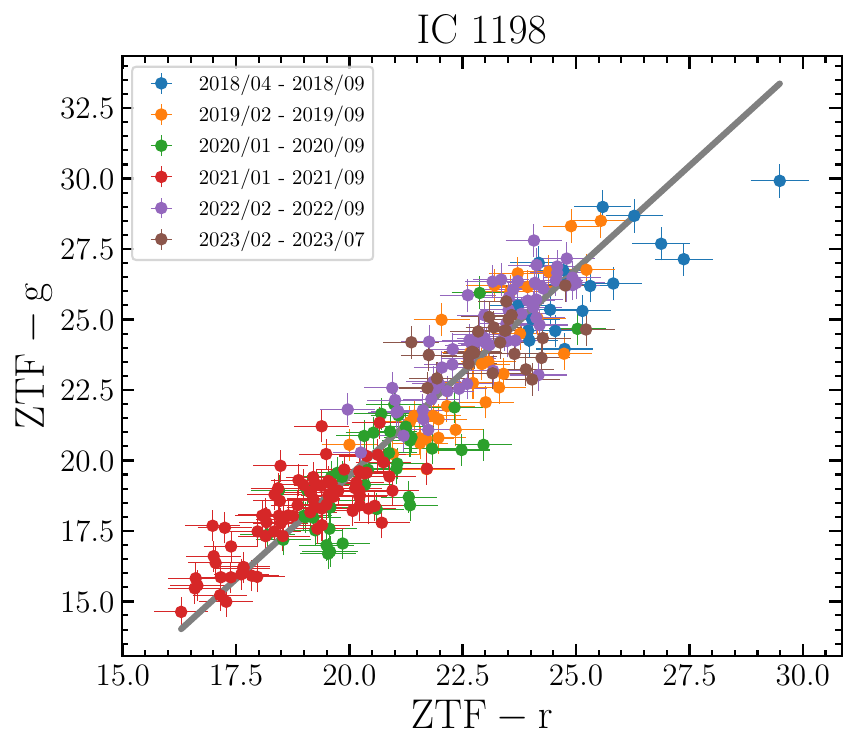}
\includegraphics[width=0.245\textwidth]{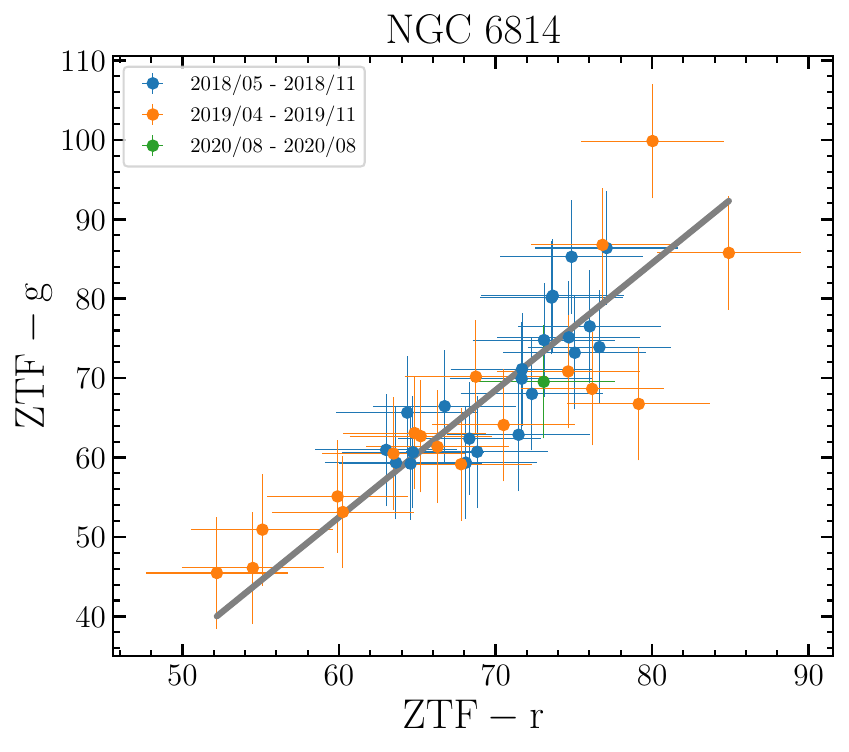}
\includegraphics[width=0.245\textwidth]{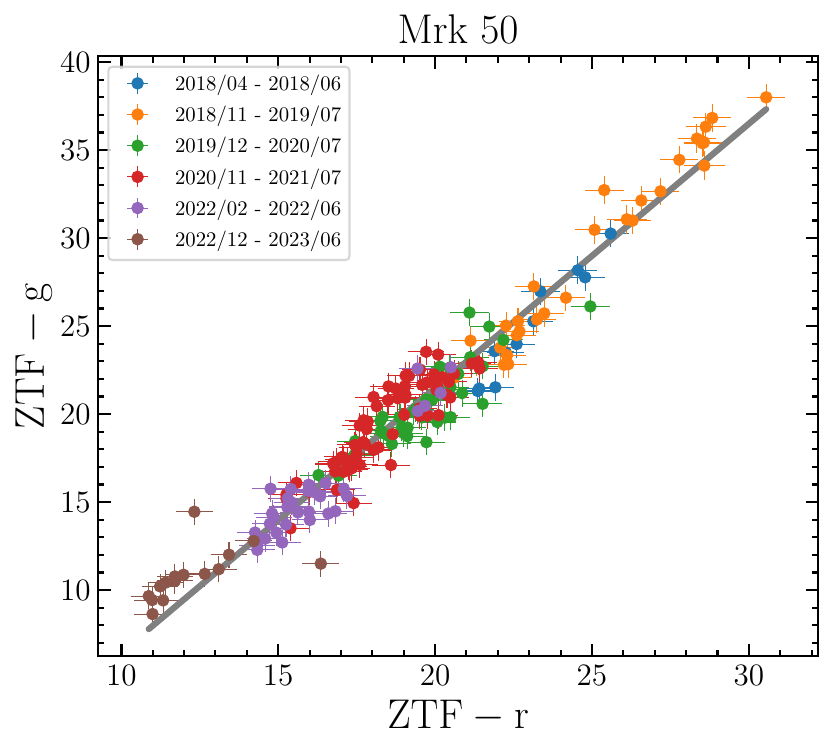}
\includegraphics[width=0.245\textwidth]{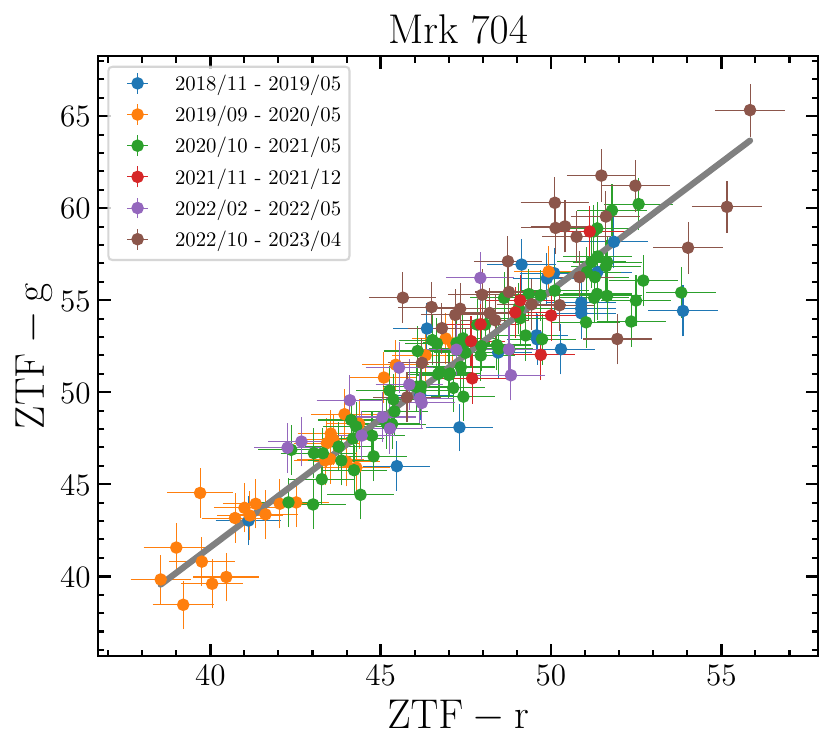}
\includegraphics[width=0.245\textwidth]{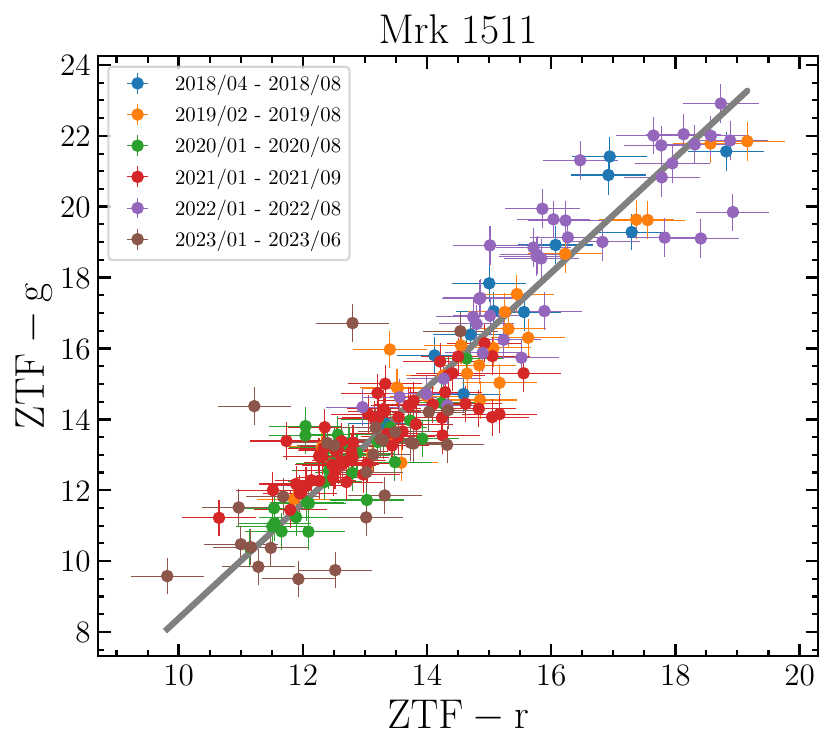}
\includegraphics[width=0.245\textwidth]{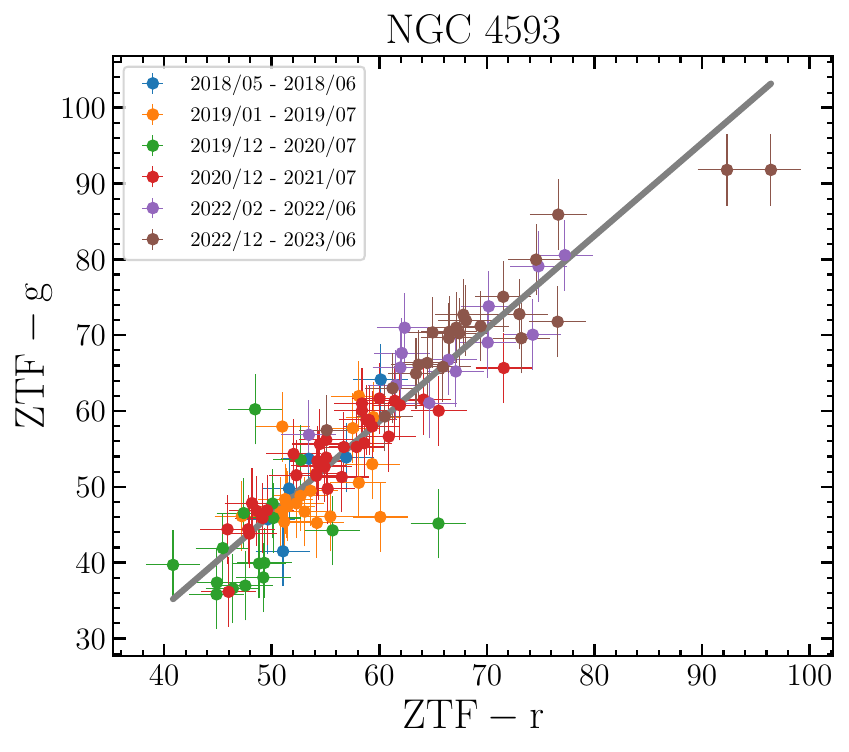}
\includegraphics[width=0.245\textwidth]{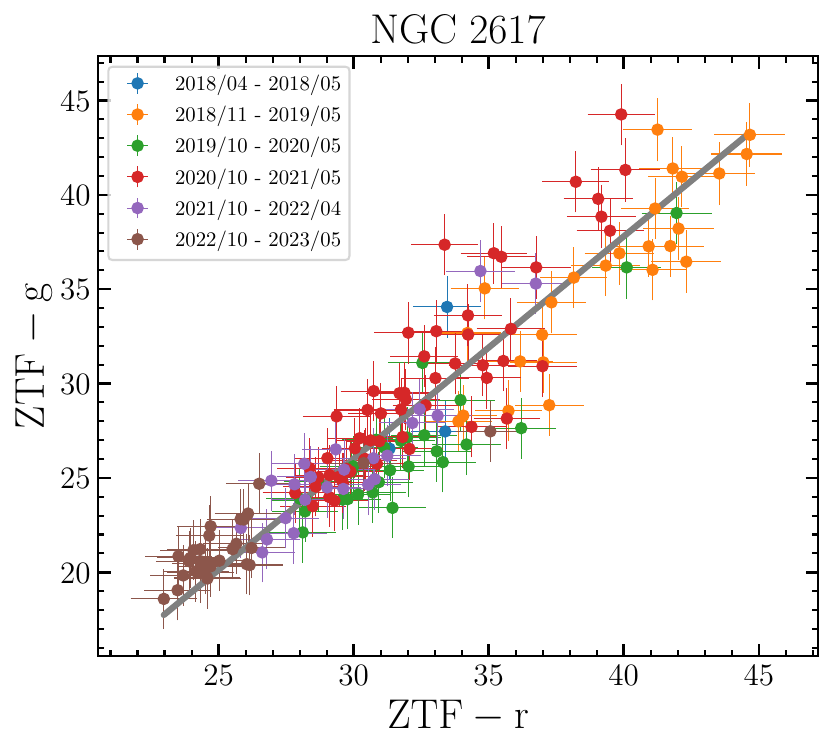}
\includegraphics[width=0.245\textwidth]{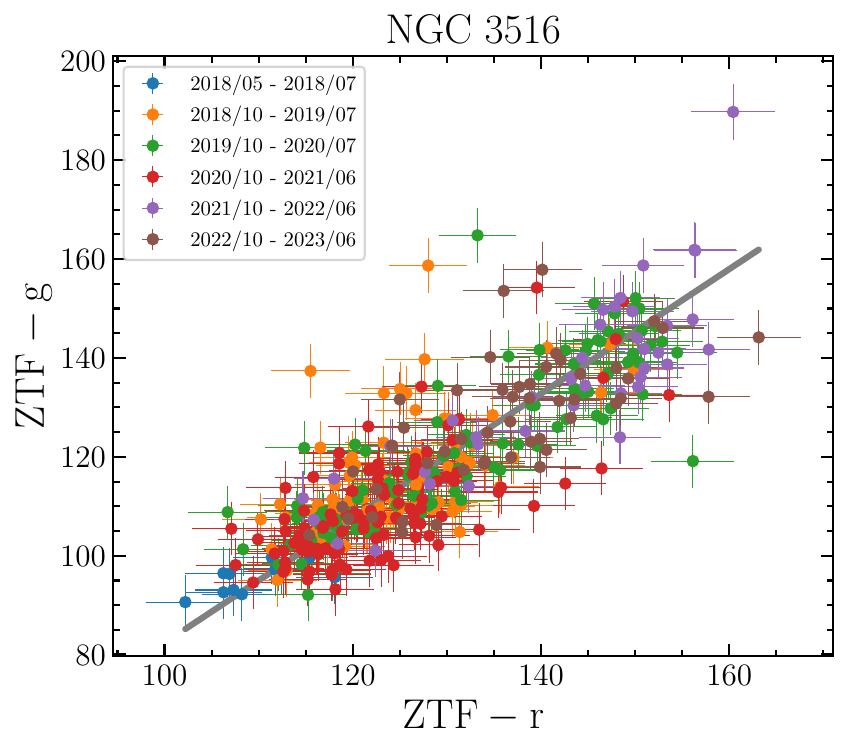}
\includegraphics[width=0.245     \textwidth]{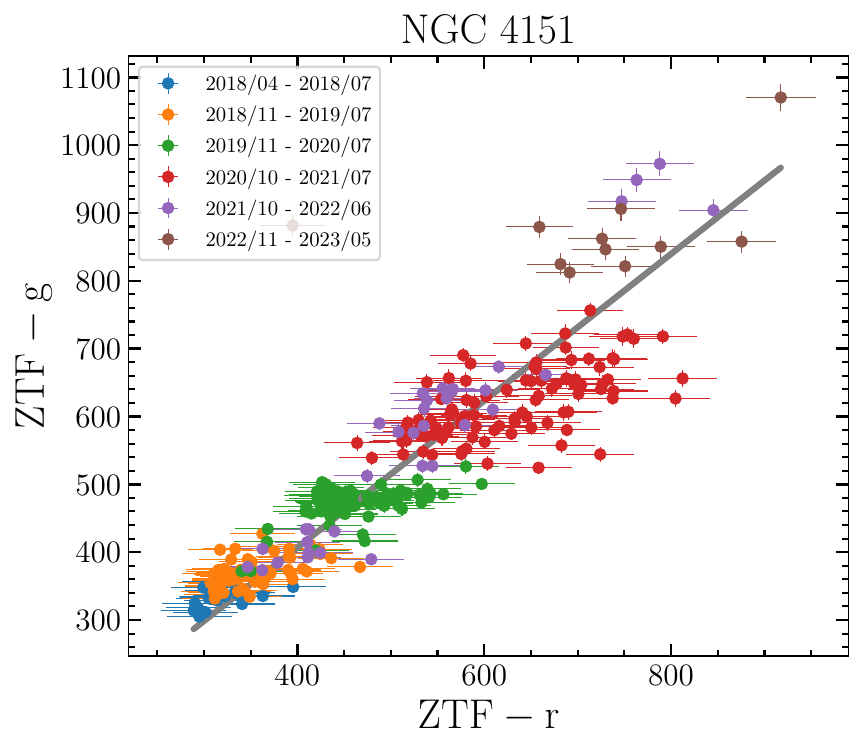}
\caption{The flux-to-flux diagrams between ZTF-g and ZTF-r bands. The data samples are from LAMP2008, LAMP2011, and Lijiang projects. 19 out of 22 AGNs were included in the analysis (3 AGNs lacking data). The colored points represent different observational times, and the grey line represents the linear fitting result.
}
\label{fig:grslope}
\end{figure*}

Here, we implemented two distinct approaches to investigate potential changes in the reddening within AGNs. The first approach utilizes flux-flux diagrams constructed from multi-band light curves. These diagrams enable us to measure the slopes between light curves in two bands, providing a means to assess extinction as outlined by \citet{Winkler1997}. By tracking the evolution of these slopes over time, we are able to identify any shifts in reddening assuming that the AGN power-law index is constant. 

For this initial approach, we utilized observational data from the ZTF, which conducts time-domain observations using a 48-inch telescope across the ZTF-g, ZTF-r, and ZTF-i bands \citep{Bellm2019}. Due to the relatively sparse data in the ZTF-i band, we focused our analysis on light curves from the ZTF-g and ZTF-r bands. We paired ZTF-g and ZTF-r band light curves with time intervals of less than 0.8 days and converted them to fluxes based on the AB magnitude system. Then, we corrected for Galactic extinction based on dust maps \citep{Schlafly2011} and band-specific extinction coefficients \citep{Yang2023}. Using the Least Squares Method, we fitted these paired light curves with a linear regression. As shown in Figure~\ref{fig:grslope}, the highly linear relationship and lack of curvature in the flux-flux diagram for 19 AGNs (with 3 lacking sufficient data) suggests negligible reddening variations over month-long timescales.

\begin{figure*}[!ht]%
\centering
\includegraphics[width=0.245\textwidth]{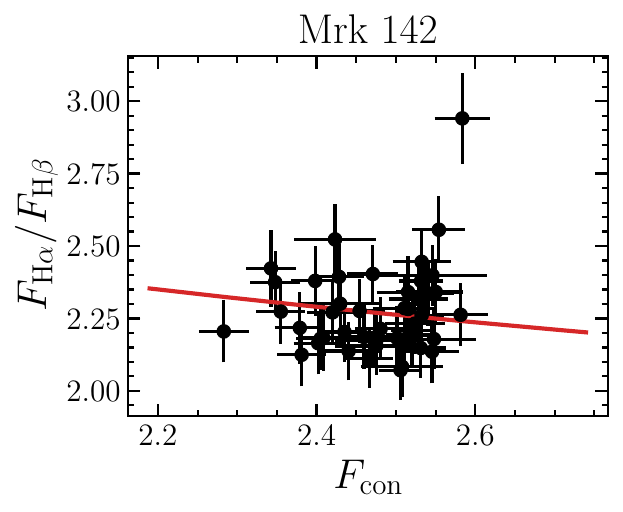}
\includegraphics[width=0.245\textwidth]{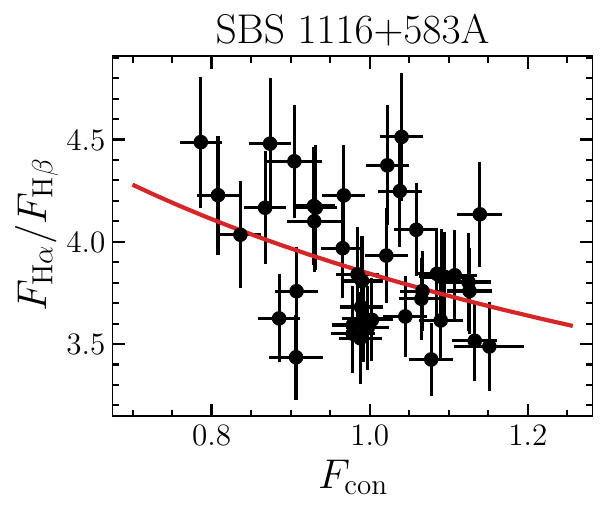}
\includegraphics[width=0.245\textwidth]{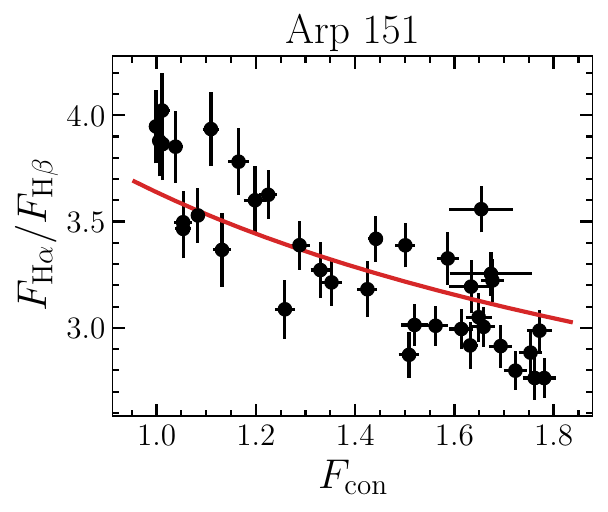}
\includegraphics[width=0.245\textwidth]{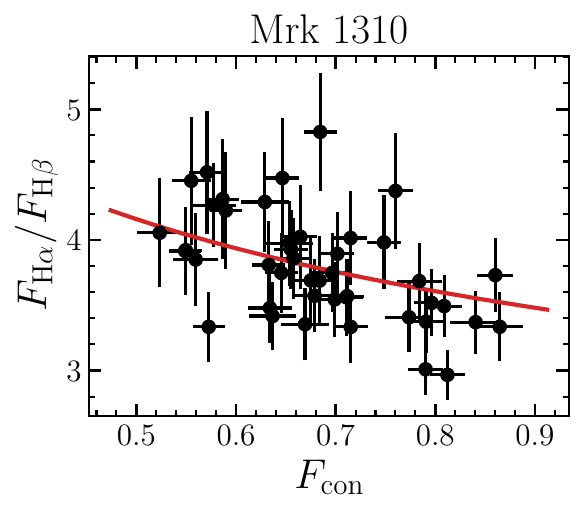}
\includegraphics[width=0.245\textwidth]{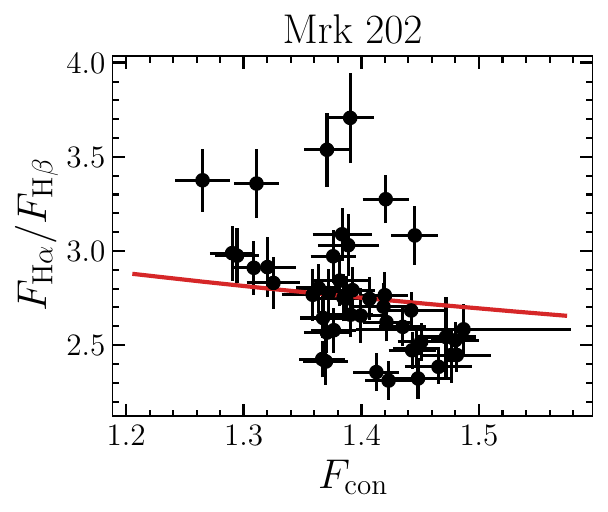}
\includegraphics[width=0.245\textwidth]{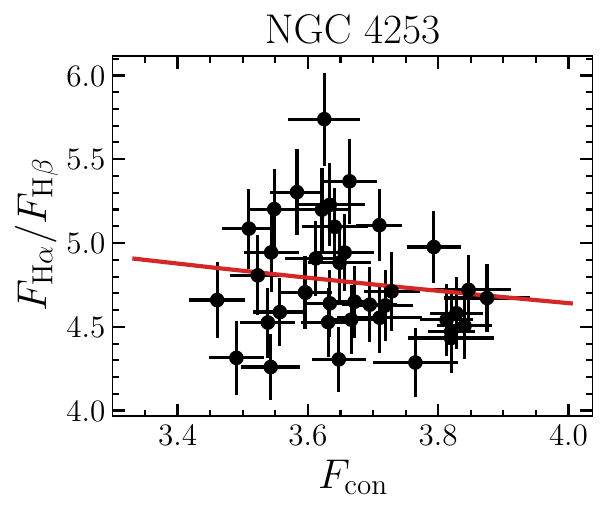}
\includegraphics[width=0.245\textwidth]{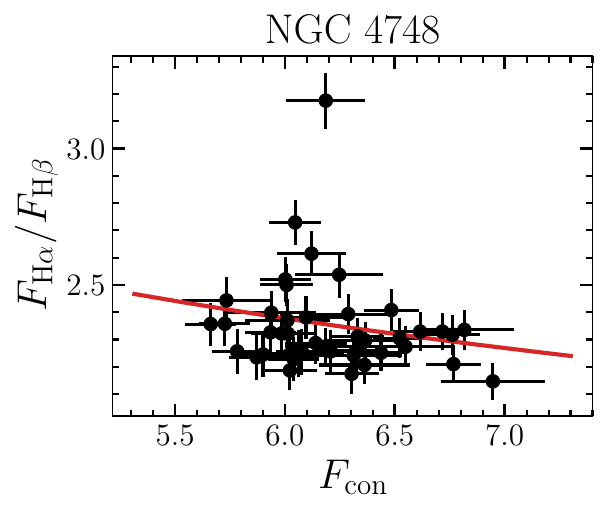}
\includegraphics[width=0.245\textwidth]{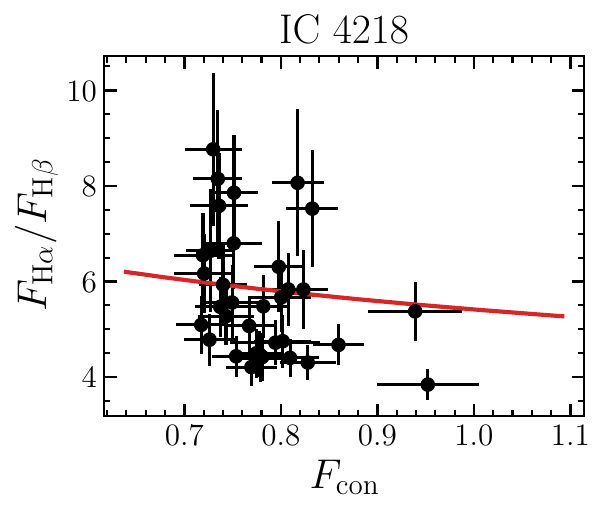}
\includegraphics[width=0.245\textwidth]{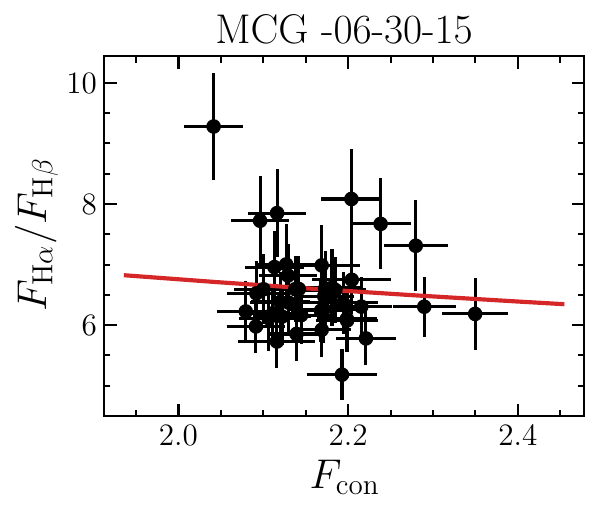}
\includegraphics[width=0.245\textwidth]{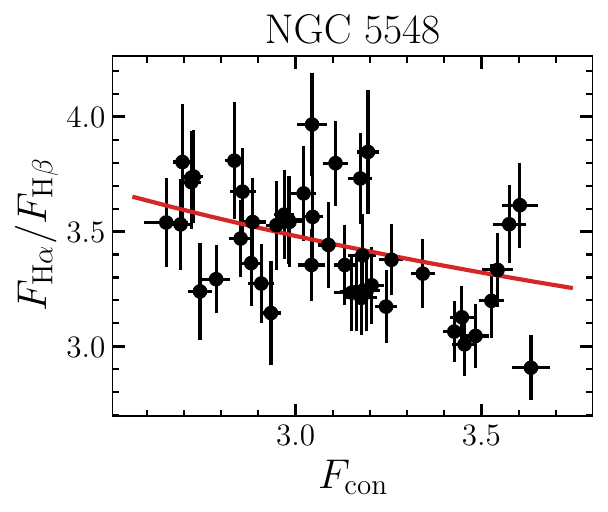}
\includegraphics[width=0.245\textwidth]{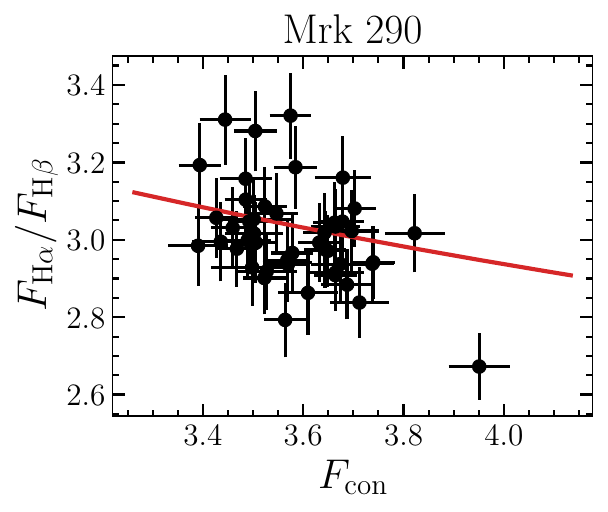}
\includegraphics[width=0.245\textwidth]{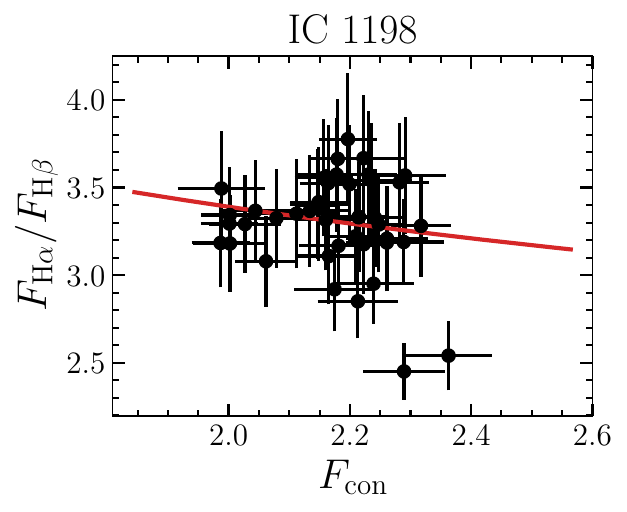}
\includegraphics[width=0.245\textwidth]{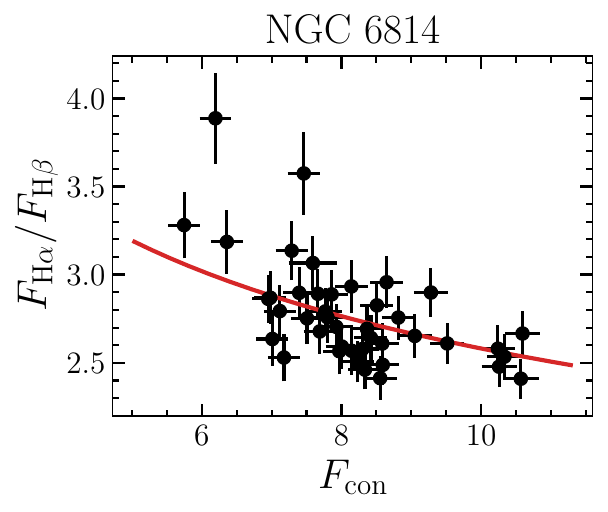}
\includegraphics[width=0.245\textwidth]{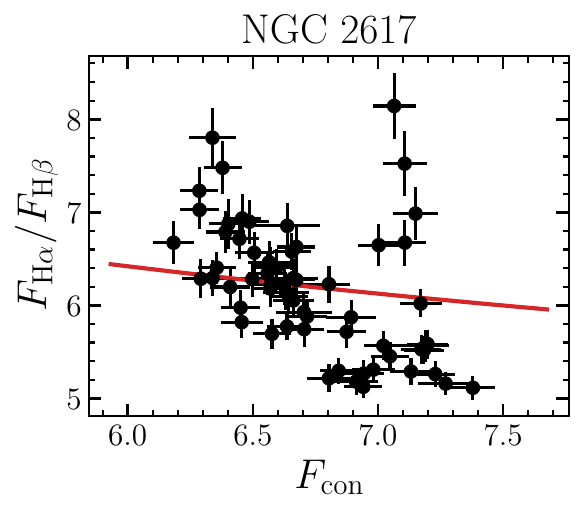}
\includegraphics[width=0.245\textwidth]{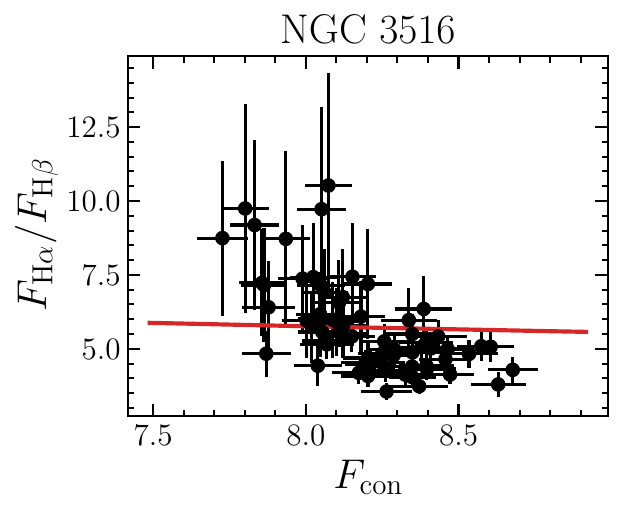}
\includegraphics[width=0.245\textwidth]{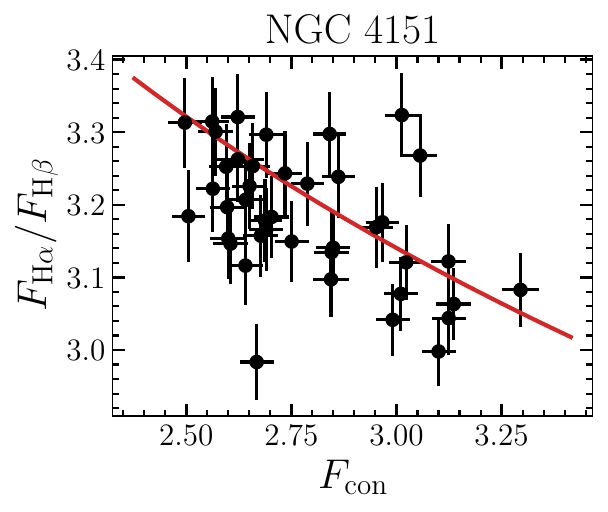}
\caption{The simulated result between emission line flux ratio and continuum. The data employed for this analysis originate from LAMP2008 and Lijiang projects. In the figures, the observed data are depicted as black points. Operating under the assumption that changes in the continuum result from dust extinction fluctuations, we have modeled the expected \fab, which is represented by the red lines.}
\label{fig:metricBD}
\end{figure*}

The second method relies on data from LAMP2008 and Lijiang samples, with the LAMP2011 dataset contributing narrow line components to the light curves of \ha\ and \hb. Under the assumption that continuum variations are driven by fluctuating dust extinction, we utilized the extinction curve from \citet{Fitzpatrick1999} to estimate the corresponding reddening and to model the expected \fab. The results are illustrated in Figure~\ref{fig:metricBD}, where the red lines represent our simulated expectations. The poor agreement between modeled and observed \fab\ (Figure~\ref{fig:metricBD}) implies that variations in extinction alone may not explain the observed changes in the continuum and the variability of \fab.

\end{document}